\documentclass[aps,prd,10pt,twocolumn,superscriptaddress,showpacs,longbibliography,nofootinbib,preprintnumbers]{revtex4-1}

\pdfoutput=1
\usepackage{graphicx}
\usepackage{amsfonts,amsmath,amssymb,bm,bbm}
\usepackage{color}
\usepackage{enumitem}
\usepackage{url}  

\usepackage{hyperref}
\usepackage[capitalize]{cleveref}
\usepackage[dvipsnames, usenames]{xcolor}
\usepackage{orcidlink}

\usepackage{graphicx}
\newcommand{\orcid}[1]{\href{https://orcid.org/#1}{\includegraphics[width=10pt]{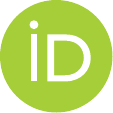}}}
\newcommand{\github}[1]{\href{https://github.com/#1}{\includegraphics[width=10pt]{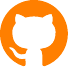}}}

\usepackage{aas_macros}

\definecolor{romared}{RGB}{142,0,28}
\usepackage{hyperref}
\hypersetup{
    pdfnewwindow=true,      
    colorlinks=true,       
    linkcolor=romared,          
    citecolor=romared,        
    filecolor=romared,      
    urlcolor=romared        
}

\def\bi{\begin{itemize}[noitemsep,leftmargin=*]
\setlength\itemsep{1em}
        }
\def\ei{\end{itemize}}

\usepackage{lipsum}
\usepackage{diagbox}

\newcommand{\pd}{\partial}  
\newcounter{ichi}
\newcounter{ni}
\newcounter{san}
\newcounter{yon}
\begin{document}
\preprint{FERMILAB-PUB-26-0016-T}

\title{IceCube's convex all-sky neutrino spectrum consistent with the magnetically powered corona scenario for active galactic nuclei}

\author{Kohta Murase \orcid{0000-0002-5358-5642}}
\affiliation{Department of Physics, The Pennsylvania State University, University Park, Pennsylvania 16802, USA}
\affiliation{Department of Astronomy and Astrophysics, The Pennsylvania State University, University Park, Pennsylvania 16802, USA}
\affiliation{Institute for Gravitation and the Cosmos, The Pennsylvania State University, University Park, Pennsylvania 16802, USA}
\affiliation{Center for Gravitational Physics and Quantum Information, Yukawa Institute for Theoretical Physics, Kyoto, Kyoto 606-8502 Japan}

\author{Shigeo S. Kimura \orcid{0000-0003-2579-7266}}

\affiliation{Frontier Research Institute for Interdisciplinary Sciences, Tohoku University, Sendai 980-8578, Japan}
\affiliation{Astronomical Institute, Graduate School of Science, Tohoku University, Sendai 980-8578, Japan}

\author{Mainak Mukhopadhyay \orcid{0000-0002-2109-5315}}
\affiliation{Astrophysics Theory Department, Theory Division, Fermi National Accelerator Laboratory, Batavia, Illinois 60510, USA}
\affiliation{Kavli Institute for Cosmological Physics, University of Chicago, Chicago, Illinois 60637, USA}
\affiliation{Department of Physics, The Pennsylvania State University, University Park, Pennsylvania 16802, USA}
\affiliation{Department of Astronomy and Astrophysics, The Pennsylvania State University, University Park, Pennsylvania 16802, USA}
\affiliation{Institute for Gravitation and the Cosmos, The Pennsylvania State University, University Park, Pennsylvania 16802, USA}

\author{Mukul Bhattacharya}
\affiliation{Department of Astronomy, Astrophysics and Space Engineering, Indian Institute of Technology Indore, Simrol, MP 453552, India}
\affiliation{Department of Physics, Wisconsin IceCube Particle Astrophysics Center, University of Wisconsin, Madison, WI 53703, USA}
\affiliation{Department of Physics, The Pennsylvania State University, University Park, Pennsylvania 16802, USA}
\affiliation{Department of Astronomy and Astrophysics, The Pennsylvania State University, University Park, Pennsylvania 16802, USA}
\affiliation{Institute for Gravitation and the Cosmos, The Pennsylvania State University, University Park, Pennsylvania 16802, USA}

\date{submitted 30 Dec 2025}

\begin{abstract}
High-energy multimessenger background analyses over the past decade have provided evidence for a population of hidden neutrino sources that are opaque to GeV--TeV gamma rays, a picture bolstered by recent observations of the nearby active galaxy NGC 1068. The coronal regions in the hearts of active galactic nuclei (AGNs) have been proposed as the most promising sites for such hidden nonthermal particle production, and NGC 1068 is expected to be the most neutrino-active galaxy for IceCube. We demonstrate that the latest all-sky neutrino spectrum, exhibiting a spectral bend around 3--30~TeV, is consistent with predictions of the magnetically powered corona scenario,  
and the models for the all-sky neutrino flux can simultaneously explain the multimessenger data from NGC 1068 within observational and modeling uncertainties. We further show, in a largely model-independent way, that the contribution from NGC 1068-like sources does not overshoot the observed medium-energy neutrino flux. Finally, we highlight the key role of the Eddington ratio, which can drive substantial variations in the predicted neutrino fluxes of nearby AGNs, and we encourage systematic multimessenger searches for the neutrino-brightest AGNs.
\end{abstract}

\maketitle

\section{Introduction}
Over the past decade, the IceCube Collaboration has established a large all-sky flux of astrophysical neutrinos in the medium-energy ($\sim1-100$~TeV) range~\cite{IceCube:2020acn,IceCube:2024fxo,IceCube:2025tgp}. 
Detailed multimessenger analyses through comparison with the nonblazar component of the extragalactic gamma-ray background~\cite{TheFermi-LAT:2015ykq} have shown the need for a population of neutrino sources that are hidden in GeV--TeV gamma rays~\cite{Murase:2015xka,Capanema:2020rjj,Fang:2022trf}. The sources are predominantly extragalactic, which has also been supported by recent observations of diffuse Galactic neutrinos~\cite{IceCube:2023ame}.
Contrary to the high-energy neutrino flux above $\sim100$~TeV~\cite{Aartsen:2013bka,Aartsen:2013jdh}, the large all-sky flux at medium energies, $E_\nu^2\Phi_\nu \sim {10}^{-7}~{\rm GeV}~{\rm cm}^{-2}~{\rm s}^{-1}~{\rm sr}^{-1}$~\cite{IceCube:2020acn}, violates the nucleon-survival bound~\cite{Waxman:1998yy} as well as the nucleus-survival bound~\cite{Murase:2010gj}, which is consistent with the argument that hidden neutrino sources are not ideal sites for ultrahigh-energy cosmic-ray (UHECR) production~\cite{Murase:2015xka}.  

The hearts of active galactic nuclei (AGNs) are the most promising sites for medium-energy neutrino production. AGNs are known to be dominant in the x-ray and gamma-ray sky, and their large black hole accretion luminosity density makes them most energetically favorable sources for the primary origin of cosmic neutrinos in the $10-100$~TeV range~\cite{Murase:2018utn}. Murase, Kimura and M\'esz\'aros 2020~\cite{Murase:2019vdl} (hereafter MKM20) suggested that ions can be energized to relativistic energies by magnetic reconnections and turbulence in magnetically powered coronae and then interact with coronal plasma and ambient photons from the disk and corona. They are naturally hidden neutrino sources, in which GeV--TeV gamma rays are cascaded down to MeV energies through electromagnetic interactions with optical, ultraviolet and x-ray photons. IceCube's association of NGC~1068 with multi-TeV neutrinos~\cite{IceCube2022NGC1068,Murase:2022azo}, together with its low flux of GeV--TeV gamma rays from the starburst origin~\cite{Ajello:2023hkh}, indeed provides evidence for the existence of such hidden cosmic-ray accelerators embedded in dense and opaque regions, and Ref.~\cite{Murase:2022dog} has shown that particle acceleration and associated neutrino production are likely to occur near supermassive black holes (SMBHs). Recent developments in observational searches~\cite{IceCube:2024dou,IceCube:2024ayt,IceCube:2025lev,IceCube:2025ssu,Abbasi:2025tas,IceCube:2026hzq,Neronov:2023aks} and theoretical turbulent-corona modeling (see recent reviews~\cite{Murase:2022feu,Padovani:2024ibi} and references therein) further support the viability of coronae as the primary sources of TeV--PeV neutrinos. It is intriguing that, in addition to NGC~1068, $\sim3\sigma$ excesses have been reported for NGC~4151 that is expected to be the second brightest neutrino active galaxy in IceCube's northern sky and the Circinus Galaxy that would be the brightest in the southern sky~\cite{Murase:2023ccp}.   

\begin{figure*}[th]
\includegraphics[width=\linewidth]{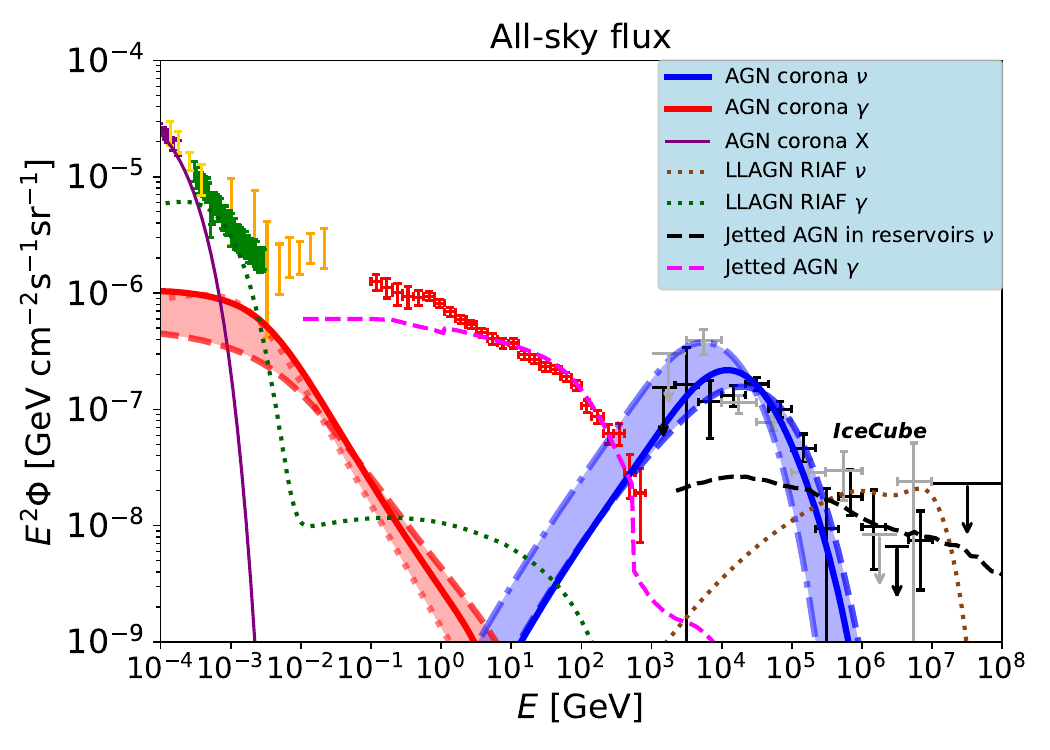}
\caption{Summary of AGN models for all-sky neutrino and electromagnetic background fluxes. The original MKM20 model (Model A)~\cite{Murase:2019vdl} is shown by the solid curves (blue, red and purple). Model B and Model C (see text for details) are optimized for the latest IceCube data of medium-energy starting events~\cite{IceCube:2025tgp} (black) and enhanced starting track event selection~\cite{IceCube:2024fxo} (gray), respectively. The x-ray and soft gamma-ray data are taken from {\it Swift} BAT~\cite{Ajello:2008xb} (purple), Nagoya balloon~\cite{1975Natur.254..398F} (yellow), SMM~\cite{1997AIPC..410.1223W} (green), COMPTEL~\cite{2000AIPC..510..467W} (orange), and {\it Fermi} LAT~\cite{Ackermann:2014usa} (red). Thick curves (AGN corona) represent neutrinos, gamma rays and x-rays from coronae of jet-quiet AGNs. Neutrinos and gamma rays including Comptonized RIAF (radiatively inefficient accretion flow) emission from low-luminosity AGNs are taken from Ref.~\cite{Kimura:2020thg} (LLAGN RIAF). Neutrino emission from jetted AGNs embedded in galaxy clusters and groups~\cite{Fang:2017zjf} and the sum of gamma rays from these objects and blazars~\cite{Ajello:2013lka} are also depicted (jetted AGN).   
\label{fig:summary}
}
\end{figure*}

More recently, the IceCube Collaboration reported a convex spectrum of the diffuse high-energy neutrino background, which is characterized by a spectral break around $3-30$~TeV energies~\cite{IceCube:2024fxo,IceCube:2025tgp}. Such a structure is highly informative about the origin of medium-energy neutrinos. Indeed, a spectral turnover is one of the predictions of the magnetically powered corona model for the all-sky neutrino flux~\cite{Murase:2019vdl}, different from shock acceleration models~\cite{Inoue:2019fil,Inoue:2022yak,Murase:2022dog}. In this work, we demonstrate that a population of jet-quiet AGNs with parameters anchored to x-ray and optical observations reproduces IceCube's convex all-flavor spectrum within the framework of the corona model (see Fig.~\ref{fig:summary} for a summary plot). In particular, we quantify the allowed model parameters including the coronal magnetic field and the cosmic-ray power. We then provide systematic predictions for scaling relations between the x-ray luminosity and the neutrino/gamma-ray luminosity, and outline prospects for establishing a population of jet-quiet AGNs as the dominant sources of medium-energy neutrinos with an emphasis on the importance of prespecifying neutrino-active AGN catalogs. Throughout this work, we use $Q/Q_x=10^{x}$ in CGS units and $M/M_x=10^{x}~M_{\odot}$. We also consider the flat $\Lambda$CDM universe with $\Omega_m = 0.3$, $\Omega_\Lambda=0.7$ and $H_0 = 70$~km/s/Mpc (or $h=0.7$).

\section{Overview of the magnetically-powered corona model}
In this section, we review the framework of the magnetically powered corona model proposed by MKM20~\cite{Murase:2019vdl}. MKM20 assumes a magnetized, turbulent corona that presumably has a two-temperature structure with $T_p > T_e$ (where $T_p$ and $T_e$ are the proton and electron temperatures, respectively). Note that large-amplitude turbulence is assumed, and only the turbulent component is considered in the original phenomenological framework. Particle acceleration by turbulence is described by the Fokker-Planck approach, which approximately deals with both resonant and nonresonant (hard-sphere diffusion) cases (see, e.g., Refs.~\cite{Schlickeiser1989,ParkPetrosian1995,Becker:2006nz,Stawarz:2008sp,Murase:2011cx,Brunetti:2016oaw,Nishiwaki:2021axb}). We also note that the magnetically powered corona model is different from early models, such as the black hole envelope and tidal disruption models~\cite{Berezinsky1977}, and the accretion shock model~\cite{Stecker:1991vm}. 

Phenomenologically, we characterize jet-quiet AGNs by the intrinsic x-ray luminosity at the $2-10$~keV band ($L_X$). 
The bolometric luminosity ($L_{\rm bol}$) is expressed in terms of the Eddington ratio,
\begin{equation}
\lambda_{\rm Edd} \equiv \frac{L_{\rm bol}}{L_{\rm Edd}},
\end{equation}
where 
\begin{equation}
L_{\rm Edd}|_M\simeq4.0\times{10}^{45}~{\rm erg}~{\rm s}^{-1}~M_{7.5},
\end{equation}
is the Eddington luminosity that is a function of the SMBH mass $M$, and $L_{\rm bol}$ includes disk, corona, and dust-torus components. We typically expect $L_{\rm bol}/L_{X}\sim 20-30$. In MKM20, we give $L_{\rm bol}$ as a function of $L_X$ through the scaling given in Fig.~1 of Hopkins et al.~\cite{Hopkins:2006fq}. 

The Eddington ratio may govern properties of both the accretion flow and the corona. For instance, low-$\lambda_{\rm Edd}$ sources exhibit harder x-ray spectra and higher coronal temperatures. The mass accretion rate onto a SMBH is characterized by 
\begin{equation}
\dot{M} \approx \frac{L_{\rm bol}}{\eta_{\rm rad} c^2},
\end{equation}
where $\eta_{\rm rad} = 0.1$ is the radiative efficiency. 

In MKM20, the coronal region is phenomenologically characterized by five principal parameters, $L_X$, $\lambda_{\rm Edd}$, ${\mathcal R}\equiv R/R_{S}$, $\alpha$, and $\beta$, where ${\mathcal R}$ is the dimensionless coronal compactness, $R$ is the coronal radius, $R_S = 2GM/c^2\simeq3.0\times10^{12}~{\rm cm}~(M/10^7~M_\odot)$ is the Schwarzschild radius, $\alpha$ is the viscosity parameter, and $\beta$ is the plasma beta. The former two parameters, $L_X$ and $\lambda_{\rm Edd}$, can be determined by multiwavelength observations of AGNs.

We consider a corona that coexists with the standard disk that is geometrically thin and optically thick, composing of multitemperature blackbody regions. The setup of MKM20 is valid when the Eddington ratio is larger than the critical value,
\begin{equation}
\lambda_{\rm crit}\approx \eta_{\rm rad}3\alpha^2\simeq3\times{10}^{-3}~\eta_{\rm rad,-1}\alpha_{-1}^2 ,  
\end{equation}
below which we need to consider radiatively inefficient accretion flows (RIAFs) for modeling neutrino and gamma-ray emission from jet-quiet AGNs \cite{Kimura:2014jba,Kimura:2019yjo,Kimura:2020thg}. 
We also assume that the system is sub-Eddington or near-Eddington, i.e., $\lambda_{\rm Edd}\lesssim{\rm a~few}$. For super-Eddington systems where the photon trapping is relevant, the disk is geometrically thicker and coronal properties are expected to be different. 

\subsection{Spectral energy distributions of jet-quiet AGNs}
Spectral energy distributions (SEDs) are important for calculating neutrino and gamma-ray spectra. AGN typically display two main components in their electromagnetic output: the multitemperature blackbody emission from the accretion disk and the nonthermal high-energy tail attributed to Compton scattering by energetic electrons in the corona. In the SED model of MKM20, both these components are characterized by $L_X$ and $\lambda_{\rm Edd}$. 

The coronal x-ray spectrum is modeled as a power law with an exponential cutoff. The photon index and exponential cutoff energy depend on $\lambda_{\rm Edd}$~\cite{Ricci:2017dhj} via
\begin{equation}
\Gamma_X \approx 0.167 \log(\lambda_{\rm Edd}) + 2.0
\end{equation}
and
\begin{equation}
\varepsilon_{X,\rm cut} \approx -74\log(\lambda_{\rm Edd}) + 1.5\times10^2\ {\rm keV},
\end{equation}
and the coronal electron temperature is set by
\begin{equation}
T_e \approx \frac{\varepsilon_{X,\rm cut}}{2k_B},
\end{equation}
and the Thomson optical depth in the corona with a coronal scale height of $H$ (that is a function of $R$) is
\begin{equation}
\tau_T=\sigma_T n_e H \approx 10^{(2.16 - \Gamma_X)/1.06}\left( \frac{k_B T_e}{\rm keV} \right)^{-0.3},
\end{equation}
from which we obtain the electron (and positron) density in the corona, and $\sigma_T$ is the Thomson cross section and $n_e$ is the coronal number density of electrons. 

The disk SEDs are based on the average SED templates in Ref.~\cite{Ho:2008rf} with a cutoff at
\begin{equation}
\varepsilon_{\rm disk,cut} \approx 2.8\,k_B T_{\rm disk} \approx 1.37 k_B \left( \frac{L_{\rm bol}}{144\pi\sigma_{\rm SB} \eta_{\rm rad} R_S^2} \right)^{1/4},
\end{equation}
where $\sigma_{\rm SB}$ is the Stefan-Boltzmann constant and the disk inner edge is assumed to be $3R_S$.
The ultraviolet and x-ray parts are interpolated with an exponential cutoff, ensuring conservative photon densities for calculations of neutrino and gamma-ray spectra. To avoid the contamination from dust torus emission, we restrict our disk-corona SEDs to photons above 2~eV. The dust component in the infrared band can be important for the attenuation of gamma rays only at very high energies.   

In this work, the radiation energy density in our emission region is approximated to be 
\begin{equation}
U_{\rm disk+corona}\approx\frac{(1+\tau_T)(L_{\rm disk}+L_{\rm corona})}{2\pi {\mathcal R}^2 R_s^2 c},
\end{equation}
where $L_{\rm disk}$ is the total luminosity of optical and ultraviolet photons from the accretion disk and $L_{\rm corona}$ is the total luminosity of x rays radiated from the corona. Note that $L_{\rm disk}+L_{\rm corona}<L_{\rm bol}$ is satisfied. 

\subsection{Corona Regions}
X-ray emission from AGNs is known to be variable, which indicates that the coronal regions should be as compact as $\mathcal R\lesssim10$. In this work, we assume that the coronal compactness, ${\mathcal R}$, ranges from 3 to 30, which is supported by the observation of NGC~1068~\cite{Murase:2022dog,Ajello:2023hkh,Das:2024vug}. 
We treat $\mathcal R$ as one of the model parameters, but it may be a function of the Eddington ratio. It has also been suggested that ${\mathcal R}$ decreases with $\lambda_{\rm Edd}$, which may be attributed to the smooth transition from a strongly disk-dominated state in bright AGNs to the collapse of the disk into RIAFs in low-luminosity AGNs~\cite{Nemmen:2023kpa,Hagen:2024lyg,Kang:2024coc}.

We also assume that the corona can be regarded as a (quasi-)steady state, which is characterized by plasma advected with the infall velocity, $V_{\rm fall }\approx \alpha V_K$, where $V_K\approx\sqrt{GM/({\mathcal R} R_S)}=c/\sqrt{2{\mathcal R}}$ is the Keplerian velocity. MKM20 regard the corona as hot accretion inflows with a scale height of $H\approx{\mathcal R}R_S/\sqrt{3}\simeq(5.5\times{10}^{13}~{\rm cm})~{\mathcal R}_1M_{7.5}$. We also note that one can replace $V_{\rm fall}$ with the outflow velocity $V_{\rm out}$ when a corona can be regarded as hot outflows. Although the outflow may fail due to fallback accretion~\cite{Inoue:2022yak}, we note that $V_{\rm out}\sim V_K$ is typically $\sim \alpha^{-1}$ times faster than the infall velocity.  

Black-hole coronae are characterized by effectively hot electrons with $k_BT_e\sim10-100$~keV, which are energized by either random particle motions or kinetic motions from magnetic reconnections (e.g., Ref.~\cite{Groselj:2026nix}). For the majority of x-ray AGNs, coronal optical depth of $\tau_T\sim0.1-1$ is typically needed to reproduce x-ray data with the Comptonization mechanism. This allows us to estimate the proton density ($n_p$) via
\begin{equation}
n_p\approx \frac{\tau_T}{\zeta_e\sigma_T H}\simeq(1.1\times{10}^{10}~{\rm cm})~\tau_{T,-0.4}\zeta_e^{-1}{\mathcal R}_1^{-1}M_{7.5}^{-1}, 
\end{equation} 
where $\zeta_e$ is the possible pair loading factor. In MKM20, $\zeta_e=1$ is assumed considering electron-ion plasma, which is relevant for neutrino emission produced via inelastic $pp$ interactions~\cite{Murase:2019vdl}.   

The coronal magnetic field is very uncertain. We consider a magnetized corona whose magnetic field strength can be characterized by plasma beta, $\beta=P_{\rm th}/P_B$, where $P_{\rm th}$ is the thermal pressure and $P_B$ is the magnetic pressure. Then the magnetic field strength is parameterized as
\begin{eqnarray}
B&=&\sqrt{\frac{8\pi n_p k_BT_p}{\beta_p}}\nonumber\\
&\approx& \sqrt{\frac{4\pi \tau_T m_pc^2}{3^{1/2}\zeta_e\sigma_T {\mathcal R}^2 R_S \beta}}\nonumber\\
&\simeq&(2.6\times{10}^3~{\rm G})~\tau_{T,-0.4}^{1/2}\zeta_e^{-1/2}{\mathcal R}_1^{-1}M_{7.5}^{-1/2}\beta^{-1/2},\,\,\,\,\,\nonumber\\
&\simeq&(1.2\times{10}^3~{\rm G})~\tau_{T,-0.4}^{1/2}\zeta_e^{-1/2}{\mathcal R}_1^{-1}\lambda_{\rm Edd}^{1/2}\nonumber\\
&\times&{\left(\frac{L_{\rm bol}}{20L_X}\right)}^{-1/2}L_{X,44}^{-1/2}\beta^{-1/2},\,\,\,\,\,
\end{eqnarray}
As shown in MKM20, one may regard AGN corona as two-temperature plasma, and we assume that the proton temperature in the corona is the virial temperature, $k_BT_p \approx k_BT_{\rm vir}\equiv GMm_p/(3\mathcal RR_S)=m_pc^2/(6\mathcal R)$, such that the proton beta ($\beta_p$) is comparable to $\beta$. Note that MKM20 only utilizes a turbulent component of the coronal magnetic fields, and large-amplitude magnetic fluctuations in the plasma are implicitly assumed. If one introduces ordered magnetic fields, an additional parameter should be introduced. The Alfv\'en velocity is given by
\begin{eqnarray}\label{eq:Alfven}
V_A= c\sqrt{\frac{\sigma}{1+\sigma}}, 
\end{eqnarray}
where the magnetization parameter is
\begin{eqnarray}
\sigma=\frac{B^2}{4\pi n_p(m_p+\zeta_em_e)c^2}. 
\end{eqnarray}
In the nonrelativistic limit, we have $V_A\approx \sqrt{\sigma}c\approx {(3{\mathcal R}\beta)}^{-1/2}c \simeq 0.18~c~{\mathcal R}_1^{-1/2}\beta^{-1/2}$~(see also Ref.~\cite{Murase:2022dog}). 

Coronal plasma is assumed to be heated by magnetic dissipation (presumably through magnetic reconnections), where in general the dissipation time scale is written as $t_{\rm diss}\approx l_{\rm diss}/V_{\rm rec}$, where $l_{\rm diss}$ is macroscopic dissipation (energy-release) length scale (e.g., the size of reconnecting magnetic structures or loops), $V_{\rm rec}=\epsilon_{\rm rec}V_A$ and $\epsilon_{\rm rec}\sim0.1$ is the reconnection rate. Note that $l_{\rm diss}$ is a fraction of $H$ and it can be as small as $\sim R_S$, and MKM20 assumes $t_{\rm diss}\approx H/V_A$ for simplicity. Then, in this model, the (turbulent) magnetic luminosity originating from the magnetic dissipation is 
\begin{eqnarray}\label{eq:L_B}
L_{B}&\approx&\frac{2\pi \mathcal R^3 R_S^3 U_B}{\sqrt{3}t_{\rm diss}} \nonumber\\
&\approx&\frac{\pi \tau_T m_p c^3 R_S \sigma^{1/2}}{\sqrt{3} \zeta_e\sigma_T\beta {(1+\sigma)}^{1/2}}\nonumber\\
&\approx&\frac{\tau_T}{2\sqrt{3}\zeta_e\beta {(1+\sigma^{-1})}^{1/2}}L_{\rm Edd}|_M.
\end{eqnarray}
In the nonrelativistic limit of $V_A$, we have $t_{\rm diss}\approx \alpha\sqrt{\beta/2} t_{\rm fall}$, and we obtain the simple formula, 
\begin{eqnarray}
L_B&\approx&\frac{1}{6}\tau_T\zeta_e^{-1}{\mathcal R}^{-1/2}\beta^{-3/2} L_{\rm Edd}|_M\nonumber\\
&\simeq&(8.5\times10^{43}~{\rm~erg~s}^{-1})~\tau_{T,-0.4}\zeta_e^{-1}{\mathcal R}_1^{-1/2}\beta^{-3/2}M_{7.5}.\,\,\,\,\,\,\,\,\,\,
\end{eqnarray}

MKM20 focuses on $\beta\lesssim1-10$. This is because such low-beta plasma is required given that the coronal heating is powered by magnetic dissipation. If x-rays originate from hot electrons energized by magnetic dissipation (e.g., magnetic reconnection), then $L_X<L_B$ is required in the (quasi-)steady state, leading to
\begin{eqnarray}
\beta \lesssim 2.6~\tau_{T,-0.4}^{2/3}\zeta_e^{-2/3}{\mathcal R}_1^{-1/3}{\lambda}_{\rm Edd,-1}^{-2/3}{\left(\frac{L_{\rm bol}}{20L_X}\right)}^{2/3}.
\label{eq:betalimitX}
\end{eqnarray}
Note that the above upper limit on $\beta$ is valid only for hard x-ray emitting coronae, and the magnetic field can be much weaker for outer coronal regions that may be relevant for emissions at different wavelengths.   

In contrast, the plasma beta should have a lower limit because of the energy budget of the accretion system. Assuming that some fraction of the accretion energy is converted to the magnetic energy in the corona, we can write the magnetic luminosity in the corona as $L_B=\eta_B \dot Mc^2=(\eta_B/\eta_{\rm rad})L_{\rm bol}$, where $\eta_B$ is the efficiency of the magnetic field generation in accretion flows (including the disk). Equating this to Eq.~(\ref{eq:L_B}) and setting $\eta_B <1$, we obtain the lower limit on $\beta$ as 
\begin{eqnarray}
\beta \gtrsim 0.016~\eta_{\rm rad,-1}^{2/3}\tau_{T,-0.4}^{2/3}\zeta_e^{-2/3}{\mathcal R}_1^{-1/3}\lambda_{\rm Edd}^{-2/3}.
\label{eq:betalimitEdd}
\end{eqnarray}
Therefore, the corona cannot be too magnetized at least globally. The proton magnetization is limited as
\begin{eqnarray}
\sigma_p &=&\frac{B^2}{4\pi n_pm_pc^2}\approx \frac{1}{3\mathcal{R}\beta}\nonumber\\
&\lesssim& 2~\eta_{\rm rad,-1}^{-2/3}\tau_{T,-0.4}^{-2/3}\zeta_e^{2/3}{\mathcal R}_1^{-2/3}\lambda_{\rm Edd}^{2/3},
\end{eqnarray}
where we assume that the enthalpy is dominated by the proton rest-mass energy. This upper limit is much lower than the values assumed in relativistic magnetic reconnection models with $\sigma_p\gg1$~\cite{Fiorillo:2023dts,Mbarek:2023yeq}. Although we do not exclude the existence of regions with $\sigma_p\gg1$, the proton magnetization would be more likely to be semi-relativistic for typical values of $\lambda_{\rm Edd}\sim0.01-0.1$, so Ref.~\cite{Kheirandish:2021wkm} investigated the magnetic reconnection model for $\sigma_p\lesssim1$. 
In this paper, we examine two limiting situations, $\beta\sim0.01-0.03$ and $\beta\sim1-3$ for the representative models. In particular, we use $\beta=1$ for Models A and B, and $\beta=0.01$ for Models C (see Table~\ref{tab:diffuse}). 

\subsection{Cosmic-Ray Acceleration}
%
\begin{table*}[t]
\begin{center}
\caption{Parameters of the magnetically powered corona model used for the all-sky neutrino flux at medium energies. Note that parameters of Model A are exactly the same as those of MKM20~\cite{Murase:2019vdl}.  
\label{tab:diffuse}
}
\scalebox{1.0}{
\begin{tabular}{|c||c||c|c|c||c|c||c|c||c|}
\hline Model & SED model & $q$ & $\eta_{\rm tur}$ & $f_{\rm inj}^{1\rm TeV}$ $({\hat{p}}_{\rm CR}^{44})$ & $\alpha$ & $\beta$ &  $\bar{\lambda}_{\rm Edd}$ & $ \bar{\mathcal R}$ & $d\rho_{\rm AGN}/dL_X$\\
\hline Model A & MKM20~\cite{Murase:2019vdl} & $5/3$ & $10$ & $5.0\times{10}^{-5} (1.3\%)$ & $0.1$ & $1$ & Mayers et al.~\cite{Mayers:2018hau} & $30$ & Ueda et al.~\cite{Ueda:2014tma} \\
\hline Model B & MKM20~\cite{Murase:2019vdl}  & $5/3$ & $20$ & $8.8\times{10}^{-5} (0.5\%)$ & $0.1$ & $1$ & Shen et al.~\cite{Shen:2020obl} & $30$ & Gilli et al.~\cite{Gilli:2006zi} \\
\hline Model C & MKM20~\cite{Murase:2019vdl} & $2$ & $15$ & $2.5\times{10}^{-4} (2.0\%)$ & $0.1$ & ${10}^{-2}$ & Mayers et al.~\cite{Mayers:2018hau} & $10$ & Ueda et al.~\cite{Ueda:2014tma} \\
\hline
\end{tabular}
}
\end{center}
\end{table*}
\begin{table}[t]
\begin{center}
\caption{Cosmic-ray loading factors derived from Table~\ref{tab:diffuse}. 
\label{tab:CRloading}
}
\scalebox{1.0}{
\begin{tabular}{|c|c|c|c|}
\hline \diagbox[dir=SE]{$L_{X}$ [${\rm erg}~{\rm s}^{-1}$]}{$\xi_{{\rm CR}/X}$ (Model)} 
& Model A & Model B & Model C \\
\hline ${10}^{42}$ & ${10}^{-1.03}$ & ${10}^{-1.28}$ & ${10}^{-0.37}$ \\ 
\hline ${10}^{43}$ & ${10}^{-0.97}$ & ${10}^{-1.02}$ & ${10}^{-0.53}$ \\
\hline ${10}^{44}$ & ${10}^{-0.79}$ & ${10}^{-0.81}$ & ${10}^{-0.72}$ \\
\hline ${10}^{45}$ & ${10}^{-0.68}$ & ${10}^{-0.69}$ & ${10}^{-0.91}$ \\
\hline ${10}^{46}$ & ${10}^{-0.59}$ & ${10}^{-0.57}$ & ${10}^{-1.10}$ \\
\hline
\end{tabular}
}
\end{center}
\end{table}

MKM20 considers stochastic particle acceleration by magnetohydrodynamic (MHD) turbulence in coronal regions, focusing on a turbulent magnetic component.
Although details of the coronal formation remain unknown, accretion flows are expected to be strongly turbulent due to the magnetorotational instability (MRI)~\cite{Balbus:1991ay,Balbus:1998ja}. Magnetic buoyancy can then lift magnetized structures from the disk into the low-density atmosphere, producing a patchy, magnetically dominated corona where energy is dissipated via magnetic reconnections and turbulent cascades (e.g., Ref.~\cite{Galeev:1979td}).
Stratified shearing-box and global MHD simulations indeed show buoyant magnetic loops and an overlying magnetized atmosphere/corona with intermittent dissipation ~\cite{Miller:1999ix,Fromang:2006vd}.
Buoyancy-related processes (Parker-type instabilities) and associated dynamo action may also sustain magnetic variability and expel magnetic structures into the corona~\cite{1992MNRAS.259..604T,Johansen:2008ed}. Once the magnetic structures populate the corona, turbulent footpoint motions and the Keplerian shear may build current sheets and trigger magnetic reconnections between neighboring structures. A statistical loop-ensemble framework connects such a footpoint stressing on the coronal magnetic pressure and dissipation rate~\cite{Uzdensky:2008ce}, while reconnections are expected to be intrinsically time dependent and fragmented (e.g., via plasmoid formation), yielding strong intermittency and reconnection-driven turbulence (e.g., Refs.~\cite{Lazarian:1998wd,Uzdensky:2014uda}). Resulting coronal MHD turbulence may stochastically accelerate cosmic rays via either resonant or nonresonant scatterings. Magnetic reconnections and colliding blobs in the turbulence may also lead to shock acceleration~\cite{Murase:2022dog}, which is not considered in this work.  

In general, the acceleration time for stochastic acceleration is expressed as 
\begin{equation}
t_{\rm acc}=\eta_{\rm tur}{\left(\frac{c}{V_A}\right)}^2\frac{l_{\rm tur}}{c}{\left(\frac{\varepsilon_p}{eBl_{\rm tur}}\right)}^{2-q},
\label{eq:tacc}
\end{equation}
where $l_{\rm tur}$ is the turbulence correlation length, $q$ is the index of the diffusion coefficient, $V_A$ is the Alfv\'en velocity. Note that in MKM20, $\eta_{\rm tur}$ is introduced as a phenomenological parameter describing the acceleration efficiency. In general, $l_{\rm tur}$ depends on the structure and the mechanism of turbulence. For simplicity, we assume $l_{\rm tur}=H$, and various uncertain effects such as contributions from fast-mode fluctuations, intermittency, and turbulence-driven structures are absorbed into the effective parameter $\eta_{\rm tur}$. More technically, in the traditional approach, the cosmic-ray spectrum resulting from stochastic acceleration is obtained by solving the following Fokker-Planck equation: 
\begin{equation}
\frac{\partial {\mathcal F}_p}{\partial t} = \frac{1}{\varepsilon_p^2}\frac{\partial}{\partial \varepsilon_p}\left(\varepsilon_p^2D_{\varepsilon_p}\frac{\partial {\mathcal F}_p}{\partial \varepsilon_p}+\frac{\varepsilon_p^3\mathcal{F}_p}{t_{p-\rm cool}}\right)-\frac{{\mathcal F}_p}{t_{p-\rm esc}}+\dot{\mathcal F}_{p,\rm inj},
\end{equation}
where ${\mathcal F}_p=c n_{\varepsilon_p}/(4\pi p^2)$ is the cosmic-ray energy distribution function, $D_{\varepsilon_p}$ is the energy diffusion coefficient, $t_{p-\rm cool}$ is the proton cooling time scale, $t_{p-\rm esc}=(t_{\rm fall}^{-1}+t_{p-\rm diff}^{-1})^{-1}$ is the proton escape timescale, and $\dot{\mathcal F}_{\rm p,inj}$ is the injection term. For the energy diffusion coefficient, corresponding to Eq.~(\ref{eq:tacc}), MKM20 utilizes
\begin{equation}  
D_{\varepsilon_p}=\frac{\varepsilon_p^2}{t_{\rm acc}}=\frac{c}{\eta_{\rm tur}H}\left(\frac{V_A}{c}\right)^2{\left(\frac{r_L}{H}\right)}^{q-2}\varepsilon_p^2,
\end{equation}
which is applicable to both resonant and nonresonant acceleration. 
In the quasi-linear theory, $\eta_{\rm tur}$ and $q$ are connected to the power spectrum of the MHD turbulence, $P_k\propto k^{-\tilde{q}}$, as $\eta_{\rm tur}=8\pi \int dk P_k/B^2$ and $q=\tilde{q}$, although this is not required in the current phenomenological framework (see, e.g., Refs.~\cite{Becker:2006nz,Murase:2011cx,Brunetti:2016oaw,Nishiwaki:2021axb} for related discussions). More practically, in the critically balanced Alfv\'enic turbulence, most power resides in the perpendicular modes, and gyroresonant scattering by Alfv\'en and slow modes can be inefficient, as found in recent particle-in-cell (PIC) \cite{Zhdankin:2018lhq,Comisso:2019frj} and MHD simulations \cite{Lynn:2014dya,Kimura:2018clk,Sun:2021ods}. A single value of $D_{\varepsilon_p}$ may not capture the inhomogeneity, and the generalized Fokker-Planck approach better describes the power-law tail of the cosmic-ray spectrum (see Refs.~\cite{Lemoine:2019ofp,Lemoine:2022rpj,Lemoine:2023wsw,Lemoine:2024roa} for details including possible effects of cosmic-ray feedback). The conventional approach still gives an approximate description of a steady-state spectrum with $\varepsilon_p^4 {\mathcal F}_p \propto \varepsilon_p^{3-q}$, with a spectral pileup due to strong radiative cooling. The nonresonant case has been widely treated as the hard-sphere limit in the literature (e.g., Refs.~\cite{ParkPetrosian1995,Becker:2006nz,Stawarz:2008sp,Murase:2011cx,Brunetti:2016oaw,Nishiwaki:2021axb,Fiorillo:2024akm,Kawashima:2025wzq}), where $\eta_{\rm tur}$ and $q$ are regarded as phenomenological parameters. Although Alfv\'en and slow modes are dominant in MRI turbulence due to its subsonic nature \cite{Kimura:2018clk}, Alfv\'en modes could be converted into fast modes in strongly magnetized plasma~\cite{Takamoto:2017vhf}, and the role of fast modes in particle acceleration can be more significant~\cite{Yan:2002qm,Cho:2005mb,Teraki:2019qam,Demidem:2019jzn}. In addition, some simulations of solar coronae indicate that impulsive reconnection events excite large-scale fast-mode waves \cite{2015ApJ...800..111Y,2024ApJ...977..235M}, and shocks~\cite{Groselj:2026nix} may also generate fast modes. Given that such details are uncertain, we allow for possibilities that the coronal turbulence contains a non-negligible compressive or quasi-isotropic component. 

In this work, within the phenomenological framework of MKM20, we consider three representative acceleration models assuming large-amplitude turbulence. For Models A and B, we consider $q=5/3$, which is motivated by resonant acceleration. We also consider the hard-sphere diffusion limit, $q=2$, which is motivated by nonresonant acceleration. Model C is motivated by test-particle simulations in simulated MRI turbulence~\cite{Kimura:2018clk,Kimura:2026}. However, as we will see below, the phenomenological consequences for neutrinos and gamma rays are degenerate and indistinguishable based on the current data. 

The proton diffusion timescale is given by $t_{p-\rm diff}=H^2/D_R$, where we use the spatial diffusion coefficient,
\begin{equation}
D_R=\frac{cH}{\eta_R}{\rm{max}}\left[\left(\frac{r_L}{l_{\rm coh}}\right)^{1/2},~\left(\frac{r_L}{l_{\rm coh}}\right)^{2-q}\right]. 
\end{equation}
Here $l_{\rm coh}\approx H/6$ is the turbulent coherence length (where the factor 1/6 comes from the turbulence power spectral index of 1.5, based on MHD simulations~\cite{Kawazura:2024usv}). For $q=5/3$, we use $\eta_R=16/\eta_{\rm tur}$ as in MKM20, while for $q=2$ we utilize $\eta_R=40$ based on test-particle simulations with MHD simulations for hot accretion flows~\cite{Kimura:2026}. Note that with the current formulation $\eta_R$ in $D_R$ does not have to be connected to $\eta_{\rm tur}$ in $D_{\varepsilon_p}$. Also, the energy dependence and normalization of the length scale are different between MHD and PIC simulations, which should be understood as an uncertainty of the model. Nevertheless, our key results on neutrino and gamma-ray spectra are largely unaffected because other cooling time scales are typically more important in coronae.

In our phenomenological framework, the energy fraction carried by cosmic rays is controlled by $f_{\rm inj}$ and $\eta_{\rm tur}$. 
We set $\dot{\mathcal{F}}_{\rm inj}\propto f_{\rm inj}L_X \varepsilon_{p,\rm inj}^{-3}\delta(\varepsilon_p-\varepsilon_{p,\rm inj})$, where $f_{\rm inj}$ and $\varepsilon_{p,\rm inj}$ are the injection efficiency and injection energy, respectively. We fix $\varepsilon_{p,\rm inj}=1$ TeV, and scale $f_{\rm inj}$ to obtain the normalization of cosmic-ray component. With this setup, the total pressure of cosmic rays, $P_{\rm CR}$, is proportional to $f_{\rm inj}$. Although $P_{\rm CR}$ also depends on $\varepsilon_{p,\rm inj}$, we can always rescale these parameters. The scaling relation to provide the same value of $P_{\rm CR}$ is expressed by $f_{\rm inj}\propto \varepsilon_{p,\rm inj}$, which are tabulated in Table \ref{tab:diffuse}.

The cosmic-ray luminosity of the MKM20 model is written as
\begin{eqnarray}
L_{\rm CR} 
&\equiv& \xi_{{\rm CR}/X} L_X \nonumber\\
&\approx& \frac{2\pi {\mathcal R}^3R_S^33P_{\rm CR}}{\sqrt{3} t_{p-\rm loss}^*}\nonumber\\
&\approx& \frac{2 t_{\rm fall}}{t_{p-\rm loss}^*} \hat{p}_{\rm cr} L_{\rm th},
\end{eqnarray}
where $\xi_{{\rm CR}/X}$ is the so-called cosmic-ray loading factor against the intrinsic $2-10$~keV luminosity, $L_{\rm th}\equiv (2\pi/\sqrt{3}) \mathcal{R}^3R_S^3(3/2)P_{\rm vir}/t_{\rm fall}$ is the luminosity of advected thermal plasma, and $t_{p-\rm loss}^*$ is the effective proton loss time that can often be comparable to the loss time around the maximum energy of protons. The derived values of $\xi_{{\rm CR}/X}$ are described in Table~\ref{tab:CRloading}. 
Alternatively, the cosmic-ray luminosity can be connected with the magnetic dissipation power as $L_{\rm CR}\equiv\epsilon_{\rm CR}L_{B}$, and we have
\begin{eqnarray}
\epsilon_{\rm CR}\approx \frac{3t_{\rm diss}}{t_{p-\rm loss}^*}
\beta\hat{p}_{\rm cr} < 1.    
\end{eqnarray}
For our model parameters described in Tables~\ref{tab:diffuse} and \ref{tab:CRloading}, we will see that the hierarchy of $L_{\rm CR}\lesssim L_{\rm th} \lesssim L_{B}$ is always satisfied (see also Sec.~\ref{sec:1068implication}). Note that $L_{\rm CR}\lesssim L_{\rm th}$ is not necessarily required in general, but for $\eta_{\rm CR}\equiv L_{\rm CR}/(\dot{M}c^2)$ the virial theorem leads to $\eta_{\rm CR}<(1/40){\mathcal R}_1^{-1}$~\cite{Murase:2020lnu}.

\subsection{Neutrino and gamma-ray production}\label{sec:AMES}
High-energy cosmic rays interact with nucleons in the coronal plasma, leading to the $pp$ production of mesons (mostly pions). They also interact with photons produced by the optically thick disk and Comptonized photons from the coronae, which leads to photomeson production. These processes generate high-energy neutrinos via decay channels such as $\pi^+\to\mu^+\nu_\mu\to\nu_\mu\bar{\nu}_\mu\nu_ee^+$. For a cosmic-ray proton with $\varepsilon_p$, the typical neutrino energy is $\varepsilon_\nu\sim(0.03-0.05)\varepsilon_p$ for both $pp$ and $p\gamma$ interactions~\cite{Murase:2013rfa,Murase:2015xka}. In addition to these meson production channels, the Bethe-Heitler pair production ($p+\gamma\rightarrow p+e^++e^-$) is also effective especially for protons with $\varepsilon_p\sim0.03-3$~PeV. MKM20 shows that this can significantly suppress the meson production efficiency in the medium-energy range with $\varepsilon_\nu\sim10-100$~TeV, especially for luminous AGNs.

For a given (quasi)steady-state cosmic-ray spectrum, we consistently calculate differential energy densities of neutrinos, gamma-rays, and electrons-positrons. Using \textsc{AMES} (Astrophysical Multimessenger Emission Simulator), we solve the following kinetic equations~\cite{Murase:2019vdl,Murase:2022dog,Zhang:2023ewt,Murase:2023chr}: 
\begin{eqnarray}\label{eq:cascade}
\dot{n}_{\varepsilon_e}^e &=& \dot{n}_{\varepsilon_e}^{(\gamma\gamma)} 
- \frac{n_{\varepsilon_e}^{e}}{t_{e-\rm esc}}
- \frac{\pd}{\pd \varepsilon_e} [(P_{\rm IC}+P_{\rm syn}+P_{\rm bre}+P_{\rm Cou}) n_{\varepsilon_e}^e] \nonumber\\ 
& & + \dot{n}_{\varepsilon_e}^{\rm inj}\nonumber\\ 
\dot{n}_{\varepsilon_\gamma}^\gamma &=& -\frac{n_{\varepsilon_\gamma}^{\gamma}}{t_{\gamma \gamma}} 
- \frac{n_{\varepsilon_\gamma}^{\gamma}}{t_{\rm esc}} -\frac{n_{\varepsilon_\gamma}^{\gamma}}{t_{\rm matter}} 
+ \dot{n}_{\varepsilon_\gamma}^{(\rm IC)}
+ \dot{n}_{\varepsilon_\gamma}^{(\rm syn)}
+ \dot{n}_{\varepsilon_\gamma}^{(\rm bre)} \nonumber\\ 
& & + \dot{n}_{\varepsilon_\gamma}^{\rm inj}\nonumber\\ 
\end{eqnarray}
where $n_{\varepsilon_i}^i\equiv dn^i/d\varepsilon_i$ and $\varepsilon_i$ is particle energy for a particle species with $i$. Following Refs.~\cite{Murase:2019vdl,Murase:2023chr}, we calculate energy loss rates of electrons (and positrons) for inverse-Compton radiation ($P_{\rm IC}$), synchrotron radiation ($P_{\rm syn}$), relativistic bremsstrahlung ($P_{\rm bre}$), and Coulomb collisions ($P_{\rm Cou}$), respectively. 
Also, $t_{\gamma\gamma}$ is the two-photon annihilation time, $t_{\rm matter}$ is the energy loss time scale due to interactions with matter (i.e., Compton scatterings and the Bethe-Heitler pair production due to interactions with atoms), $t_{\rm esc}\approx r/c$ is the escape time of neutral particles such as neutrinos and photons (in the optically thin limit), where ${\mathcal V}=(4\pi/3)r^3=2\pi {\mathcal R}^2 R_S^2 H$. Note that for electrons and positrons we consider the advection escape instead of adiabatic cooling ($P_{\rm ad}$), and we use $t_{e-\rm esc}=t_{\rm fall}$. The calculation is performed for a steady-state cosmic-ray spectrum. Note that Ref.~\cite{Murase:2022dog} updated the treatments of photomeson production, Bethe-Heitler pair production, and inelastic $pp$ interactions at low-energies, and the differential energy injection rates used here take full account of details of secondary production compared to MKM20.
Finally, the differential neutrino and photon luminosities are calculated by 
\begin{equation}
\varepsilon_i L_{\varepsilon_i}= \frac{(\varepsilon_i^2 n_{\varepsilon_i}){\mathcal V}}{t_{\rm esc}}. 
\end{equation}

Details of neutrino and gamma-ray production in AGN coronae are described in MKM20 and Ref.~\cite{Murase:2022dog}. Here it is useful to clarify the relative importance of $pp$ interactions. Using Eqs.~(1) and (2) of MKM20, assuming $\Gamma_X\sim2$, we obtain  
\begin{eqnarray}
\frac{f_{pp}}{f_{p\gamma}} 
&\sim& \frac{(\tau_T/0.5){\mathcal R}_{1.5}M_{8}}{\zeta_e L_{X,44}}{\left(\frac{\varepsilon_\nu}{70~\rm TeV}\right)}^{-1}\nonumber\\
&\sim&0.4\frac{\tau_{T,-0.4}{\mathcal R}_{1}}{\zeta_e \lambda_{\rm Edd,-1}}
{\left(\frac{L_{\rm bol}}{20L_X}\right)}
{\left(\frac{\varepsilon_\nu}{70~\rm TeV}\right)}^{-1}, 
\end{eqnarray}
where $f_{pp}$ and $f_{p\gamma}$ are effective optical depths for $pp$ and $p\gamma$ interactions, respectively. Thus, the contribution of $p\gamma$ interactions is expected to be more important at higher energies and/or for AGNs with higher Eddington ratios.    

In addition, for NGC~1068 that is known to be Compton thick, we consider external matter attenuation due to the Compton scattering and the Bethe-Heitler pair production, considering a column density of $N_H=10^{25}~{\rm cm}^{-2}$ as well as infrared emission from the dust torus~\cite{Murase:2022dog}.

\subsection{All-sky multimessenger intensities}\label{sec:diffuseformalism}
To evaluate the line-of-sight integral contribution from jet-quiet AGNs with coronae to the all-sky multimessenger fluxes, we convolve the source model with the observed x-ray luminosity function across redshifts, and integrate up to $z=5$. Note that the corona model simultaneously predicts the diffuse MeV gamma-ray background, and the correlation between neutrino and MeV gamma-ray fluxes will offer a powerful cross check for future multimessenger studies. 

The all-sky neutrino flux from the sources is calculated by the following formula, 
\begin{eqnarray}\label{eq:allsky}
\Phi_\nu &=& \frac{c}{4\pi H_0} \int^{z_{\rm max}} dz\,
\frac{1}{\sqrt{(1+z)^3\Omega_m + \Omega_\Lambda}}
\int dL_X\, \int dM\,\nonumber\\ 
&\times&  \frac{d^2\rho_{\rm AGN}}{dL_X dM}(z) \, 
\frac{L_{\varepsilon_\nu}(L_X,M)}{\varepsilon_\nu},\nonumber\\
 &=& \frac{c}{4\pi H_0} \int^{z_{\rm max}} dz\,
\frac{1}{\sqrt{(1+z)^3\Omega_m + \Omega_\Lambda}}
\int dL_X\, \int d\lambda_{\rm Edd}\,\nonumber\\ 
&\times&  \frac{d^2\rho_{\rm AGN}}{dL_X\lambda_{\rm Edd}}(z) \, 
\frac{L_{\varepsilon_\nu}(L_X,\lambda_{\rm Edd})}{\varepsilon_\nu},\nonumber\\
&\approx& \frac{c}{4\pi H_0} \int^{z_{\rm max}} dz\,
\frac{1}{\sqrt{(1+z)^3\Omega_m + \Omega_\Lambda}}
\int dL_X\, \nonumber\\ 
&\times& \frac{d\rho_{\rm AGN}}{dL_X}(z) \, 
\frac{L_{\varepsilon_\nu}(L_X,\bar{\lambda}_{\rm Edd})}{\varepsilon_\nu},
\end{eqnarray}
where $d\rho_{\rm AGN}/dL_X$ is the comoving x-ray luminosity function, $L_{\varepsilon_\nu}(L_X,\lambda_{\rm Edd})$ is the neutrino luminosity per energy, and $\varepsilon_\nu = (1+z)E_\nu$. 

MKM20 adopts the x-ray luminosity function parametrized by Ueda et al.~\cite{Ueda:2014tma}, which relies on the hard x-ray observation. The functional form is given by
\begin{equation}
\frac{d\rho_{\rm AGN}}{d \log L_X}(L_X,z)= A \left[ \left(\frac{L_X}{L_*}\right)^{\gamma_1} + \left(\frac{L_X}{L_*}\right)^{\gamma_2} \right]^{-1} e(z,L_X)\,,
\end{equation}
where $L_*$ is the break luminosity, and $\gamma_1$ and $\gamma_2$ and power-law indices. We choose $A=2.91 \times 10^{-6}~{(h/0.7)}^3~{\rm Mpc}^{-3}$, $L_* = 10^{43.97}~{\rm erg}~{\rm s}^{-1}$, $\gamma_1 = 0.96$, and $\gamma_2=2.71$. The generalization to an arbitrary redshift $z$ is achieved using a luminosity-dependent evolution factor $e(z, L_X)$ defined as
\begin{eqnarray}
\label{eq:zevol}
e(z,L_X)=\begin{cases}
(1+z)^{p_1}\,, {\rm for}~z \leq z_{c_1}(L_X)\\
(1+z_{c_1})^{p_1}\left( (1+z)/(1+z_{c_1}) \right)^{p_2}\,, \nonumber\\
{\rm for}~z_{c_1}(L_X)<z\leq z_{c_2}(L_X)\\
(1+z_{c_1})^{p_1} \left((1+z_{c_2})/(1+z_{c_1}) \right)^{p_2} \nonumber\\
\times \left( (1+z)/(1+z_{c_2}) \right)^{p_3}\, {\rm for}~z>z_{c_2}\,,
\end{cases}\,
\end{eqnarray}
and
\begin{align}
\label{eq:cutoffz}
z_{c_1} &= \begin{cases}
z^*_{c_1} \left( L_X/L_{a_1} \right)^{\alpha_1}\,, L_X \leq L_{a_1}\\
z^*_{c_1}\,, L_X > L_{a_1}
\end{cases}\,,\\
z_{c_2} &= \begin{cases}
z^*_{c_2} \left( L_X/L_{a_2} \right)^{\alpha_2}\,, L_X \leq L_{a_2}\\
z^*_{c_2}\,, L_X > L_{a_2}
\end{cases}\,.\nonumber
\end{align}
In the above equations, $z_{c_1}$ and $z_{c_2}$ denote the cutoff redshifts where the power-law index changes from $p_1$ to $p_2$ and from $p_2$ to $p_3$, respectively. The luminosity-dependent index is $p_1(L_X) = p_1^* + \beta_1 \left( \log L_X - \log L_X^p\right)$, where $L_X^p = 10^{44}~{\rm erg}~{\rm s}^{-1}$, $\beta_1 = 0.84$, $p_1^*=4.78$, $p_2=-1.5$, and $p_3=-6.2$. We also fix $z_{c_1}^*=1.86$, $z_{c_2}^* =3.0$, $\alpha_2=-0.1$, and choose the luminosity thresholds, $L_{a_1}=10^{44.61}~{\rm erg}~{\rm s}^{-1}$ and $L_{a_2}=10^{44}~{\rm erg}~{\rm s}^{-1}$. 

We also use the soft x-ray luminosity function provided by Ref.~\cite{Hasinger:2005sb}, with a conversion to the hard x-ray luminosity function suggested by Gilli et al.~\cite{Gilli:2006zi}. The primary difference is in the parameterization of the redshift evolution factor $e(z, L_X)$. For this case, $e(z, L_X)$ has a single cutoff redshift $z_{c_1}$ (see Eq.~\ref{eq:cutoffz} for $z_{c_1}$), where the power-law index changes from $p_1$ to $p_2$.
The luminosity-dependent power-law index $p_1$ has the same definition as above, while $p_2 (L_X) = p_2^* + \beta_2 \left(\log L_X-\log L_X^p\right)$, and we choose $p_2^* = -1.5$ and $\beta_2 = 0.6$. The other parameters used for this x-ray luminosity function are $A = 2.62 \times 10^{-7}~{(h/0.7)}^3~{\rm Mpc}^{-3}$, $L_* = 10^{43.94}~{\rm erg}~{\rm s}^{-1}$, $\gamma_1 = 0.87$, $\gamma_2 = 2.57$, $p_1^* = 4.7$, $z_{c_1}^* = 1.42$, $\alpha_1 = 0.21$, $L_{a_1} = 10^{44.67}~{\rm erg}~{\rm s}^{-1}$, and $\beta_1=0.7$. The above prescription from Ref.~\cite{Hasinger:2005sb} provides the luminosity function for soft x rays ($0.5-2$~keV). This can then be converted to include hard x rays ($2-10$~keV) by introducing a luminosity-dependent correction factor $R(L_{{\rm sf}X})$~\cite{Gilli:2006zi}, where $R(L_{{\rm sf}X})$ is defined as the ratio between obscured Compton thin ($21<\log N_H <24$) and unobscured AGN ($\log N_H<24$) given by
\begin{equation}
R(L_{{\rm sf}X}) = R_S \exp \left( -\frac{L_{{\rm sf}X}}{L_C} \right) + R_Q\left[ 1- \exp \left( -\frac{L_{{\rm sf}X}}{L_C} \right)\right]\,,
\end{equation}
where the ratio in the low-luminosity (Seyfert) regime and the ratio in the high-luminosity (quasar) regime are given by $R_S$ and $R_Q$, respectively. The division between these two regimes is characterized by the $0.5-2$~keV characteristic luminosity, We choose $R_S=4.0$, $R_Q=1.0$, and $L_C = {10}^{43.5}~{\rm erg}~{\rm s}^{-1}$. The hard x-ray luminosity function is then given by
\begin{equation}
\frac{d\rho_{\rm AGN}}{d \log L_X}(L_X,z) = \eta_{S\rightarrow H} (L_{{\rm sf}X})\frac{d\rho_{\rm AGN}
(L_{{\rm sf}X},z)}{d \log L_{{{\rm sf}X}}} \left(1+R(L_{{\rm sf}X})\right)\,,
\end{equation}
where $\eta_{S\rightarrow H} (L_{{\rm sf}X})$ is the luminosity-dependent conversion factor from the soft to hard x-ray band. We choose a characteristic value, $\eta_{S\rightarrow H}\sim 1/1.7$, and ignore the luminosity dependence for simplicity. In either case of the adopted x-ray luminosity function, the all-sky flux evaluated by Eq.~(\ref{eq:allsky}) mainly comes from AGNs located at $z\sim 1-2$, and yields the redshift evolution factor, $\xi_z \sim 3$~\cite{Murase:2016gly,Waxman:1998yy}. 

The corona model of MKM20 depends on the Eddington ratio $\lambda_{\rm Edd}$ or the SMBH mass $M$ for a given $L_X$. Ideally, we need to take into account the distribution of $\lambda_{\rm Edd}$. However, because this is highly uncertain at present~\cite{Zou+2024}, it is reasonable to adopt the fiducial values of $\lambda_{\rm Edd}$ based on the possible $L_X-M$ relation. The SMBH mass is related to $L_X$ by Fig.~15 of Ref.~\cite{Mayers:2018hau}. MKM20 adopts 
\begin{equation}
\bar{\lambda}_{\rm Edd}
\approx 0.074\left(\frac{L_{\rm bol}}{20~L_X} \right)
\left( \frac{20 L_X}{10^{44}~{\rm erg~s}^{-1}} \right)^{0.254}.
\label{EddMayers1}
\end{equation}
Note that the updated result by Mayers et al.~\cite{Mayers:2018hau} gives 
\begin{equation}
\bar{\lambda}_{\rm Edd}
\approx 0.089\left(\frac{L_{\rm bol}}{20~L_X} \right)
\left( \frac{20 L_X}{10^{44}~{\rm erg~s}^{-1}} \right)^{0.42},
\label{EddMayers2}
\end{equation}
although our results are not affected significantly. 
On the other hand, Eq.~(29) of Shen et al.~\cite{Shen:2020obl} leads to 
\begin{equation}
\bar{\lambda}_{\rm Edd}
\approx 0.015{\left(\frac{L_{\rm bol}}{10^{44}~{\rm erg}~{\rm s}^{-1}}\right)}^{0.469},
\label{EddShen}
\end{equation}
which gives lower values of the Eddington ratio compared to those from x-ray observations.

\section{Interpretations of All-Sky Multimessenger Fluxes}
%
\begin{figure*}[t]
\includegraphics[width=0.325\linewidth]{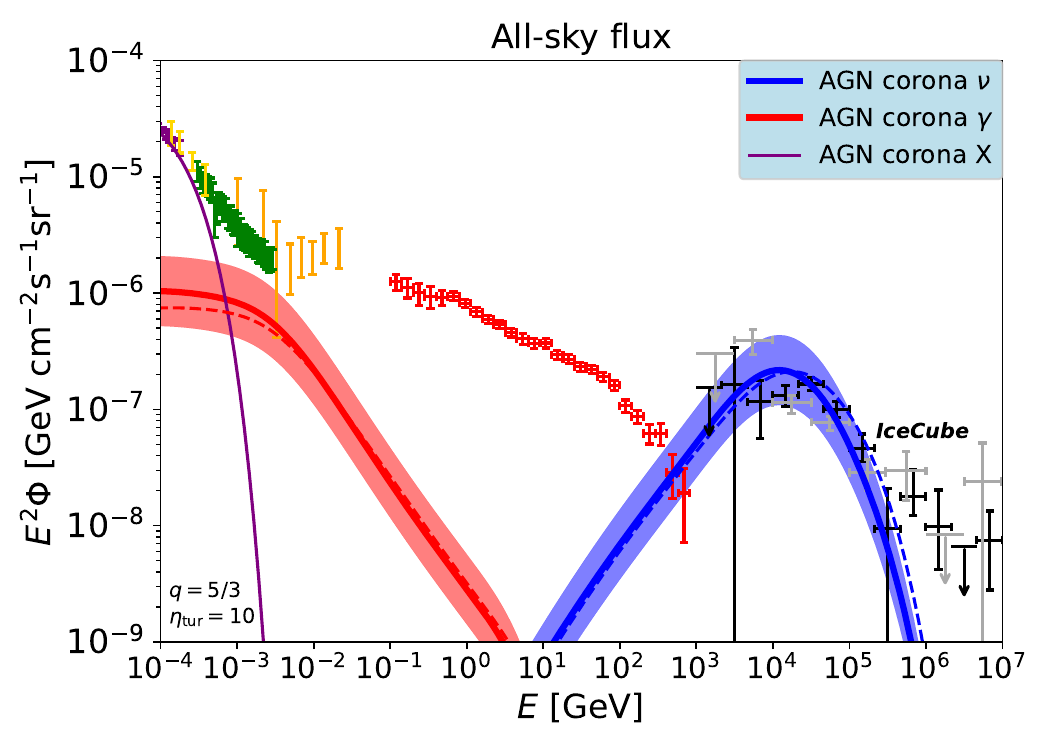}
\includegraphics[width=0.325\linewidth]{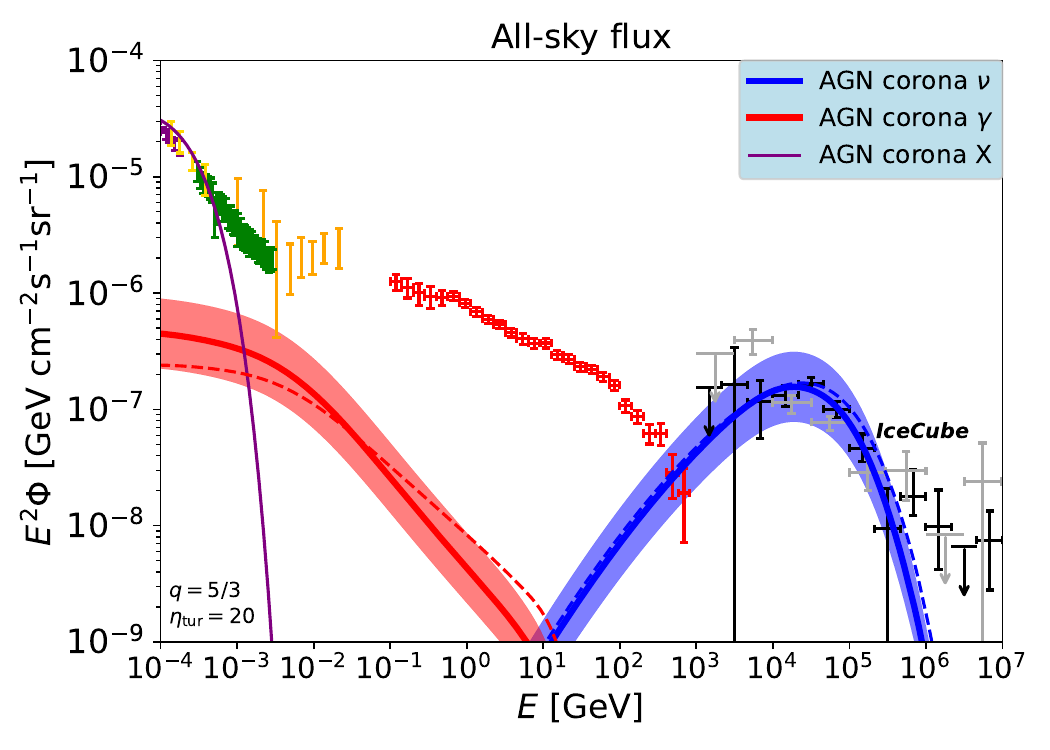}
\includegraphics[width=0.325\linewidth]{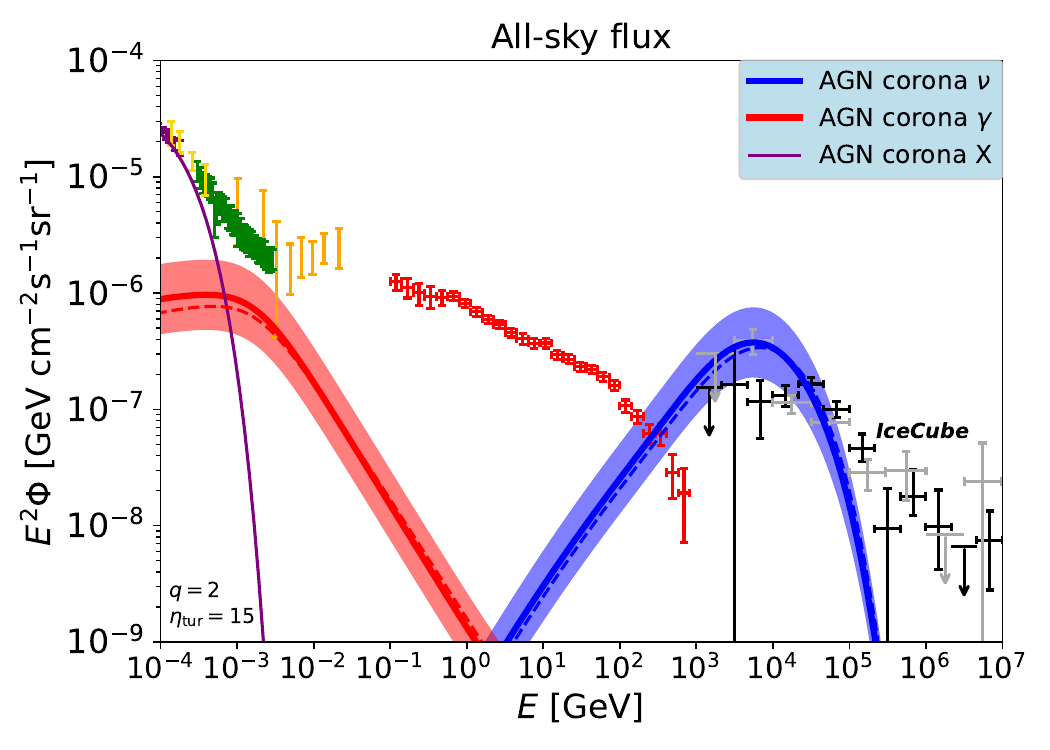}
\caption{All-sky neutrino and gamma-ray spectra for Model A (left panel), Model B (middle panel), and Model C (right panel). Note that the results for Model A are essentially the same as those presented in MKM20 except improved treatments of microphysics. The extragalactic neutrino, x-ray and gamma-ray background data are the same as those in Fig.~\ref{fig:summary}. 
\label{fig:diffuse}
}
\end{figure*}

Our numerical results on the all-sky multimessenger fluxes are shown in Fig.~\ref{fig:diffuse}. From Table~\ref{tab:diffuse}, we see that if all jet-quiet AGNs are neutrino active, a small energy fraction ($\sim1-10$\%) of the coronal energy budget channeled into nonthermal protons suffices to account for the all-sky IceCube flux at tens of TeV in neutrino energies. This is consistent with the findings of MKM20 but this conclusion is confirmed more robustly. Another generic prediction is a convex neutrino spectrum whose high-energy portion is dominated by lower-luminosity AGNs, as now hinted at by the latest IceCube data. This point will be discussed in detail below. While all the three models share these characteristics, there is large degeneracy within the phenomenological framework, and particle acceleration models (e.g., resonant or nonresonant acceleration) cannot be discriminated from the current data.  

Analytically, the all-neutrino flux is approximated to be~\cite{Murase:2019vdl}
\begin{eqnarray}
E_\nu^2\Phi_\nu &\approx& \frac{ct_H}{4\pi} \varepsilon_\nu Q_{\varepsilon_\nu} \xi_z \nonumber\\
&\approx& \frac{ct_H}{4\pi} \left(\frac{\frac{3K}{4[1+K]}f_{\rm mes}}{1+f_{\rm mes}+f_{\rm BH}}\right) \left(\frac{\xi_{{\rm CR}/X}L_X\rho_{\rm AGN}\xi_z}{{\mathcal C}_{\rm CR}}\right) \,\,\,\,\,\,\,\,\,\,\,\,\nonumber\\
&\simeq&1.0\times{10}^{-7}~{\rm GeV}~{\rm cm}^{-2}~{\rm s}^{-1}~{\rm sr}^{-1}~\left(\frac{\xi_z}{3}\right)\nonumber\\
&\times&\left(\frac{2K}{1+K}\right)\left(\frac{15f_{\rm mes}}{1+f_{\rm BH}+f_{\rm mes}}\right)\nonumber\\
&\times&{\left(\frac{\xi_{{\rm CR}/X,-1}{\mathcal C}_{\rm CR}^{-1}L_X\rho_{\rm AGN}}{2\times{10}^{46}~{\rm erg}~{\rm Mpc}^{-3}~{\rm yr}^{-1}}\right)}.
\label{eq:diffusenu}
\end{eqnarray}
which is consistent with the numerical results shown in Fig.~\ref{fig:diffuse}.
Here $K=1$ and $K=2$ for $p\gamma$ and $pp$ interactions, respectively, $\xi_z\sim3$ due to the redshift evolution of the AGN luminosity density~\cite{Waxman:1998yy,Murase:2016gly}, ${\mathcal C}_{\rm CR}$ is the conversion factor from bolometric to differential luminosities, and $\xi_{{\rm CR}/X}\equiv L_{\rm CR}/L_X$ is the cosmic-ray loading parameter defined against the x-ray luminosity, where $P_{\rm CR}/P_{\rm th}\sim0.01$ corresponds to $\xi_{{\rm CR}/X}\sim0.1$ in our model (see Table~\ref{tab:CRloading}). Note that $\xi_{{\rm CR}/X}$ relies on the observations because $L_X$ is defined as the intrinsic x-ray luminosity at the $2-10$~keV band. One should keep in mind that $L_{\rm corona}$ is always larger than $L_X$~\cite{Das:2024vug}, and the smallness of $L_{\rm CR}/L_{\rm corona}$ is consistent with our model requirements of $L_{\rm CR} < L_{B}$ and $L_{\rm corona} < L_{B}$. We also note that $L_{\rm CR}/L_{\rm bol}$ is further smaller because of $L_{\rm corona} \sim 0.1 L_{\rm bol}$.   

\subsection{Origin of a spectral break}\label{sec:break}
%
\begin{figure*}[th]
\includegraphics[width=0.325\linewidth]{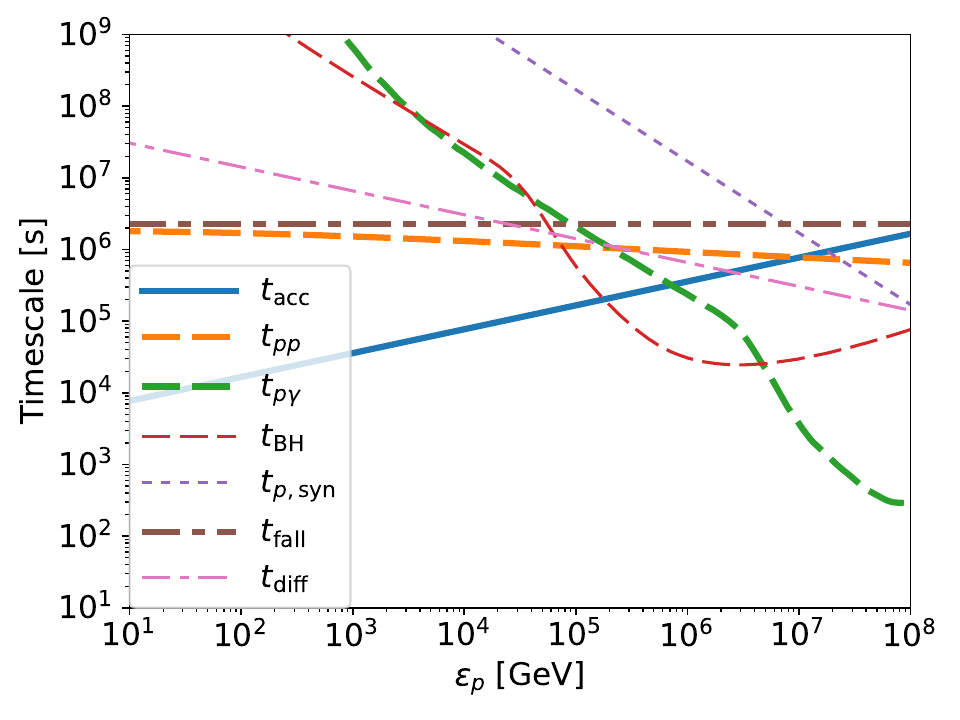}
\includegraphics[width=0.325\linewidth]{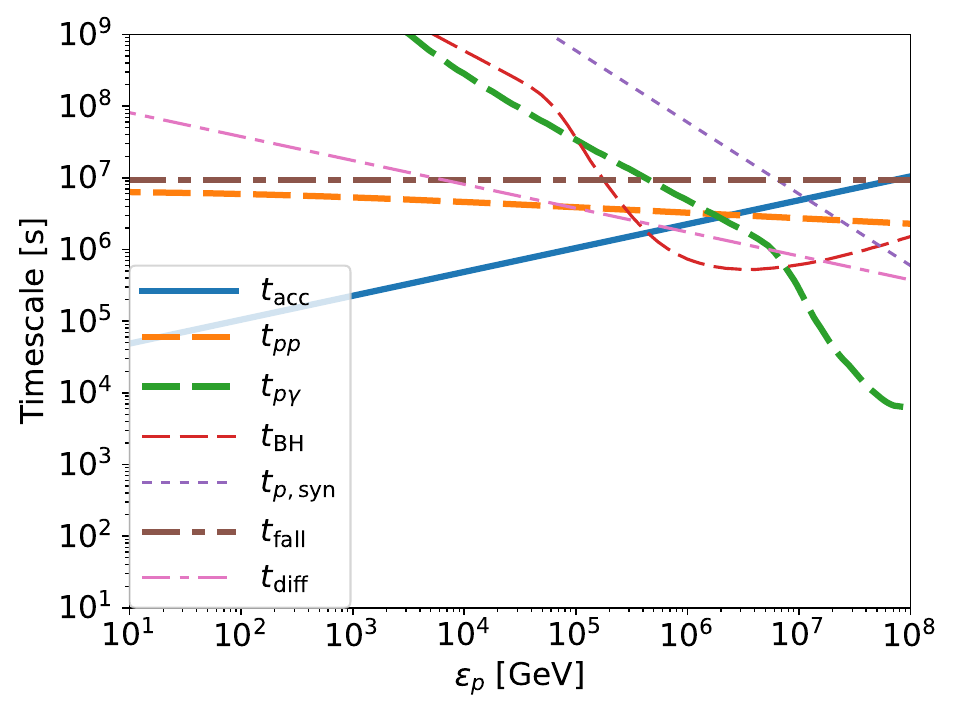}
\includegraphics[width=0.325\linewidth]{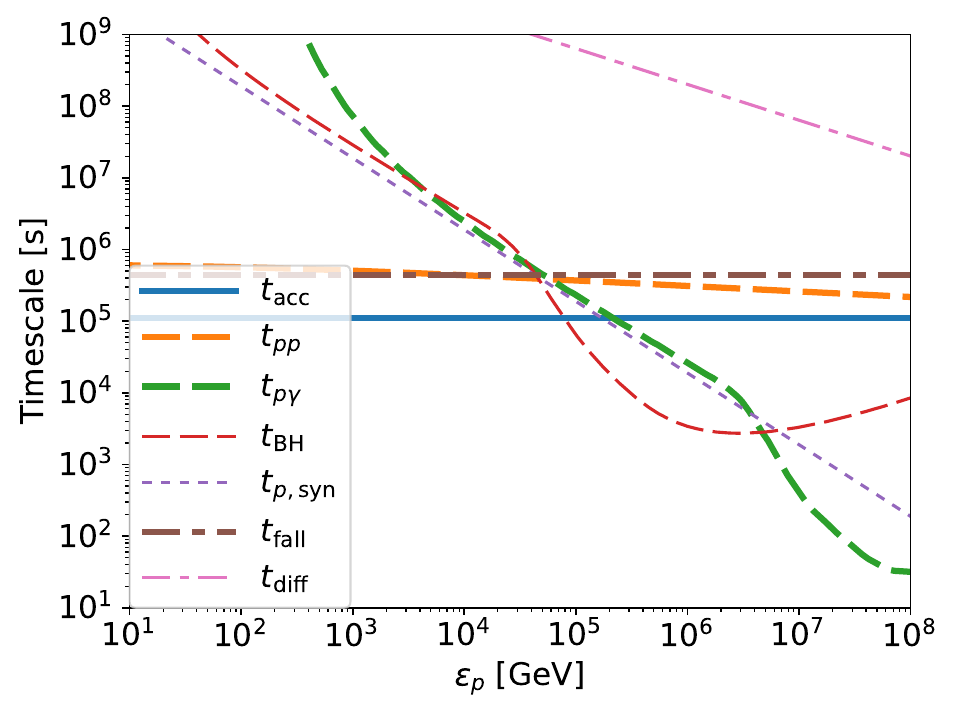}
\caption{Cosmic-ray proton cooling time scales for coronae of jet-quiet AGNs with $L_X=10^{44}~{\rm erg}~{\rm s}^{-1}$ for Model A (left panel), Model B (middle panel), and Model C (right panel).
\label{fig:time}
}
\end{figure*}
\begin{figure*}[th]
\includegraphics[width=0.48\linewidth]{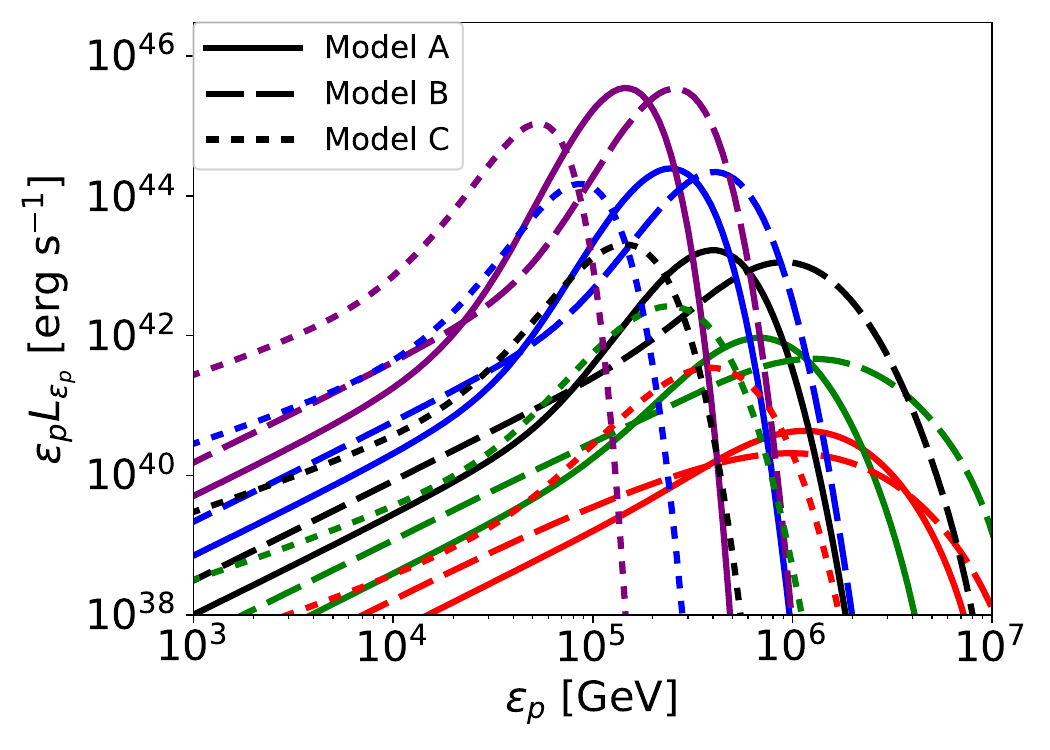}
\includegraphics[width=0.48\linewidth]{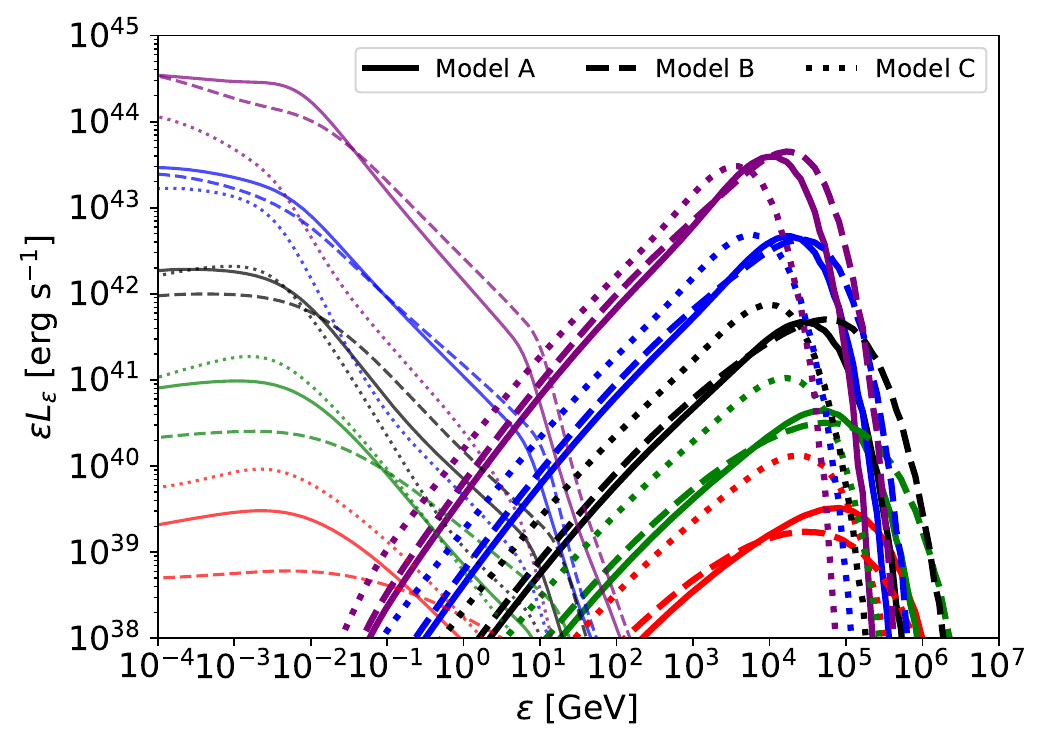}
\caption{Left panel: Differential cosmic-ray luminosities in for $L_X={10}^{42}$, ${10}^{43}$, ${10}^{44}$, ${10}^{45}$, and ${10}^{46}~{\rm erg}~{\rm s}^{-1}$ (from bottom to top). 
Right panel: Differential neutrino (thick) and gamma-ray (thin) luminosities in for $L_X={10}^{42}$, ${10}^{43}$, ${10}^{44}$, ${10}^{45}$, and ${10}^{46}~{\rm erg}~{\rm s}^{-1}$ (from bottom to top).
\label{fig:luminosity}
}
\end{figure*}

\begin{figure*}[th]
\includegraphics[width=0.325\linewidth]{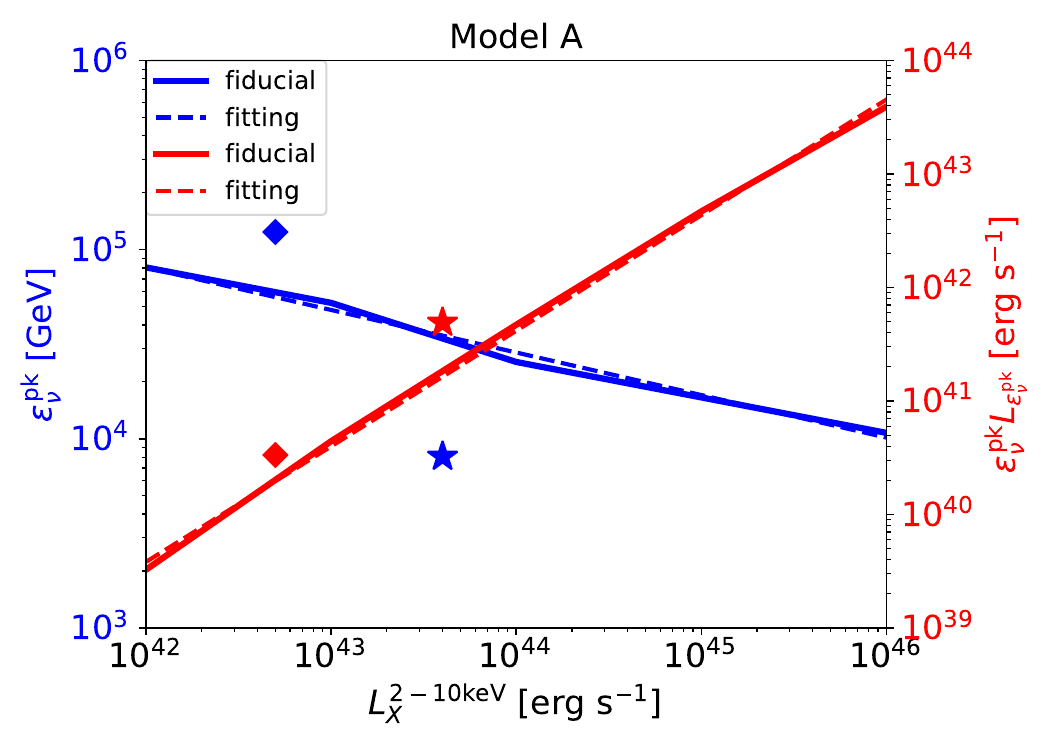}
\includegraphics[width=0.325\linewidth]{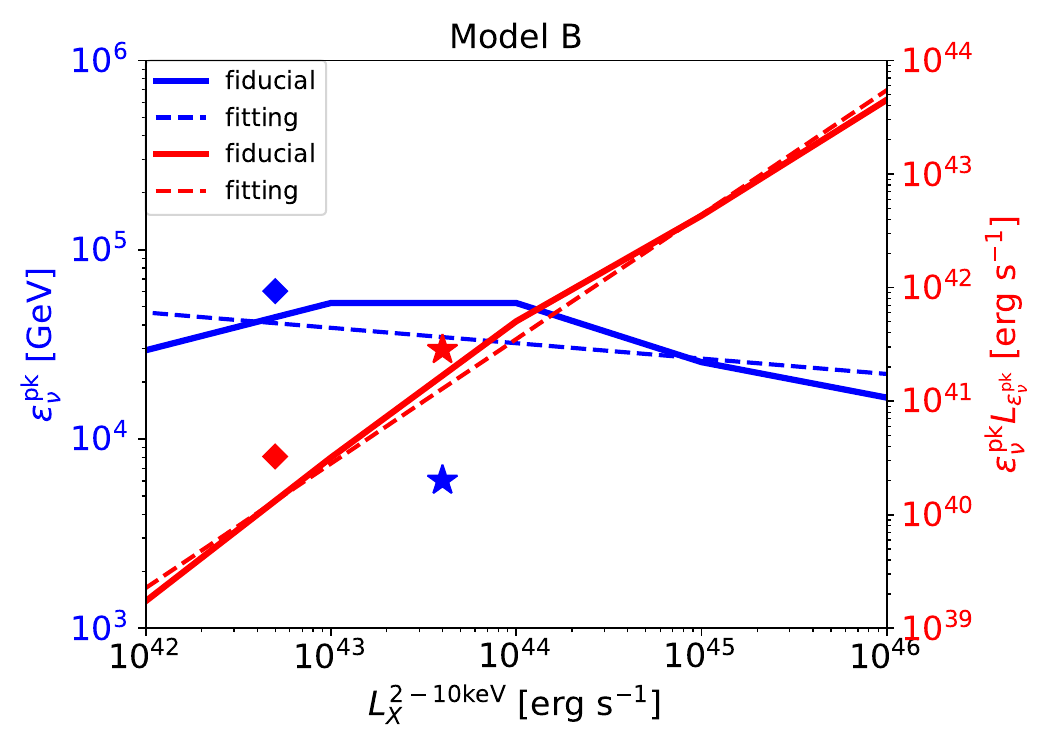}
\includegraphics[width=0.325\linewidth]{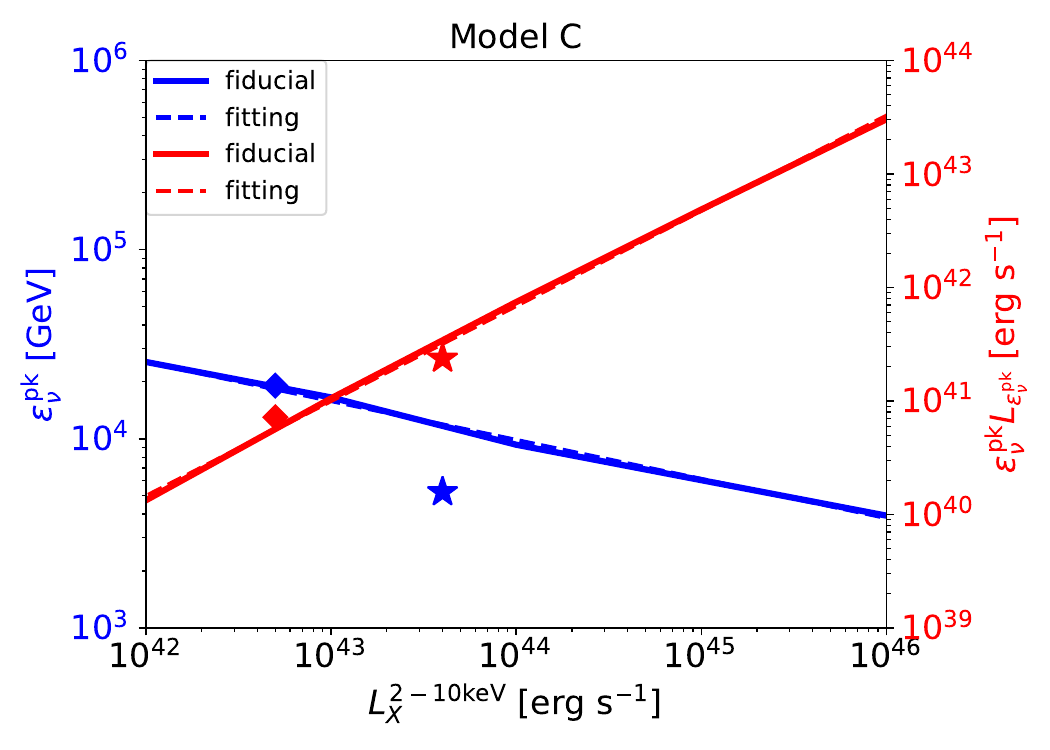}
\caption{X-ray luminosity dependence of the peak neutrino luminosity ($\varepsilon_\nu^{\rm pk}L_{\varepsilon_\nu^{\rm pk}}$) and the peak neutrino energy $\varepsilon_\nu^{\rm pk}$ for Model A (left panel), Model B (middle panel), and Model C (right panel). NGC~1068 and NGC~4151 are indicated by stars and diamonds, respectively.  
\label{fig:peak}
}
\end{figure*}

The latest IceCube shower analyses have reported evidence for a spectral break (or curvature) at $\sim 30~\mathrm{TeV}$ with $\gtrsim 4\sigma$ significance~\cite{IceCube:2025tgp}, and complementary starting track analyses are consistent with a convex spectrum in the $\sim 3~\mathrm{TeV}$ range~\cite{IceCube:2024fxo}. While further experimental studies are necessary to establish the existence of such a spectral bend, we note that such a spectral turnover is theoretically required by an energetics argument for the cumulative medium-energy neutrino background. It is possible to show that a simple, featureless extension to GeV energies is unlikely. 

Ref.~\cite{Murase:2018utn} summarizes luminosity densities of various astrophysical sources in the universe. Among the known source classes, galaxies and AGNs are likely to be dominant in the energy budget of cosmic rays. Their cosmic-ray luminosity densities satisfy
\begin{eqnarray}
Q_{\rm CR}=\int d\varepsilon_p Q_{\varepsilon_p} \lesssim 3\times {10}^{46}~{\rm erg}~{\rm Mpc}^{-3}~{\rm yr}^{-1}. 
\end{eqnarray}
The simple power-law fit with enhanced starting track events in IceCube gives $s_\nu=2.58$~\cite{IceCube:2024fxo}, whereas the analysis with medium-energy starting events leads to $s_\nu=2.55$~\cite{IceCube:2025tgp}. Assuming $\varepsilon_\nu Q_{\varepsilon_\nu} \propto \varepsilon_p^{2-s_\nu} \propto \varepsilon_p^{-0.55}$, from Eq.~(\ref{eq:diffusenu}) and the above requirement we have 
\begin{align}
\varepsilon_p^b \gtrsim 20~{\rm GeV}
\Bigg[
&\left(\frac{E_\nu^2\Phi_\nu|_{100~{\rm TeV}}}
{10^{-7}~{\rm GeV~cm^{-2}~s^{-1}~sr^{-1}}}\right)
\left(\frac{3}{\xi_z}\right)
\nonumber\\
&\times
\left(\frac{1+K}{2K}\right)
\left(\frac{1+f_{\rm BH}+f_{\rm mes}}{f_{\rm mes}}\right)
\nonumber\\
&\times
\left(\frac{Q_{\rm CR}}
{3\times10^{46}~{\rm erg~Mpc^{-3}~yr^{-1}}}\right)
\Bigg]^{20/11}.
\label{eq:breaklimit}
\end{align}
Thus, the fact that the all-sky medium-energy neutrino flux is large suggests that the sources have a spectral break in their population-averaged intrinsic cosmic-ray spectra\footnote{In general, the spectral break may be caused by the superposition of sources with different maximum energies, and it does not have to be inherent to the acceleration mechanism.}. 

MKM20 predicts the existence of a spectral turnover, which is evident in all of the three models presented in this work (see Figs.~\ref{fig:summary} and \ref{fig:diffuse}). 
Actually, a spectral curvature is a generic consequence of viable corona models in which the Bethe-Heitler suppression is important. Using $f_{\rm BH}+f_{\rm mes}+1\sim15 f_{\rm mes}$ in Eq.~(\ref{eq:breaklimit}), we obtain 
$\varepsilon_p^b \gtrsim 3~{\rm TeV}$, leading to $\varepsilon_\nu^b \gtrsim 100~{\rm GeV}$. More specifically, in the MKM20 model, the convex neutrino spectrum can naturally arise from several physical reasons. First, the effective photomeson production optical depth, $f_{p\gamma}\approx t_{\rm fall}/t_{p\gamma}$, increases with $\varepsilon_p$, as seen from Fig.~\ref{fig:time}. As a result, the spectrum of neutrinos from $p\gamma$ interactions can be hard up to certain energy at which the system is calorimetric for cosmic-ray protons, and indeed $f_{p\gamma}\gtrsim1$ is satisfied at $\varepsilon_p\gtrsim30-100$~TeV for $L_X\sim10^{44}~{\rm erg}~{\rm s}^{-1}$ (see Eq.~1 of MKM20). Second, the Bethe-Heitler pair production due to interactions with disk photons suppresses meson production at $\varepsilon_p\sim0.1-1$~PeV, which may cause a turnover in neutrino spectra (cf. Ref.~\cite{Murase:2008sp} for a similar feature in the context of gamma-ray bursts). Third, the stochastic acceleration mechanism leads to a hard spectrum with a cutoff that is related to the maximum energy of cosmic-ray protons. In particular, as shown in Fig.~\ref{fig:luminosity}, luminous objects lead to a pileup feature that is known to be characterized by a power law with a Maxwellian bump (see also Ref.~\cite{Stawarz:2008sp}). This also assures that the results on neutrino and gamma-ray spectra are rather insensitive to the possible existence of a power-law tail in the cosmic-ray spectrum due to the intermittency of turbulence.  
These combination can produce convex neutrino spectra without imposing an {\it ad hoc} phenomenological break, while being consistent with the numerical all-sky neutrino spectra shown for representative model realizations (see Fig.~\ref{fig:diffuse}).

Given that convex cosmic-ray and neutrino spectra are expected at different x-ray luminosities (see Fig.~\ref{fig:luminosity}), the all-sky neutrino contribution above the spectral turnover largely comes from AGNs with $L_X\lesssim{10}^{44}~{\rm erg}~{\rm s}^{-1}$. This situation can be qualitatively understood as follows. From Eq.~(\ref{eq:tacc}), for example, the acceleration time scales as $t_{\rm acc}\propto L_X^{0.5} \varepsilon_p^{-0.3}$ for $q=5/3$ and $t_{\rm acc}\propto L_X^{0.6} \varepsilon_p^{0}$ for $q=2$.  
The maximum energy of cosmic-ray protons is estimated by $t_{\rm acc}=t_{p-\rm cool}$. In Model A, the dominant cooling process is the Bethe-Heitler pair production process. In Model B, this is dominant for $L_X\gtrsim10^{44}~{\rm erg}~{\rm s}^{-1}$, although $pp$ interactions are relevant for lower luminosities. In Model C, the Bethe-Heitler process is dominant for $L_X\gtrsim10^{43}~{\rm erg}~{\rm s}^{-1}$, while proton synchrotron emission is relevant for lower luminosities. Although energy dependence of the Bethe-Heitler process is nontrivial (see Fig.~\ref{fig:time}), in the limited energy range of our interest, we may approximate the scaling as $t_{\rm BH} \propto L_X^{0.1} \varepsilon_p^{-2}$. 
These allow us to naively expect, e.g., $\varepsilon_\nu^{\rm max}\approx0.05\varepsilon_p^{\rm max}\propto L_X^{-0.24}$ for $q=5/3$. 
Indeed, as shown in Fig.~\ref{fig:peak}, we have $\varepsilon_\nu^{\rm pk}\approx28~{\rm TeV}~{(L_X/10^{44}~{\rm erg}~{\rm s}^{-1})}^{-0.225}$ for Model A. For Models B and C, we obtain $\varepsilon_\nu^{\rm pk}\approx33~{\rm TeV}~{(L_X/10^{44}~{\rm erg}~{\rm s}^{-1})}^{-0.081}$ and $\varepsilon_\nu^{\rm pk}=10~{\rm TeV}~{(L_X/10^{44}~{\rm erg}~{\rm s}^{-1})}^{-0.206}$, respectively.  
Here $\varepsilon_\nu^{\rm pk}$ is the peak neutrino energy, at which $\varepsilon_p L_{\varepsilon_p}$ reaches the maximum. Note that conceptually the peak energy is set by the maximum energy but they can be quantitatively different. When the energy dependence of $t_{\rm acc}$ is weak as expected in stochastic acceleration, the peak in the energy spectrum of cosmic rays can be $\sim10$ times higher than the maximum proton energy inferred from equating the acceleration time with the escape or cooling time~\cite{Becker:2006nz,Kimura:2014jba,Murase:2023ccp} (see also Figs.~\ref{fig:time} and \ref{fig:luminosity}). 

One of the model predictions (see Sec.~\ref{sec:scaling} for details) is $L_\nu \propto L_X$, which leads to $E_\nu F_{E_\nu} \propto F_X$, where $F_X$ is the intrinsic x-ray flux. In general, the all-sky neutrino flux is obtained by integrating the flux distribution with flux weights, and we have 
\begin{eqnarray}
E_\nu^2 \Phi_\nu &=& \frac{1}{4\pi}\int dF_X \left(\frac{dN}{dF_X}\right) E_\nu F_{E_\nu} \nonumber\\ 
&\propto& F_X^{-0.8} F_X \propto \varepsilon_\nu^{-1},
\end{eqnarray}
where the results of Ueda et al.~\cite{Ueda:2014tma} are applied to the flux range of our interest, and the differential source number count per unit x-ray flux is given by
\begin{equation}
\frac{dN}{dF_X} = \int_0^{z_{\rm max}} dz \frac{d\rho_{\rm AGN}}{d \log L_X} \frac{d \log L_X}{dF_X} \frac{dV_c}{dz}\,,
\end{equation}
where $dV_c/dz$ is the comoving volume element.
For medium-energy starting events, while the simple power-law fit gives $s_\nu=2.55$, the broken power-law fit leads to a larger index of $s_\nu=2.84$ for the higher-energy portion of the neutrino spectrum~\cite{IceCube:2025tgp}. This may be compatible with high-energy starting-event analyses that give $s_\nu=2.87$~\cite{IceCube:2020wum}. Such a steep spectrum at high energies is consistent with the model expectation. However, one should not overinterpret the above estimate, which can be used only for qualitative understandings. In the magnetically powered corona model, radiative losses and interaction efficiencies are strongly energy and luminosity dependent. In particular, in more luminous (more compact) systems the Bethe-Heitler pair and/or photomeson production losses can more efficiently limit the maximum proton energy and soften the emergent neutrino and gamma-ray spectra, whereas lower-luminosity AGNs can more easily accelerate particles to higher energies and can be more gamma-ray transparent. 

As a result, the magnetically powered corona model predicts that lower-luminosity objects contribute increasingly to the all-sky neutrino flux at higher energies, naturally leading to another spectral component. Such another higher-energy component can arise from low-luminosity AGNs, as shown in Ref.~\cite{Kimura:2020thg} (see also Fig.~\ref{fig:summary}). Intriguingly, the RIAF model for low-luminosity AGNs also predicts $L_\nu \propto L_X$~\cite{Kimura:2019yjo}, and the scaling relation predicted by the corona model could be extended to lower luminosities. Alternatively, as shown in MKM20, gamma-ray transparent sources such as galaxy clusters and groups (with large cosmic-ray reservoirs and long cosmic-ray confinement times, and hadronuclear production in the intracluster medium) can become competitive or even dominant, depending on the cosmic-ray loading of jet-loud AGNs~\cite{Murase:2008yt,Kotera:2009ms,Fang:2017zjf}. These provide another physically motivated route to the extension of IceCube's diffuse neutrino spectrum to PeV energies and beyond with a possible spectral hardening in the $\sim0.1-1$~PeV range.  

It may be worthwhile to note that these consequences of the magnetically powered corona model would be different from those of other models. For example, the accretion shock model that assumes $\beta\gtrsim10$ does not generically predict a convex neutrino spectrum in the $3-30$~TeV range~\cite{Inoue:2019fil,Anchordoqui:2021vms}. It more naturally yields softer power-law neutrino spectra with $s_\nu \gtrsim 2$ accompanied by an eventual cutoff set by the maximum proton energy, although more generally shocks themselves may play roles in strongly magnetized coronae~\cite{Murase:2022dog,Groselj:2026nix}. The confirmation of a spectral turnover that resembles the MKM20 expectation may support the magnetically powered corona interpretation and give insights into coronal physics.

\subsection{Consistency with NGC~1068}\label{sec:consistency}
The magnetically powered corona model for high-energy neutrinos~\cite{Murase:2019vdl} was proposed to explain the all-sky neutrino flux in the medium-energy range, being consistent with the observed neutrino signal from NGC~1068 by construction~\cite{IceCube2022NGC1068,Ajello:2023hkh}. 
Follow-up source-specific calculations presented for NGC~1068 have shown that, with plausible values of $L_X$, $\mathcal R$ and some other parameters, the predicted neutrino spectra can match the observed level while preserving the gamma-ray--hidden nature of the source~\cite{Kheirandish:2021wkm,Eichmann:2022lxh,Ajello:2023hkh,Fiorillo:2024akm,Lemoine:2024roa,Saurenhaus:2025ysu,Carpio:2026}. In this sense, NGC~1068 provides an existence proof that a nearby Seyfert can be neutrino active without violating GeV--TeV gamma-ray constraints~\cite{Murase:2022dog,Blanco:2023dfp,Das:2024vug}.
However, one natural question is how similar the physical parameters inferred from NGC~1068 are with those required by the modeling of the all-sky neutrino flux, and whether the same parameter space can simultaneously satisfy individual source and diffuse background constraints. 

There are two important points to discuss the general consistency between the all-sky neutrino flux and the NGC~1068 neutrino flux. First, NGC~1068, which appears to be the brightest in intrinsic x rays in IceCube's northern sky (including the near-horizon region), may also be regarded as one of the most powerful sources in the local universe. Indeed, from Ueda et al.~\cite{Ueda:2014tma}, we find that the source number density for $L_X=4\times{10}^{43}~{\rm erg}~{\rm s}^{-1}$ is $\rho_{\rm AGN}^{\rm NGC1068}\sim10^{-5}~{\rm Mpc}^{-3}$, implying that the probability to have such an object within $d\sim10$~Mpc in IceCube's northern sky is $\sim 1-\exp(-2\pi \rho_{\rm AGN}^{\rm NGC1068}d^3/3)\sim0.03$. Furthermore, NGC~1068 seems to have the high Eddington ratio, $\lambda_{\rm Edd}\sim1$, which is significantly larger than the average value, $\bar{\lambda}_{\rm Edd}$, used for modeling the all-sky neutrino flux.  
Second, the neutrino flux of NGC~1068 at $\sim10-30$~TeV, which is more relevant for discussing the observed all-sky neutrino flux at medium energies, is lower than the flux at $\sim1$~TeV. Using $E_\nu F_\nu^{\rm NGC1068}\sim(0.3-1)\times{10}^{-8}~{\rm GeV}~{\rm cm}^{-2}~{\rm s}^{-1}$ at $\sim10-30$~TeV, the neutrino luminosity at relevant energies is estimated to be $\varepsilon_\nu L_{\varepsilon_\nu}=(4\pi d_L^2)(E_\nu F_{E_\nu})\sim(0.6-2)\times 10^{41}~{\rm erg}~{\rm s}^{-1}$. As a result, the all-sky neutrino flux from NGC~1068-like AGNs is  
\begin{eqnarray}\label{eq:NGC1068diffuse}
E_\nu^2\Phi_\nu &\simeq& 6.3\times{10}^{-8}~{\rm GeV}~{\rm cm}^{-2}~{\rm s}^{-1}~{\rm sr}^{-1}~\left(\frac{\xi_z}{3}\right)\nonumber\\
&\times&\left(\frac{\rho_{\rm AGN}^{\rm NGC1068}}{{10}^{-5}~{\rm Mpc}^{-3}}\right)
{\left(\frac{\varepsilon_\nu L_{\varepsilon_\nu}^{\rm NGC1068}}{{10}^{41}~{\rm erg}~{\rm s}^{-1}}\right)}.\,\,\,\,\,\,\,\,
\end{eqnarray}
Note that the distribution of ${\lambda}_{\rm Edd}$ is not considered, and the above estimate is optimistic in the sense that only a fraction of AGNs have ${\lambda}_{\rm Edd}\sim1$. Nevertheless, Eq.~(\ref{eq:NGC1068diffuse}) shows the consistency with with the all-sky neutrino data at medium energies, so we conclude that there is no obvious tension between the all-sky flux and the NGC~1068 flux. In reality, the all-sky neutrino flux comes from AGNs with different luminosities, and the above flux could be overwhelmed by the contribution from objects unlike NGC~1068, but this is a model-dependent problem. Indeed, after MKM20, several authors reached the similar conclusion that jet-quiet AGNs can give a major contribution to the all-sky neutrino flux at medium energies~\cite{Padovani:2024tgx,Fiorillo:2025ehn,Saurenhaus:2025ysu}.  
Case studies on NGC~1068 and other objects can rather be used to reinforce the importance of accounting for systematics and intrinsic dispersion (notably in $\lambda_{\rm Edd}$) when translating x-ray observables into neutrino expectations for individual AGNs.

\subsection{Implications for the MeV Gamma-Ray Background}
One of the robust predictions of the magnetically powered corona model is an accompanying contribution to the extragalactic MeV gamma-ray background, which can be tested with planned MeV missions such as {\it AMEGO-X}~\cite{Caputo:2022xpx}, {\it e-ASTROGAM}~\cite{e-ASTROGAM:2016bph} and {\it GRAMS}~\cite{Aramaki:2019bpi}. Both hadronuclear and photomeson production interactions that generate neutrinos also inject high-energy photons and electron-positron pairs, and the Bethe-Heitler pair production process is shown to give a dominant contribution in the presence of disk photons. In compact AGN coronae, these electromagnetic secondaries undergo strong absorption and subsequent cascades through interactions with optical, ultraviolet, and x-ray photons, leading to efficient reprocessing of GeV--PeV gamma rays to lower energies. The MeV gamma-ray background flux is estimated by
\begin{eqnarray}
E_\gamma^2\Phi_\gamma &\approx& \frac{ct_H}{4\pi} (\varepsilon_\gamma {\mathcal G}^s_{\varepsilon_\gamma} Q_{\gamma}) \xi_z\nonumber\\
&\approx& \frac{ct_H}{4\pi}\left(\frac{\frac{4+K}{4[1+K]}f_{\rm mes}+f_{\rm BH}}{1+f_{\rm mes}+f_{\rm BH}}\right)\left(\frac{\xi_{{\rm CR}/X}L_X\rho_{\rm AGN}\xi_z}{{\mathcal C}_{\rm CR}}\right) \,\,\,\,\,\,\,\,\,\,\,\,\nonumber\\
&\simeq&8.0\times{10}^{-7}~{\rm GeV}~{\rm cm}^{-2}~{\rm s}^{-1}~{\rm sr}^{-1}\left(\frac{\xi_z}{3}\right)\nonumber\\
&\times&\left(\frac{\varepsilon_\gamma {\mathcal G}^s_{\varepsilon_\gamma}}{0.2}\right)\left(\frac{\frac{4+K}{4(1+K)}f_{\rm mes}+f_{\rm BH}}{1+f_{\rm BH}+f_{\rm mes}}\right)\nonumber\\
&\times&{\left(\frac{\xi_{{\rm CR}/X,-1}{\mathcal C}_{\rm CR}^{-1}L_X\rho_{\rm AGN}}{2\times{10}^{46}~{\rm erg}~{\rm Mpc}^{-3}~{\rm yr}^{-1}}\right)},
\end{eqnarray}
where $\varepsilon_\gamma {\mathcal G}^s_{\varepsilon_\gamma}$ represents the shape of the energy spectrum of cascaded gamma rays and its normalization is determined by $\int d\varepsilon_\gamma{\mathcal G}^s_{\varepsilon_\gamma}=1$. See Ref.~\cite{Murase:2012df} for an analogous definition for intergalactic electromagnetic cascades. 

Remarkably, in the magnetically powered corona model, predictions for MeV gamma-ray emission are rather robust against $\xi_B$. Even for sufficiently large values of $\xi_B$ in which the synchrotron cascade is dominant, the characteristic synchrotron energy of the electron-positron pairs appears above MeV energies, and the gamma-ray output is attenuated above a few MeV whether the cascade is governed by synchrotron emission or inverse-Compton emission~\cite{Murase:2019vdl}. However, quantitative details of the observed gamma-ray flux are affected by $\xi_B$, which can be used for constraining physical properties of the emission regions~\cite{Murase:2022dog,Das:2024vug}. In general, the synchrotron spectrum directly from Bethe-Heitler pairs reflects details of the differential cross section. While it is largely smeared out by cascades when the coronal compactness is small enough, careful calculations beyond the analytical approximation for the Bethe-Heitler process can be important~\cite{Murase:2022dog,Das:2024vug}. 

It is also important to recognize that the baseline MeV background fluxes predicted in the present framework should be regarded as lower limits. This is because inverse-Compton emission from nonthermal electrons accelerated through magnetic reconnections~\cite{Inoue:2007tn,Groselj:2026nix} and synchrotron or inverse-Compton emission from secondary electron-positron pairs reaccelerated by turbulence~\cite{Murase:2019vdl} can further enhance the MeV gamma-ray background. Combined measurements in MeV gamma rays and neutrinos should offer a decisive test of the coronal interpretation of the medium-energy neutrino background.

\section{Interpretations of Individual Sources}
One of the appealing points of the magnetically powered corona model is that within IceCube's field of view NGC~1068 is expected to be the neutrino-brightest AGN and thus ranks first. Note that this is a consequence of modeling that approximately predicts $F_\nu \propto F_X$, and it does not depend on whether future neutrino observations confirm $\sim4\sigma$ evidence for the signal from NGC~1068 or not. The exact ordering depends on the catalog, x-ray bandpass, and absorption correction due to $N_H$ (that is relevant especially for Compton-thick AGNs). Also, some x-ray sources may not be physically justified. For example, x rays from the radio galaxy Centaurus A may come from jets, and the merging galaxy NGC~6240 may have a significant contribution from starburst activity especially in the soft x-ray band. Thus, it is important to prespecify neutrino-active AGN catalogs with our best knowledge. Interestingly, when the x-ray spectrum is integrated over energies, the second brightest source in the northern sky would be NGC~4151. In this work, we discuss these two sources as examples.  

\subsection{Examples: NGC~1068 and NGC~4151}
%
\begin{table}[t]
\begin{center}
\caption{Parameters of the magnetically powered corona model for NGC~1068 and NGC~4151. The other parameters except $\lambda_{\rm Edd}$ and $\mathcal R$ are the same as those for the all-sky neutrino flux in.  
\label{tab:NGC1068NGC4151}
}
\scalebox{1.0}{
\begin{tabular}{|c||c|c|c|}
\hline Source (Model) & $L_{X}$ [${\rm erg}~{\rm s}^{-1}$] & $\lambda_{\rm Edd}$ $(M [M_{\odot}])$ & $\mathcal R$ \\
\hline NGC 1068 (A) & $4^{+6}_{-3}\times10^{43}$ & $1.9^{+4.4}_{-1.6}$ $(6\times{10}^{6})$ & $3$\\
\hline NGC 1068 (B) & $4^{+6}_{-3}\times10^{43}$ & $1.9^{+4.4}_{-1.6}$ $(6\times{10}^{6})$ & $3$\\
\hline NGC 1068 (C) & $4^{+6}_{-3}\times10^{43}$ & $1.1^{+2.6}_{-0.9}$ $({10}^{7})$ & $10$ \\
\hline
\hline NGC 4151 (A) & $5^{+5.0}_{-2.4}\times10^{42}$ & $4.7^{+6.4}_{-2.6}\times{10}^{-2}$ $(1.7\times{10}^{7})$ & $30$\\
\hline NGC 4151 (B) & $5^{+5.0}_{-2.4}\times10^{42}$ & $4.7^{+6.4}_{-2.6}\times{10}^{-2}$  $(1.7\times{10}^{7})$ & $10$\\
\hline NGC 4151 (C) & $5^{+5.0}_{-2.4}\times10^{42}$ & $7.9^{+11.1}_{-4.3}\times{10}^{-2}$ $({10}^{7})$ & $10$ \\
\hline
\end{tabular}
}
\end{center}
\end{table}

\begin{figure*}[th]
\includegraphics[width=0.325\linewidth]{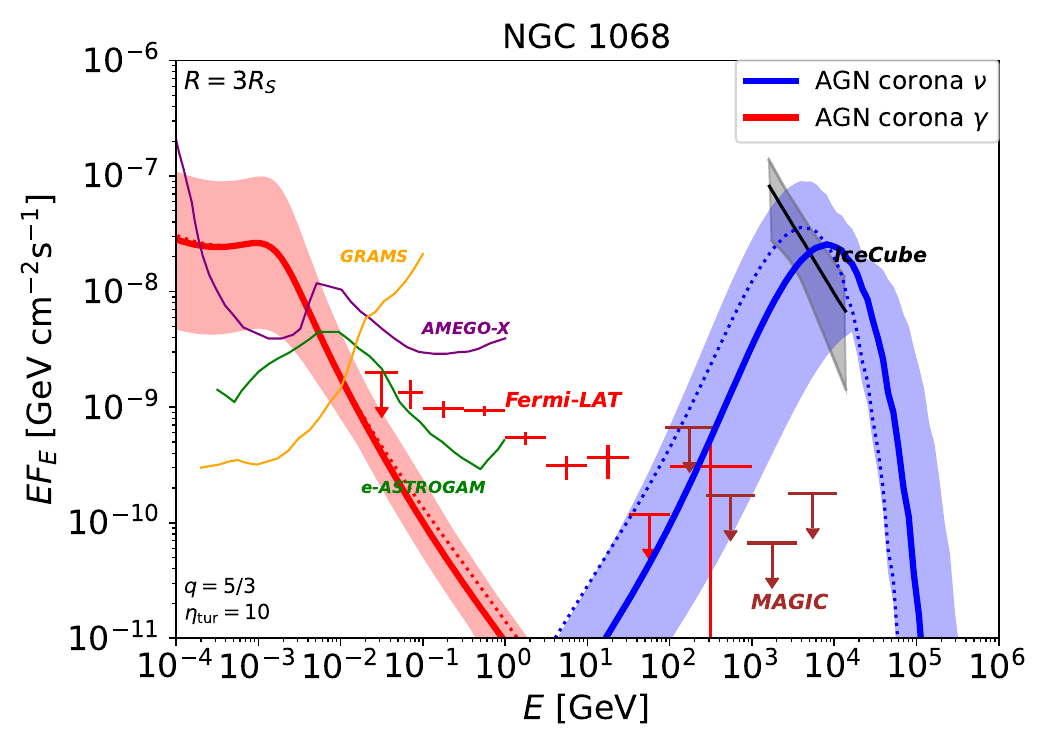}
\includegraphics[width=0.325\linewidth]{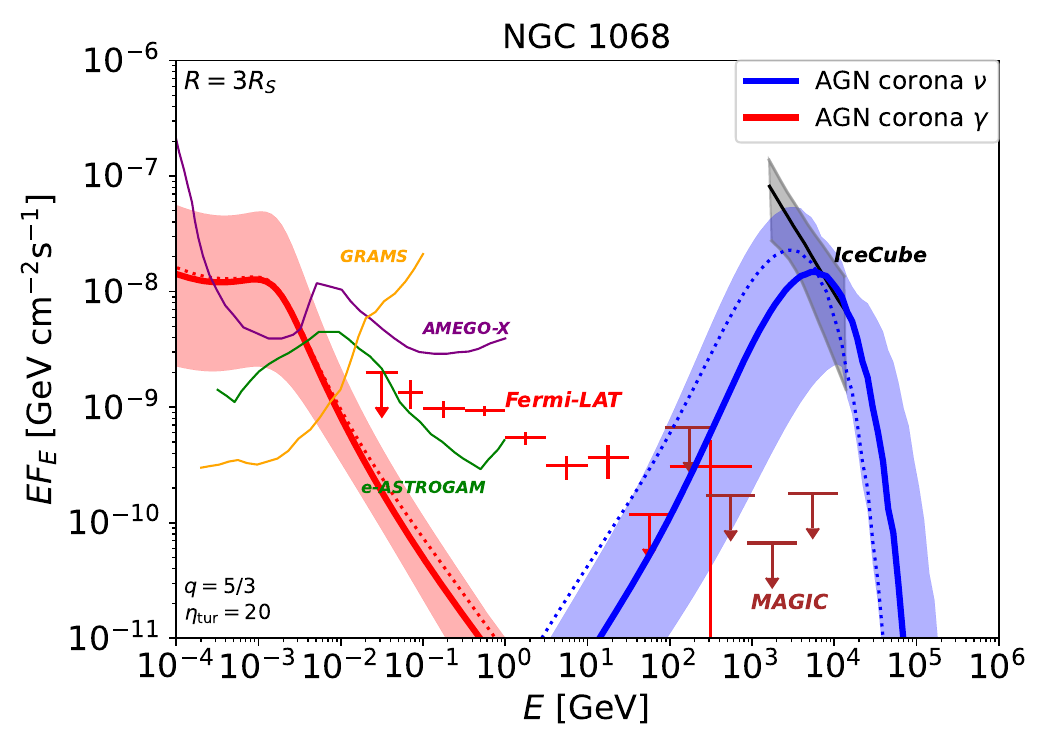}
\includegraphics[width=0.325\linewidth]{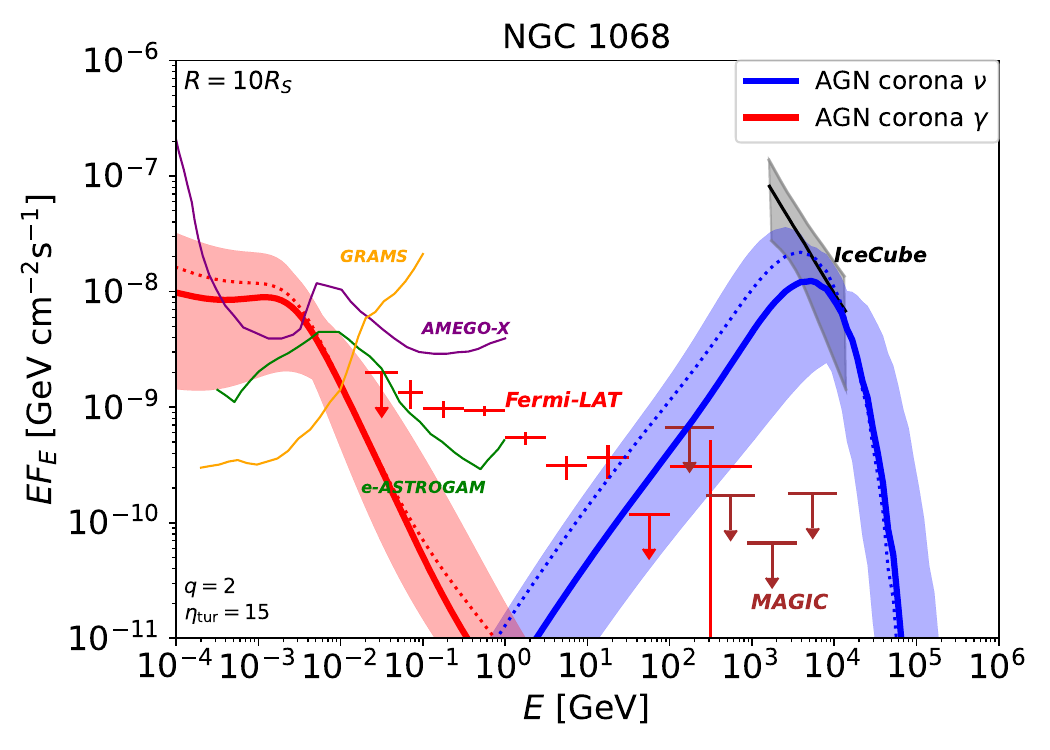}
\caption{Neutrino and gamma-ray spectra of NGC 1068 for Model A (left), Model B (middle), and Model C (right). For the solid curves, the cosmic-ray pressure ratios resulting from the all-sky neutrino flux models are $\hat{p}_{\rm cr}=0.05$ (left), $0.05$ (middle), and $0.1$ (right). For the dotted curves, $\hat{p}_{\rm cr}$ and $\beta$ are tuned as $0.05$ and $3$ (left), $0.05$ and $3$ (middle), and $0.1$ and $0.03$ (right), respectively. The neutrino and gamma-ray are taken from IceCube~\cite{IceCube2022NGC1068} and {\it Fermi} LAT~\cite{Ajello:2023hkh}, respectively. 
We also show sensitivities for {\it e-ASTROGAM}~\cite{e-ASTROGAM:2016bph} and {\it GRAMS}~\cite{Aramaki:2019bpi} with an effective exposure time of 1~yr, together with the {\it AMEGO-X} sensitivity for the 3~yr mission~\cite{Caputo:2022xpx}.
\label{fig:NGC1068}
}
\end{figure*}

\begin{figure*}[th]
\includegraphics[width=0.325\linewidth]{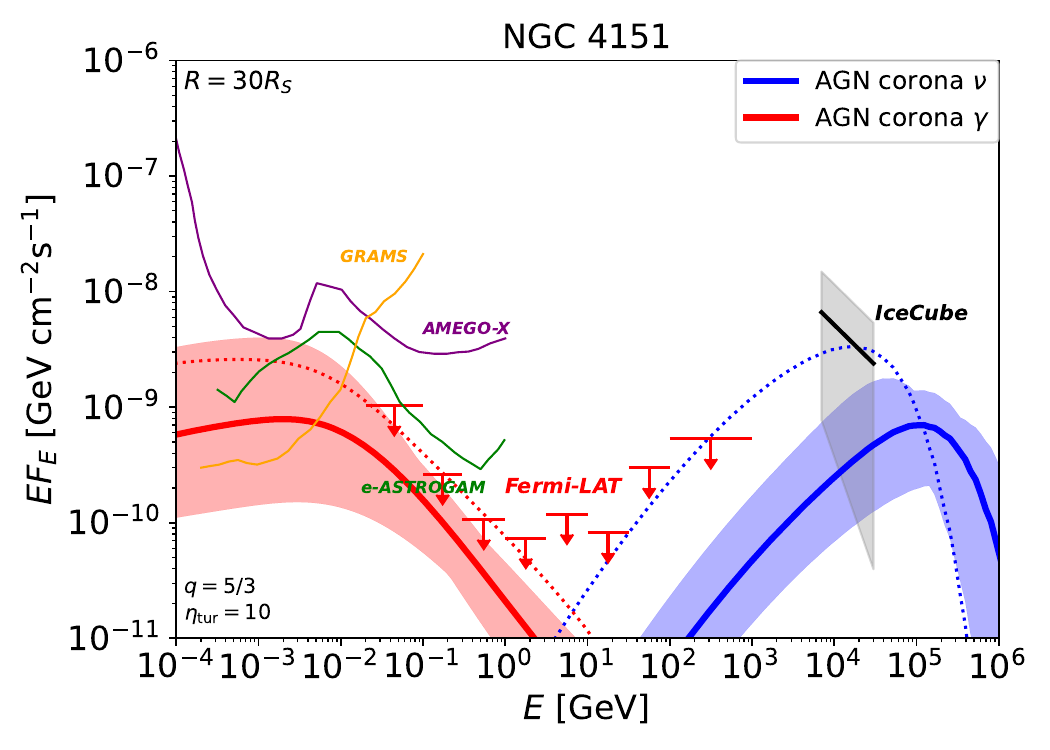}
\includegraphics[width=0.325\linewidth]{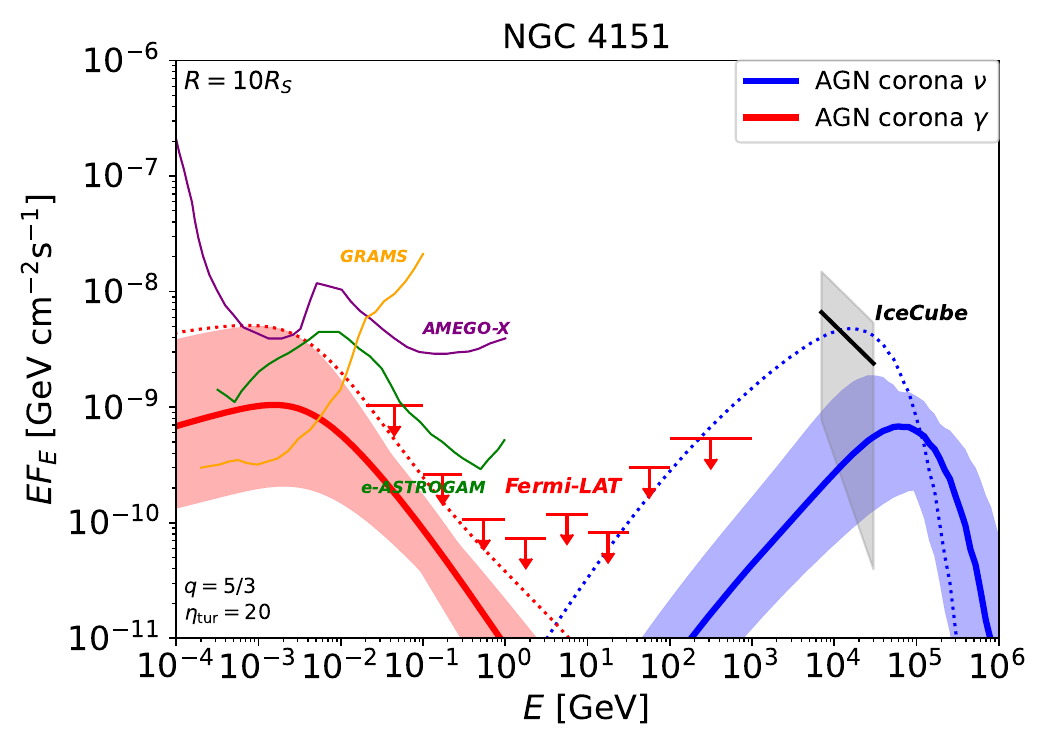}
\includegraphics[width=0.325\linewidth]{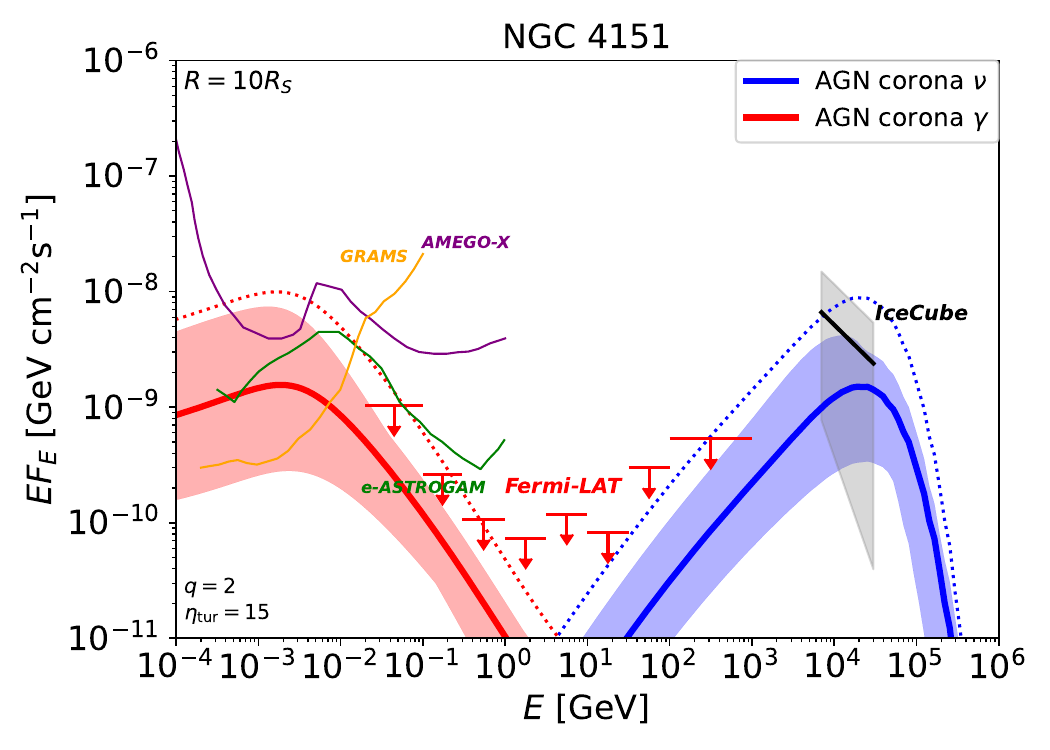}
\caption{Neutrino and gamma-ray spectra of NGC 1068 for Model A (left), Model B (middle), and Model C (right). For the solid curves, the cosmic-ray pressure ratios resulting from the all-sky neutrino flux models are $\hat{p}_{\rm cr}=0.05$ (left), $0.05$ (middle), and $0.1$ (right). For the dotted curves, $\hat{p}_{\rm cr}$ and $\beta$ are tuned as $0.05$ and $3$ (left), $0.05$ and $3$ (middle), and $0.1$ and $0.03$ (right), respectively. The neutrino and gamma-ray are taken from IceCube~\cite{Murase:2023ccp} and {\it Fermi} LAT~\cite{Murase:2023ccp}, respectively. 
\label{fig:NGC4151}
}
\end{figure*}

\begin{figure*}[th]
\includegraphics[width=0.48\linewidth]{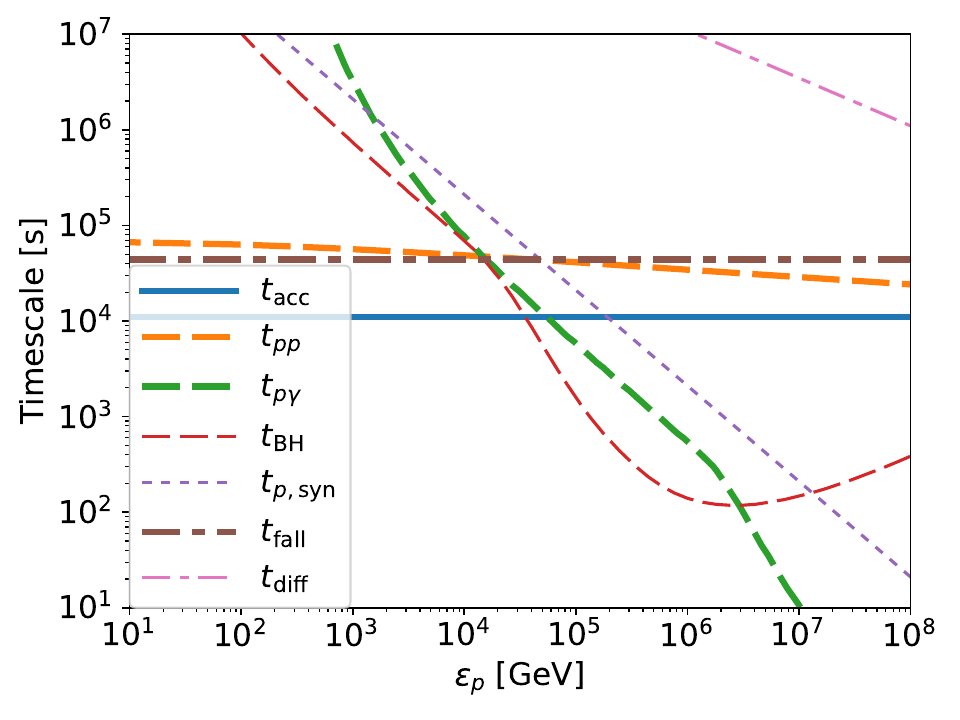}
\includegraphics[width=0.48\linewidth]{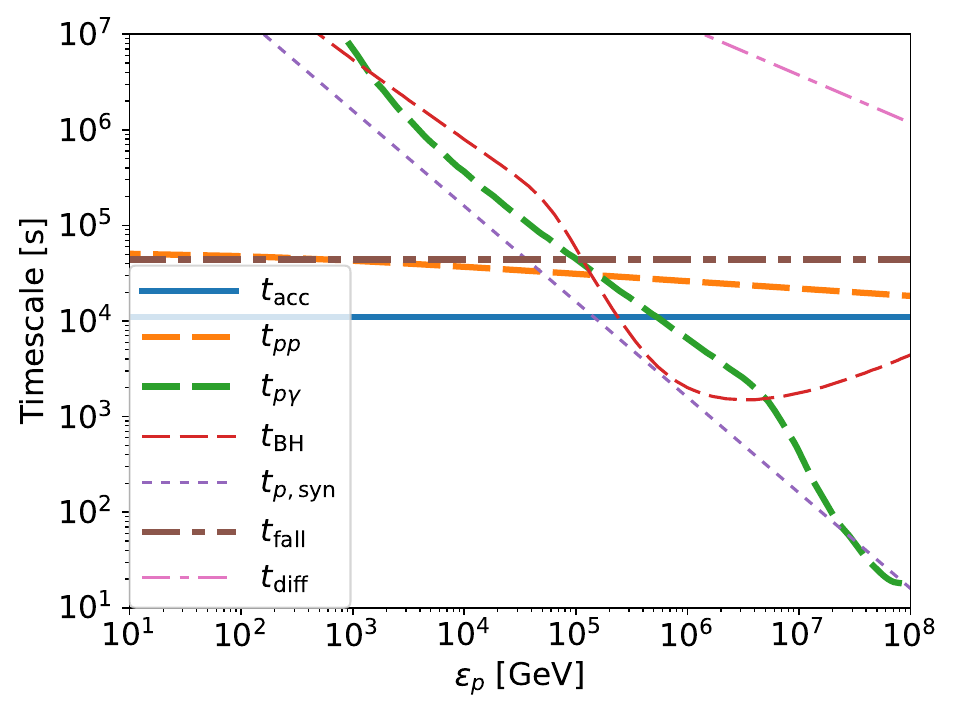}
\caption{Cosmic-ray proton cooling time scales for NGC~1068 (left panel) and NGC~4151 (right panel) in Model C. 
\label{fig:timeNGC}
}
\end{figure*}

The IceCube Collaboration has reported $\sim 4\sigma$ evidence for neutrinos from the nearby Seyfert galaxy NGC 1068, making it a benchmark target for testing the magnetically powered corona model. To illustrate how average physical parameters for the all-sky flux are compatible with multimessenger data of individual objects, we also present spectra for NGC~1068 and NGC~4151 (cf. Refs.~\cite{Murase:2022dog,Blanco:2023dfp,Das:2024vug} for generic studies). For each source, we consider the allowed ranges of the intrinsic $2-10$~keV x-ray luminosity and Eddington ratios, and vary the coronal compactness, as summarized in Table~\ref{tab:NGC1068NGC4151}, while keeping other model parameters same as those used in the all-sky calculations.

For NGC~1068, based on Ref.~\cite{Marinucci:2015fqo}, we take $L_X=4\times10^{43}~{\rm erg~s^{-1}}$ (with the uncertainty range indicated in Table~\ref{tab:NGC1068NGC4151}) and $\lambda_{\rm Edd}$ of order unit, where for $d=10$~Mpc we have SMBH masses of $M=6\times{10}^6~M_\odot$ and ${10}^7~M_\odot$, depending on the self-gravity of the disk~\cite{Lodato:2002cv}. NGC~1068 is a near-Eddington system with $\lambda_{\rm Edd}\sim1$, which motivates us to consider smaller values of the coronal compactness, and we consider ${\mathcal R}\sim3-10$. Fig.~\ref{fig:NGC1068} shows the resulting neutrino and gamma-ray spectra for three representative model variants. The predicted neutrino flux can reproduce the level and spectral shape implied by IceCube data within uncertainties, while the emergent gamma-ray signal remains consistent with {\it Fermi} LAT observations and TeV constraints because the coronal region is opaque and electromagnetic cascades reprocess the high-energy emission. The dotted curves in Fig.~\ref{fig:NGC1068} demonstrate that modest tuning of the cosmic-ray pressure (parameterized by $\hat{p}_{\rm cr}$ in the figure caption) and/or the plasma beta can bring the predicted flux into closer agreement with the inferred neutrino signal. 
The required cosmic-ray pressure is $\sim 2$--$5\%$, and our results are consistent with the viable parameter space shown in Ref.~\cite{Ajello:2023hkh}.
For compact coronae with ${\mathcal R}\sim3$, photohadronic interactions including the Bethe-Heitler pair production are important, and the cooling-induced pileup helps the peak neutrino energy be lower while keeping the neutrino flux at the observed level. It may be useful to contrast our representative choices with the $\eta_{\rm tur}=70$ model considered in MKM20. In that case, acceleration is less efficient and the neutrino output largely comes from hadronuclear interactions. Although this case may be closer to the best-fit parameters~\cite{Carpio:2026}, given that we have not reached the discovery level, we should keep in mind that current uncertainties are too large to discuss details. We also note that uncertainties in the intrinsic x-ray luminosity are large in Compton-thick AGNs, and resulting impacts on the neutrino flux could be even larger~\cite{Kheirandish:2021wkm}. 

NGC~4151 is known to be variable, and the x-ray luminosity may have $L_X \sim (2-20)\times 10^{42}~\mathrm{erg~s^{-1}}$~\cite{Yang:2001ce}. Although Ref.~\cite{Murase:2023ccp} adopts $L_X=2.6\times 10^{42}~\mathrm{erg~s^{-1}}$, motivated by Ref.~\cite{Gianolli:2023zji}, we use $L_X=5\times 10^{42}~\mathrm{erg~s^{-1}}$ and allow $L_X=10^{43}~\mathrm{erg~s^{-1}}$. For SMBH masses, we use the results of Refs.~\cite{Bentz+2022} and \cite{Kormendy:2013dxa}, in which inferred values of $\lambda_{\rm Edd}$ are at the few percent level, being not far from $\bar{\lambda}_{\rm Edd}$. Thus, we ${\mathcal R}\sim10-30$ as listed in Table~\ref{tab:NGC1068NGC4151}.Fig.~\ref{fig:NGC4151} presents the corresponding neutrino and gamma-ray predictions. Although a discovery-level neutrino excess has not been established for this source, revealing neutrino emission from this source is crucial for the purpose of testing proposed models. First, NGC~4151 is a Compton-thin AGN, and uncertainties from the absorption correction are smaller than those in NGC~1068, and the coronal geometry has been constrained by recent {\it IXPE} observations~\cite{Gianolli:2023zji}. Second, upper limits on GeV gamma rays are stringent~\cite{Murase:2023ccp}, which enable us to quantitatively test the hidden behavior, including the energy fraction carried by cosmic rays. For example, the accretion shock model with $s_\nu\gtrsim2$ extended down to GeV energies does not account for the observed level of the IceCube flux~\cite{Murase:2023ccp}. Even in the stochastic acceleration scenario, smaller values of the coronal compactness are favored, which could require larger parameter variations than in NGC~1068.   

We stress that this work does not present a source-by-source detailed parameter fit. We instead adopt representative coronal parameters motivated by the all-sky neutrino flux data, and highlight qualitative behaviors through simple rescalings (e.g., via $\hat{p}_{\rm cr}$) to illustrate the level of tuning required. A comprehensive treatment, including a systematic exploration of the coronal compactness, acceleration efficiency, and population uncertainties, will be presented in Ref.~\cite{Carpio:2026}.

Finally, our baseline calculations have assumed (quasi)steady-state emission. However, Seyfert galaxies such as NGC~4151 may drastically vary on timescales of decades~\cite{2007OAP....20..160O}, and the possibility of changing-look behaviors implies that the long-term averaged coronal state may differ from any single-epoch measurement. Consequently, rankings based on the intrinsic x-ray fluxes (and the associated neutrino predictions) may evolve when averaged over long baselines, and time-domain information with x-ray monitoring may be important for optimizing stacking strategies and interpreting apparent source-to-source differences.

\subsection{Implications for the Coronal Plasma}\label{sec:1068implication}
M22~\cite{Murase:2022dog} showed that the emission regions of high-energy neutrinos from NGC~1068 is likely to be located in the vicinity of the SMBH. Especially in the $p\gamma$ scenario, thanks to the electromagnetic output from the Bether-Heitler pair production process, the coronal compactness has to be constrained to be ${\mathcal R}\lesssim15-30$~\cite{Das:2024vug}. In addition, the latest gamma-ray analysis on the {\it Fermi} LAT data~\cite{Ajello:2023hkh} suggests $\xi_B\equiv U_{B}/U_{\gamma}\gtrsim0.1$~\cite{Das:2024vug}, where $U_\gamma\approx L_{\rm bol}/(2\pi {\mathcal R}^2 R_S^2 c)$, because inverse-Compton cascade emission can more easily overshoot the {\it Fermi} LAT data in the GeV--TeV range. Thus, the multimessenger data constrain the plasma beta as
\begin{eqnarray}
\beta \approx \left(\frac{\tau_T}{2\sqrt{3}\zeta_e\lambda_{\rm Edd}}\right)\xi_B^{-1}
\lesssim 1~\tau_{T,-0.4}\zeta_e^{-1}\lambda_{\rm Edd}^{-1}.
\label{eq:betalimitMM}
\end{eqnarray}
The corresponding lower limit on the magnetic field strength is
\begin{eqnarray}
B&\gtrsim&\sqrt{\frac{0.4L_{\rm bol}}{{\mathcal R}^2 R_S^2 c}}
\nonumber\\
&\simeq&5.5\times10^{3}~{\rm~G}~{\mathcal R}_1^{-1}\lambda_{\rm Edd}{\left(\frac{L_{\rm bol}}{20 L_X}\right)}^{-1/2}L_{X,43.6}^{-1/2}.\,\,\,\,\,
\end{eqnarray}
We stress that Eqs.~(\ref{eq:betalimitX}) and (\ref{eq:betalimitMM}) are independent, but the latter limit on $B$ does not rely on whether the corona is powered by magnetic fields. This leads to a conclusion that the corona is strongly magnetized if the neutrino emission region is associated with an x-ray-emitting corona. 
Interestingly, for NGC~1068, the cascade limit is comparable to the limit given by Eq.~(\ref{eq:betalimitX}), which implies $L_B\gtrsim L_X$ (within an order of magnitude), being consistent with the picture of a magnetically powered corona.  

On the other hand, from Eq.~(\ref{eq:L_B}) (that comes from $\eta_B<1$), we obtain the upper limit on $B$ as 
\begin{eqnarray}
B&\lesssim&4.6\times10^{4}~{\rm~G}~\eta_{\rm rad,-1}^{-1/3}\tau_{T,-0.4}^{1/6}\zeta_e^{-1/6}\mathcal{R}_{1}^{-5/6}\nonumber\\
&\times&\lambda_{\rm Edd}^{5/6}{\left(\frac{L_{\rm bol}}{20 L_X}\right)}^{-1/2}L_{X,43.6}^{-1/2}.
\end{eqnarray}
Thus, if the emission region of high-energy neutrinos from NGC~1068 coincides with the x-ray emitting region, the global magnetic field is constrained to be in the range of 
\begin{equation}
5~{\rm kG}~{\mathcal R}_1^{-1} \lesssim B \lesssim 50~{\rm kG}~{\mathcal R}_1^{-5/6}\zeta_e^{-1/6},
\end{equation}
because of $\lambda_{\rm Edd}\sim1$ and $L_X\sim4\times{10}^{43}~{\rm erg}~{\rm s}^{-1}$. However, we stress that we do not exclude the possibility that the coronal region is extended or outflowing with weaker magnetic fields, as indicated from some models accounting for millimeter observations \cite{Eichmann:2022lxh,Hankla:2025bqc}. More detailed multizone modeling is necessary to understand broadband spectra, which is beyond the scope of this work.   

The required cosmic-ray pressure in all of our models for NGC~1068 is $\hat{p}_{\rm cr}\lesssim5-10\%$ (see Fig.~\ref{fig:NGC1068}), which leads to
\begin{equation}
\frac{U_{\rm CR}}{U_{B}}=3\hat{p}_{\rm cr}\beta \lesssim 0.1-0.3.
\end{equation}
Thus, the energy density of cosmic rays is always subdominant, consistent with the assumption of magnetic energization. (Note that $\hat{p}_{\rm cr}\leq0.5$ should be satisfied more conservatively~\cite{Murase:2020lnu}.) Given that MKM20 assumes $t_{\rm diss}\approx 0.1~\alpha_{-1}\sqrt{\beta/2}t_{\rm fall}$, we find $t_{p-\rm loss}^*\gtrsim t_{\rm diss}$ (see Fig.~\ref{fig:timeNGC}), which leads to $\epsilon_{\rm cr}\lesssim0.1-1$. Note that the satisfaction of such a necessary energy budget requirement had also been examined in the models considered in MKM20, including the $\eta_{\rm tur}=70$ model with $\hat{p}_{\rm cr}=0.3$. 

On the other hand, the minimum cosmic-ray luminosity can be estimated from the observed neutrino luminosity, which can also be confronted with the x-ray luminosity and/or bolometric luminosity. As shown in M22, in the stochastic acceleration scenario that predicts hard spectra with $s\lesssim2$ at energies below the maximum proton energy, $L_{\rm CR}\lesssim L_X$ can be satisfied given that the meson production is efficient enough. However, for $s\gtrsim2$ down to GeV energies, which is expected in the shock acceleration scenario, we obtain $L_{\rm CR}\gtrsim [4(1+K)/(3K)]10 (\varepsilon_\nu L_{\varepsilon_\nu})|_{\varepsilon_\nu=1~\rm TeV}$, and $L_{\rm CR} \gtrsim L_X$ may be required in the presence of strong suppression of neutrino production due to the Bethe-Heitler pair production. In general, it may be challenging to expect such a high power of nonthermal dissipation (e.g., by cosmic rays) that is larger than the thermal dissipation power (e.g., by heating).     

The Larmor radius of cosmic-ray protons in the relevant energy scale is 
\begin{eqnarray}
r_L=\frac{\varepsilon_p}{eB}
&\simeq&5.7\times{10}^{7}~{\rm cm}~\left(\frac{\varepsilon_p}{100~\rm TeV}\right)\tau_{T,-0.4}^{-1/2}\zeta_e^{1/2}{\mathcal R}_1\nonumber\\
&\times&\lambda_{\rm Edd}^{-1/2}{\left(\frac{L_{\rm bol}}{20 L_X}\right)}^{1/2}L_{X,43.6}^{1/2}\beta^{1/2},
\end{eqnarray}
which is much smaller than the injection scale of MHD turbulence. Although details would depend on the driving mechanism, the largest perpendicular size of the eddies would be limited by the horizon scale, $\sim R_{S}\simeq3.0\times10^{12}~{\rm cm}~M_{7}$, or the scale height $\sim H\simeq1.7\times10^{13}~{\rm cm}~{\mathcal R}_1M_{7}$. On the other hand, the Larmor radius should be much larger than the dissipation scale. The electron inertial length is
\begin{eqnarray}
\lambda_{e-\rm pl}
&=&\frac{c}{\omega_{e-\rm pl}}
\approx\sqrt{\frac{m_ec^2}{4\pi n_ee^2}}\nonumber\\
&\simeq&2.3~{\rm cm}~\tau_{T,-0.4}^{-1/2}{\mathcal R}_1^{1/2}\lambda_{\rm Edd}^{-1/2}{\left(\frac{L_{\rm bol}L_{X,43.6}}{20 L_X}\right)}^{1/2}.\,\,\,\,\,\,\,\,\,\,\,\,
\end{eqnarray}
For ions, we have 
\begin{eqnarray}
\lambda_{p-\rm pl}
&=&\frac{c}{\omega_{p-\rm pl}}
\approx\sqrt{\frac{m_pc^2}{4\pi n_p e^2}}\nonumber\\
&\simeq&98~{\rm cm}~\tau_{T,-0.4}^{-1/2}\zeta_e^{1/2}{\mathcal R}_1^{1/2}\lambda_{\rm Edd}^{-1/2}{\left(\frac{L_{\rm bol}L_{X,43.6}}{20 L_X}\right)}^{1/2}\,\,\,\,\,\,\,\,\,\,\,\,
\end{eqnarray}
and the proton thermal Larmor radius is
\begin{equation}
\lambda_{p-\rm cy}=\frac{v_{p-{\rm th}}}{\omega_{\rm ci}}\approx\sqrt{\frac{kT_p m_pc^2}{e^2B^2}}\approx\sqrt{\frac{\beta}{2}}\lambda_{p-\rm pl}, 
\end{equation}
which is smaller than $\lambda_{p-\rm pl}$ for low-beta plasmas with $\beta\lesssim2$. 
The wave number spectrum of turbulence has a break around ${\rm min}[\lambda_{p-\rm cy}^{-1},\lambda_{p-\rm pl}^{-1}]$, above which the kinetic and/or Hall effects should be relevant, although the compressive turbulence could play a role in general. We may also expect that the MHD turbulence is well-developed, maintaining two-temperature plasma with $T_p > T_e$. When the critical balance is achieved, the eddy turnover time, $t_{\rm eddy}\sim l_{\perp}/\delta v_{\perp}$ (where $l_\perp$ ($l_\parallel$ ) is the perpendicular (parallel) scale of the turbulence and $\delta v_\perp$ is the velocity fluctuation), is comparable to $t_{A}\sim l_{\parallel}/V_A\approx (l_{\parallel}/H)t_{\rm diss}$. In such a strong turbulence limit (in which large-amplitude magnetic fluctuations are not necessarily achieved), the linear transfer time, $t_{kk}\approx k^2/D_{kk}$ is comparable to the eddy turnover time, where $D_{kk}$ is the diffusion coefficient in wave number space. Then, one can see that Coulomb relaxation time scales for protons are longer than $t_{\rm eddy}\sim t_{kk}$~\cite{Murase:2019vdl}.  

Based on these discussions, we may expect that stochastic acceleration in x-ray emitting coronae is a viable scenario for cosmic-ray acceleration and associated multimessenger emission from NGC~1068. From Eq.~(14) of M22, the acceleration time is estimated to be 
\begin{equation}
t_{\rm acc}\simeq2.2\times{10}^5~{\rm s}~{\mathcal R}_1^{2}\lambda_{\rm Edd}^{-1}{\left(\frac{L_{\rm bol}L_{X,43.6}}{20 L_X}\right)}\beta\eta_{\rm tur,1.3}{\left(\frac{\varepsilon_p}{eBH}\right)}^{2-q}, 
\end{equation}
where $V_A$ is assumed to be subrelativistic. For $V_A\rightarrow c$, we also have $t_{\rm acc}\simeq7.3\times{10}^3~{\rm s}~\allowbreak{\mathcal R}_1\lambda_{\rm Edd}^{-1}[L_{\rm bol}/(20 L_X)]L_{X,43.6}\eta_{\rm tur,1.3}{[\varepsilon_p/(eBH)]}^{2-q}$, although $\eta_{\rm tur}$ may have an implicit dependence on $\sigma$ and some other parameters. 
The numerical results with Eq.~\ref{eq:Alfven} are also shown in Fig.~\ref{fig:timeNGC}. 
As first demonstrated in Ref.~\cite{Kheirandish:2021wkm}, the maximum proton energy can be constrained by the neutrino data, by which one can constrain $\eta_{\rm tur}$. For a spectral index of $s_\nu\lesssim2$, the maximum proton energy is limited to be $\varepsilon_p^{\rm max}\lesssim100$~TeV. For $q=2$, from Fig.~\ref{fig:timeNGC}, we find $\beta\eta_{\rm tur}\sim 0.01-1$. One can easily see that there is strong degeneracy in the parameters regarding particle acceleration. For example, within the allowed parameter space of $\eta_{\rm tur}$ and $\beta$, we can readily keep similar values of $t_{\rm acc}$ for $\varepsilon_p^{\rm max}\sim10-100$~TeV between $q=2$ and $q=5/3$ cases, by noting $\eta_{\rm tur}^{q=2} \beta^{q=2} \sim 0.01\eta_{\rm tur}^{q=5/3}{(\beta^{q=5/3})}^{7/6}\allowbreak
{(\varepsilon_p^{\rm max}/30~{\rm TeV})}^{1/3}\allowbreak\lambda_{\rm Edd}^{1/6}{(20 L_X/L_{\rm bol})}^{1/6}L_{X,43.6}^{-1/6}{(\zeta_e/\tau_{T,-0.4})}^{1/6}$. It also implies that resonant and nonresonant acceleration scenarios cannot be discriminated from the current data.

\section{Implications for Future Searches}
\subsection{Predictions of the Scaling Relations}\label{sec:scaling}
Given that the system is nearly calorimetric, it has been shown that the corona and RIAF models theoretically predict $L_\nu \propto L_X$~\cite{Murase:2016gly,Murase:2019vdl,Kimura:2019yjo} (see also Refs.~\cite{Kun:2024meq,Ambrosone:2024zrf}). There are some details that can affect the scaling. First, neutrino production is suppressed by the Bethe-Heitler pair-production process especially for high-luminosity AGNs. Second, $L_{\rm CR}\propto L_{X}$ is also approximate because it is affected by details of cosmic-ray acceleration mechanisms including injection processes. This work focus on predictions of the MKM20 model.  
%
\begin{figure*}[th]
\includegraphics[width=0.325\linewidth]{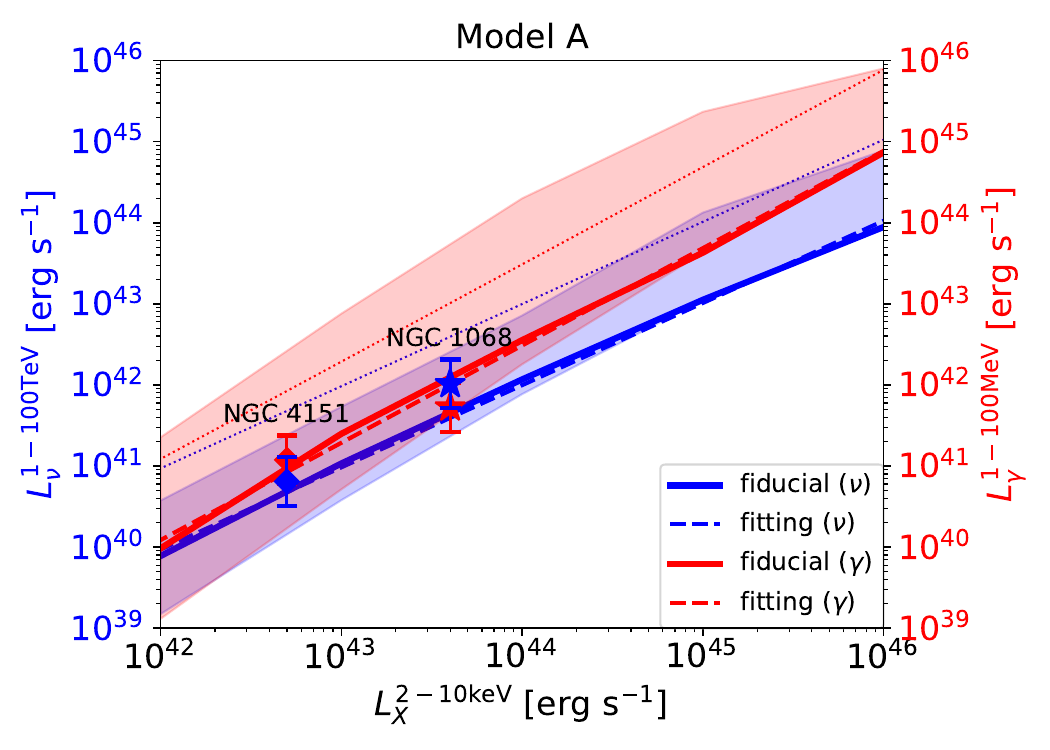}
\includegraphics[width=0.325\linewidth]{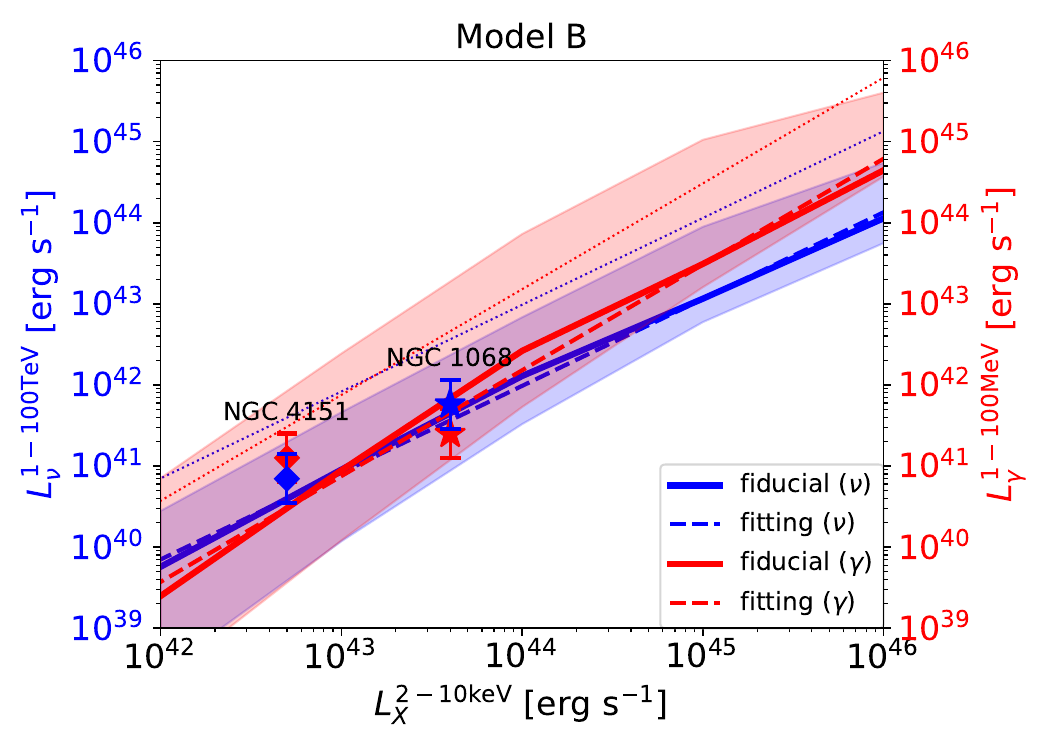}
\includegraphics[width=0.325\linewidth]{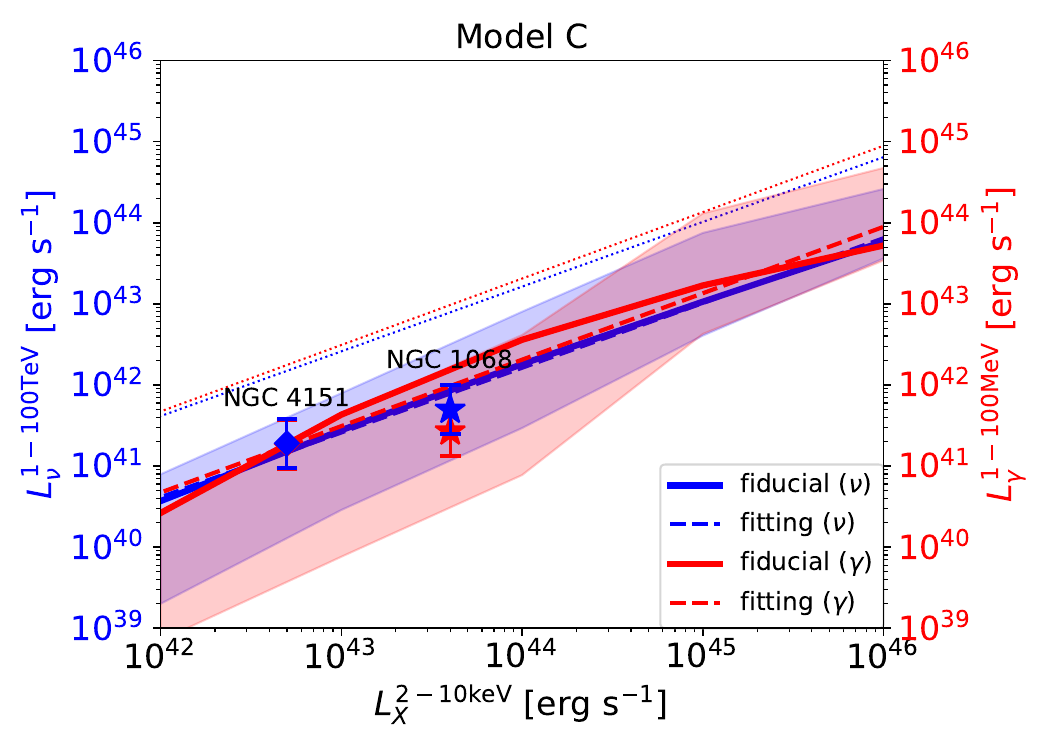}
\caption{X-ray luminosity dependence of the $1-100$~TeV neutrino luminosity ($L_\nu$) and the $1-100$~MeV gamma-ray luminosity ($L_\gamma$) for Model A (left panel), Model B (middle panel), and Model C (right panel). NGC~1068 and NGC~4151 are indicated by stars and diamonds, respectively. The fitting parameters are given in Table~\ref{tab:luminosity}. Dotted lines correspond to the cases where the normalization is enhanced by 10 (see text for details).  
\label{fig:scaling}
}
\end{figure*}
%

In Fig.~\ref{fig:luminosity} for Models A (left), B (middle), and C (right), the $L_X$-dependence of the neutrino luminosity in the $1-100$~TeV range ($L_\nu^{1-100\ \rm TeV}$) and the gamma-ray luminosity in the $1-100$~MeV range ($L_\gamma^{1-100\ \rm MeV}$) are shown in blue on the left axis and in red on the right axis, respectively. Importantly, as indicated by shaded regions, we consider systematics coming from the Eddington ratio by changing $\lambda_{\rm Edd}$ from ${\rm min}[\lambda_{\rm crit},L_{\rm bol}/L_{\rm Edd}|_{M=10^{10} M_{\odot}}$] to $1$ and ${\mathcal R}$ from 3 to 30. The solid curves represent the numerical results for the fiducial parameters adopted in each model. Note that these are theoretical predictions of the models accounting for the all-sky neutrino flux rather than those optimized for NGC~1068 and other nearby objects. The neutrino and gamma-ray luminosities of NGC~1068 and NGC~4151, corresponding to each model, are shown with stars and diamonds, respectively, with the error bars accounting for a factor of $2$ uncertainty at a given x-ray luminosity. 

In general, scaling relations can be analytically expressed as
\begin{eqnarray}
\label{eq:scale_rel}
\left(\frac{L_\nu}{10^{42}~{\rm erg}~{\rm s}^{-1}} \right) &=& {\mathcal K}_\nu {\left(\frac{L_X}{10^{44}~{\rm erg}~{\rm s}^{-1}} \right)}^{\gamma_{\nu{\rm lw}}},\\
\left(\frac{L_\gamma}{10^{42}~{\rm erg}~{\rm s}^{-1}} \right) &=& {\mathcal K}_\gamma {\left(\frac{L_X}{10^{44}~{\rm erg}~{\rm s}^{-1}} \right)}^{\gamma_{\gamma{\rm lw}}}, 
\end{eqnarray}
where the values of the best fit parameters are provided in Table~\ref{tab:luminosity}, and the 
curves using Eq.~\eqref{eq:scale_rel} and Table~\ref{tab:luminosity} are shown as dashed lines in Fig.~\ref{fig:luminosity}. Dotted lines represent the cases where the normalization is multiplied by 10, which will be used in the next subsection to demonstrate a situation where only a fraction of jet-quiet AGNs are neutrino active, while contributing the dominant fraction of the all-sky neutrino flux at medium energies.
 
We note that within model uncertainty the brightest object in intrinsic $2-10$~keV x rays, NGC~1068, is consistent with the predictions of the all-sky neutrino flux modeling for the $1-100$~TeV neutrino and $1-100$~MeV gamma-ray luminosities. The scaling relations between the intrinsic $2-10$~keV x-ray luminosity and the neutrino and gamma-ray luminosities also hold for NGC~4151. This has vital implications for stacking searches using catalogs. A catalog would involve a population based search for neutrino emission from x-ray bright AGNs. To concretely identify the sources that contribute most significantly with minimum possible bias, one would need to extend the x-ray data for the AGNs beyond the $2-10$~keV band of hard x rays. This can be done by combining data from other x-ray observatories like {\it NuSTAR} ($15-55$~keV band) or {\it Swift} ($2-10$~keV, $20-50$~keV, $14-150$~keV, and $14-195$~keV bands) to compute the intrinsic x-ray luminosity for a wider band. Assuming the scaling relations obtained in Eq.~\eqref{eq:scale_rel} hold, this can then be used to build a catalog with implementing a completeness correction to the neutrino luminosity function (discussed in Sec.~\ref{sec:diffuseformalism}). The implications such a stacking search involving a completeness corrected catalog is discussed in the following subsection. 

\begin{table}[t]
\begin{center}
\caption{Best-fit parameters of the $L_X-L_{\nu}$ and $L_{X}-L_{\gamma}$ relations inferred from the AGN corona scenario accounting for the all-sky neutrino flux. 
\label{tab:luminosity}
}
\scalebox{1.0}{
\begin{tabular}{|c||c|c||c|c|}
\hline Model & ${\mathcal K}_\nu$ & $\gamma_{\nu{\rm lw}}$ & ${\mathcal K}_\gamma$ & $\gamma_{\gamma{\rm lw}}$\\
\hline Model A & $0.973$ & $1.013$ & $3.18$ & $1.20$\\
\hline Model B & $0.953$ & $1.07$ & $1.58$ & $1.305$\\
\hline Model C & $1.57$ & $0.798$ & $2.02$ & $0.819$\\
\hline
\end{tabular}
}
\end{center}
\end{table}

\subsection{Implications for catalog-based and stacking searches}
The corona model predicts the relationship between $F_\nu$ (or $F_\gamma$) and $F_X$, enabling us to predict the $\log N-\log F_\nu$ (or $\log N-\log F_\gamma$) distribution. The cumulative number of neutrino sources is predicted through the observed $\log N-\log F_X$ distribution as  
\begin{eqnarray}
N(>F_\nu) &=& \int_{F_\nu}^\infty dF_\nu^\prime \frac{dN}{dF_\nu^\prime}
=\int_{F_X}^\infty dF_X^\prime \frac{dN}{dF_X^\prime} \frac{dF_X^\prime}{dF_\nu^\prime}
\nonumber\\ 
&=& \int d\Omega\int_0^{z_{\rm max}} dz \frac{dV_c}{dz d\Omega} 
\int_{\log L_X^{\rm th} (F_X,z)}^\infty d\log L_X \nonumber\\
&\times&
\frac{d\rho_{\rm AGN}}{d \log L_X}\frac{dL_X}{dL_\nu}\,,
\end{eqnarray}
where the minimum x-ray luminosity detectable at redshift $z$ for a given threshold $F_X$ is given by $L_X^{\rm th} (F_X,z) = 4 \pi d^2_L(z) F_X$ and $d_L$ is the luminosity distance. In general, for a given band, the luminosity $L$ and observed flux $\tilde{F}$ is related as, $L = 4\pi d_L^2 \tilde{F}$, where and $\tilde{F}$ takes into account the redshift dependent correction. In this work, we may ignore the correction because only nearby sources contribute to catalog-based and stacking searches. 
The conversion of $F_\nu$ and $F_\gamma$ (or equivalently $L_\nu$ and $L_\gamma$) to $F_X$ (or $L_X$) is achieved simply by using the scaling relations presented in Eq.~\eqref{eq:scale_rel} with the best fit parameters described in Table~\ref{tab:luminosity}.
\begin{figure*}[tb]
\includegraphics[width=0.325\linewidth]{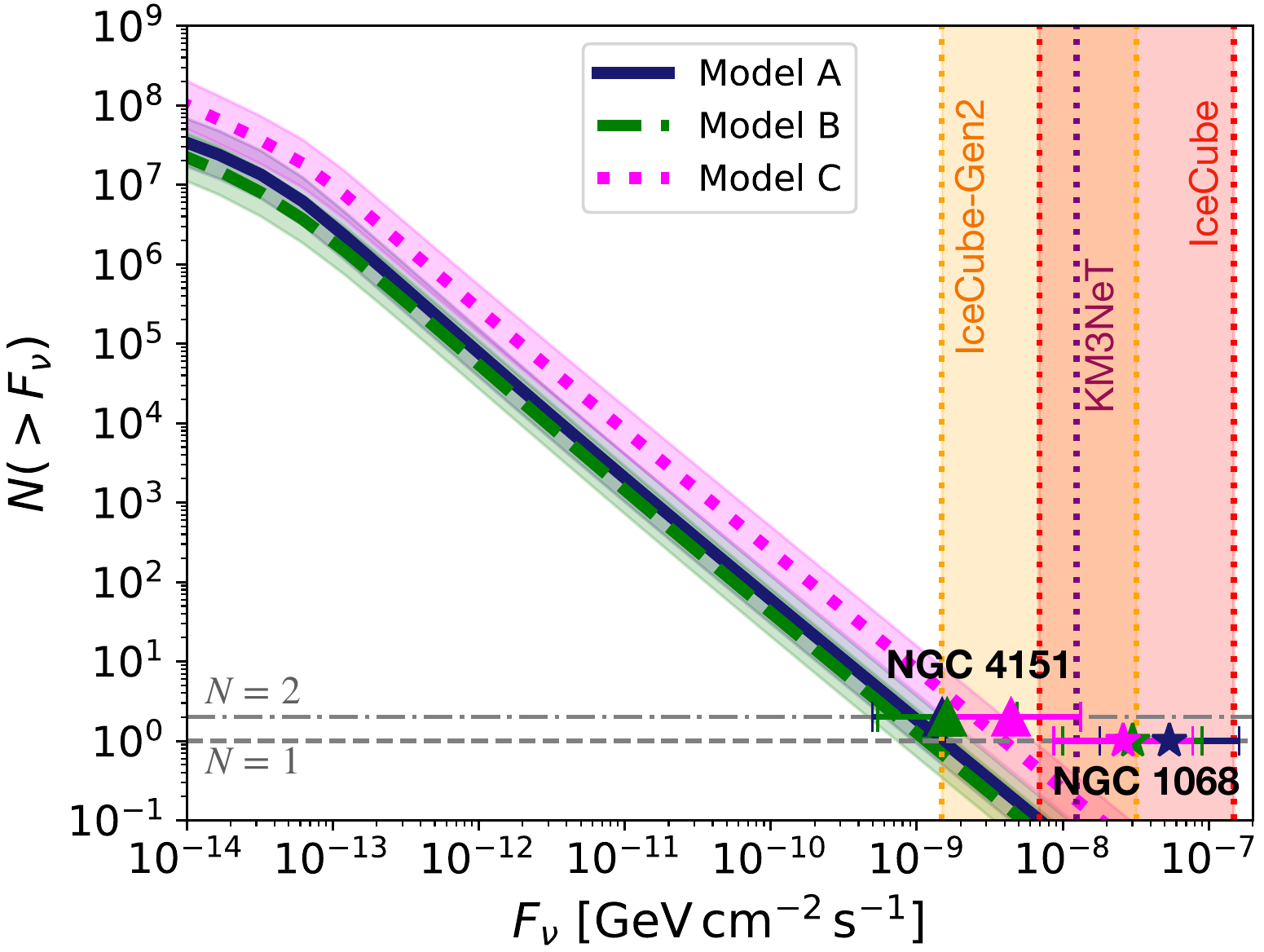}
\includegraphics[width=0.325\linewidth]{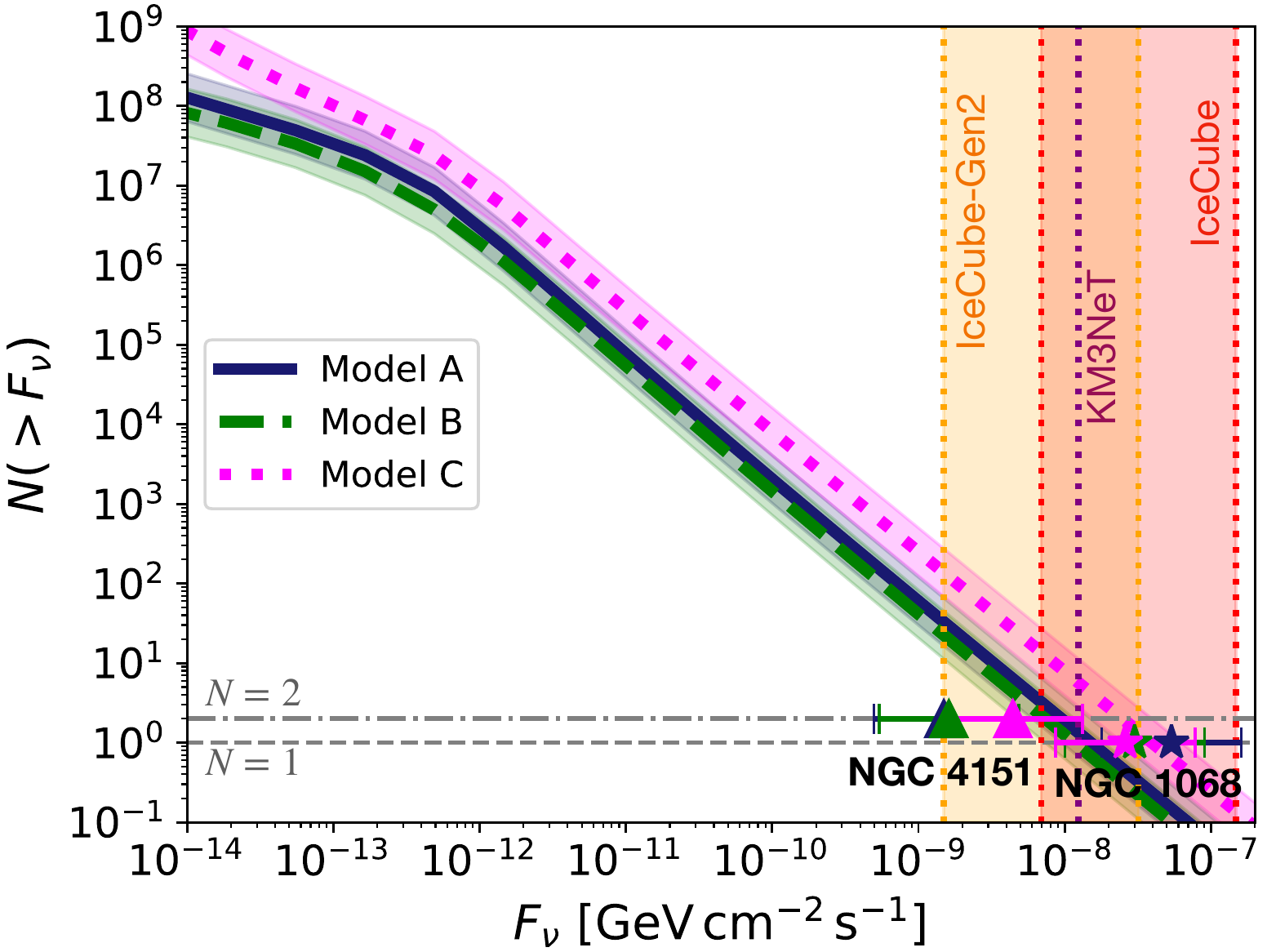}
\includegraphics[width=0.325\linewidth]{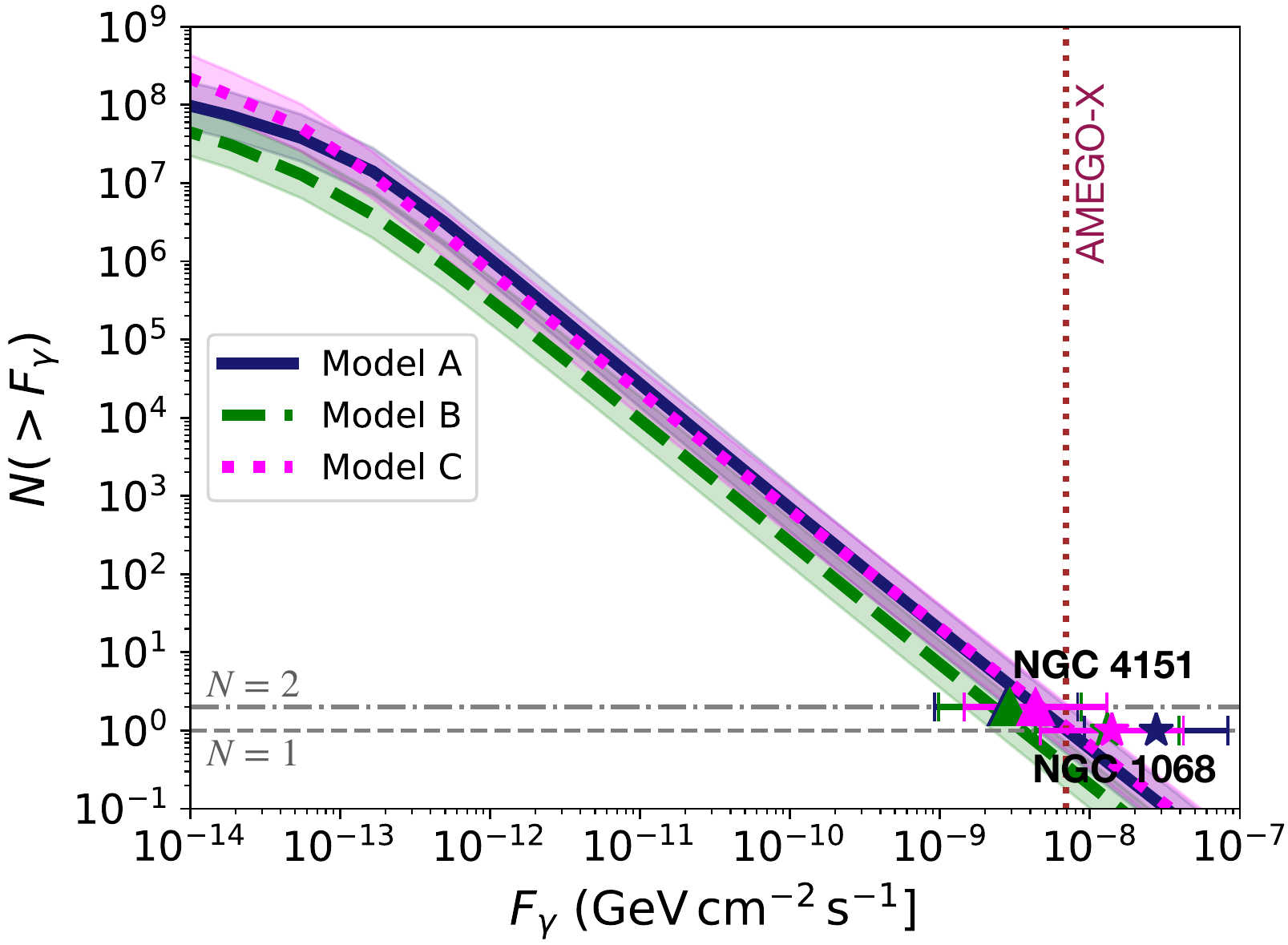}
\caption{Cumulative number-flux distribution of neutrino and gamma-ray emitting AGNs predicted from the AGN corona scenario accounting for the all-sky neutrino flux. 
The curves represent the total number of sources $N$ brighter than a given neutrino or gamma-ray flux $F_\nu$ or $F_\gamma$ respectively, integrated over redshift and all sky, for different models. A factor of $2$ shaded uncertainty band is also shown around the curves. The neutrino flux from NGC~1068 ($N=1$) and NGC~4151 ($N=2$) based on the different models are shown by stars and triangles respectively. The $5\sigma$ discovery potential (see Eq.~\ref{eq:discpot}) for various current and upcoming neutrino (IceCube, KM3NeT, and IceCube-Gen2) and gamma-ray ({\it AMEGO-X}) detectors are shown as vertical lines and shaded bands (see text for details). 
\emph{Left:} For neutrinos, \emph{Middle:} For neutrinos assuming $10\%$ of the AGNs are neutrino bright, and \emph{Right:} For gamma rays.
\label{fig:cum_numFLux}
}
\end{figure*}

In Fig.~\ref{fig:cum_numFLux} we show the $\log N-\log F$ distribution given a threshold neutrino ($F_\nu$) or gamma-ray ($F_\gamma$) flux. The results show the total number of sources ($N$) brighter than a given $F_\nu$ or $F_\gamma$, respectively. A shaded band of a factor of 2 uncertainty is also shown, considering the corresponding uncertainty in the normalization of the all-sky neutrino flux. 
The results for our fiducial models presented in this work are shown in the left and right panels of Fig.~\ref{fig:cum_numFLux}. For low neutrino fluxes of $F_\nu \sim 10^{-14}-10^{-13}~{\rm GeV}~{\rm cm}^{-2}{\rm s}^{-1}$ the number of sources $N(>F_\nu)$ starts asymptoting to $\sim 10^7 - 10^8$ sources. Note that $F_\nu \sim 10^{-14}-10^{-13}~{\rm GeV}{\rm cm}^{-2}{\rm s}^{-1}$ corresponds to neutrino fluxes from typical AGNs with $L_X\sim10^{44}~{\rm erg}~{\rm s}^{-1}$ located at $z\sim1-2$. The predicted neutrino fluxes from the brightest source NGC~1068 ($N=1$) and the second brightest source NGC~4151 ($N=2$) are indicated along with a factor of 3 uncertainty estimated from Figs.~\ref{fig:NGC1068} and \ref{fig:NGC4151}. As evident from these panels, while NGC~4151 lies on the band, NGC~1068 tends to deviate from the bands. However, this just reflects the fact that Seyferts like NGC~1068 are relatively rare in the local universe as discussed in  
Sec.~\ref{sec:consistency} is consistent with a Poisson fluctuation with a probability of $\sim10$\%.  

The middle panel of Fig.~\ref{fig:cum_numFLux} represents a scenario where ${\mathcal K}_\nu$ is multiplied by a factor of 10 (dotted lines in Fig.~\ref{fig:luminosity}). This corresponds to the situation where roughly $10\%$ of all the jet-quiet AGNs are assumed to be neutrino bright. In the MKM20 model, coronal plasma is required to be collisionless for protons to get accelerated, and the energy fraction carried by cosmic rays may also depend on the pair loading $\zeta_e$. Thus, it is possible that only a fraction of jet-quiet AGNs are neutrino bright. The most striking feature of this plot is the fact that due to the higher neutrino luminosity, all the model predictions encapsulate the neutrino flux from NGC~1068. 

In the left and middle panels of Fig.~\ref{fig:cum_numFLux}, we also show the $5\sigma$ discovery potential for IceCube, KM3NeT, and IceCube-Gen2 in an energy interval defined as
\begin{equation}
\label{eq:discpot}
F_{\nu,5\sigma} = \tilde{F}_{\nu,5\sigma}|_{100\ {\rm TeV}} \int_{1\ {\rm TeV}}^{100\ {\rm TeV}} dE_\nu\ E_\nu \left( \frac{E_\nu}{1\ {\rm TeV}} \right)^{-s_\nu}. 
\end{equation}
For IceCube we choose the differential $5\sigma$ discovery potential with 14~years of $\nu_\mu$-track observations~\cite{IceCube:2025lev} ($\tilde{F}_{\nu,5\sigma}|_{100\ {\rm TeV}}$) at $100$~TeV assuming both $E_\nu^2 F_{E_\nu} \propto E_\nu^{0}$ ($s_\nu=2$) and $E_\nu^{-1}$ ($s_\nu=3$) spectra. Furthermore, we consider the variation based on declination $\sin \delta$, considering $\sin \delta \sim 0$ and $\sin \delta \sim -1$ as optimistic and pessimistic cases, respectively. The $5\sigma$ discovery potential is then computed assuming an energy interval of $E_\nu$ between 1~TeV to 100~TeV. To plot the band for IceCube we select the best and the worst cases and shade the area in between (shown in red with red dotted lines as boundaries). For IceCube-Gen2 we simply scale the sensitivity by a factor of $10^{2/3}$ (shown in orange with orange dotted lines as boundaries). However, we note that it is possible that with better detector simulations IceCube-Gen2 can perform better than what we quote. For KM3NeT we choose $E_\nu^2 F_{E_\nu} \propto E_\nu^{0}$ and select the most sensitive declination ($\sin \delta \sim 0$) for quoting the $5\sigma$ discovery potential with 10~years of $\nu_\mu$-track observations~\cite{KM3NeT:2018wnd} (shown as a purple dotted line). As shown in Ref.~\cite{Kheirandish:2021wkm}, IceCube-Gen2 and KM3NeT would eventually allow for the $5\sigma$ discovery of NGC~1068 as a point source, while with IceCube-Gen2 detecting NGC~4151 with $5\sigma$ may also be achievable.

The right panel of Fig.~\ref{fig:cum_numFLux} shows the model predictions for the fiducial case in terms of total number of sources detectable in gamma rays. We note that in this case both NGC~1068 and NGC~4151 are on the bands predicted by our models. The sensitivity of AMEGO-X~\cite{Caputo:2022xpx} between 1~MeV to 10~MeV is shown as a dotted purple line, where 3~years of operation is assumed with a survey exposure fraction of $20$\%. Once again detecting MeV gamma rays from NGC~1068 is feasible with a detector like AMEGO-X~\cite{Murase:2019vdl,Ajello:2023hkh}.

\begin{figure}[tb]
\includegraphics[width=1.0\linewidth]{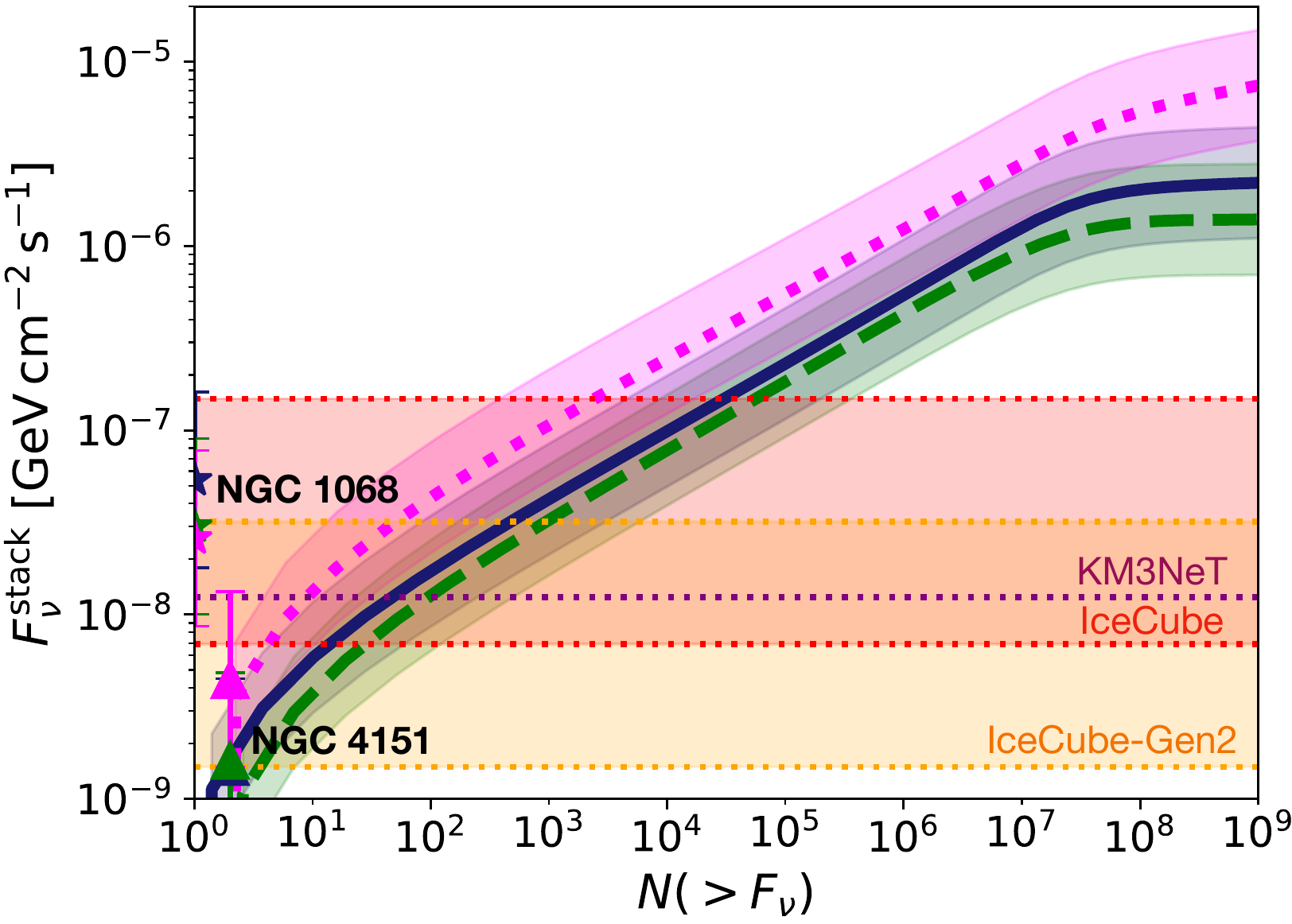}
\caption{Stacked flux $F_\nu^{\rm stack}$ of $N$ sources plotted as a function of $F_\nu$. In other words, this plot shows the result of detecting all sources above a threshold $F_\nu$ and stacking their fluxes together.
\label{fig:stack}
}
\end{figure}

It is also interesting to know how many AGN sources need to be stacked such that we reach a given stacked flux of $F_\nu^{\rm stack}$. This is shown in Fig.~\ref{fig:stack}, where we define 
\begin{equation}
F_\nu^{\rm stack} (F_\nu) = \int_{F_\nu}^{F_*} dF_\nu^\prime\ F_\nu^\prime\frac{dN}{dF_\nu^\prime}
\end{equation}
where we select $F_*$ is the neutrino flux needed for the detection of one source in the continuum limit, that is, $N(>F_*)=1$. In the figure we plot $F_\nu^{\rm stack}$ with respect to $N(>F_\nu)$ thus showing the result of detecting all sources above a threshold $F_\nu$ and stacking their fluxes together. The $5\sigma$ discovery potential bands for the detectors and the model-dependent fluxes from NGC~1068 and NGC~4151 are also shown. The lower values of $N$ represent the case when the contribution of the stacking is dominated just by the nearby bright sources, while for large values of $N$ the stacked flux plateaus because of including faint sources that add to the number of sources without contributing much to the overall stacked flux. This highlights the importance of theory priors in stacking searches. From Fig.~\ref{fig:cum_numFLux} and estimates of the signal and background we may conclude that rather than stacking everything available from the observed data as they are, it is important to take into account the signal information that connects the neutrino flux to the x-ray flux. The magnetically powered corona model enables us to identify samples of neutrino-bright AGNs that significantly contribute to the significance in quantitative analyses. It would be important to select neutrino-active galaxies {\it a priori}, based on a solid catalog of intrinsically x-ray bright AGNs that include Compton-thick AGNs. Stacking a few or dozens of AGNs from the catalog may lead to optimized detection prospects, where theoretical priors can be crucial for establishing jet-quiet AGNs as the major contributor to the all-sky neutrino flux. 

Stacking and cross-correlation analyses with x-ray AGN catalogs, such as the {\it Swift} BASS (BAT AGN Spectroscopic Survey)~\cite{BASS}, will also be useful, as discussed in Refs.~\cite{Murase:2019vdl,Kheirandish:2021wkm}. In the northern or horizontal sky of IceCube and IceCube-Gen2, the intrinsically brightest $2-10$~keV Seyferts in BASS are NGC~1068 and CGCG~164-019, and NGC~4151 is relevant once the higher-energy component is included. In the southern sky, the Circinus Galaxy is typically the brightest and hence ranks first, with NGC~4945 second~\cite{Murase:2023ccp}, and some other objects~\cite{Murase:2019vdl,Kheirandish:2021wkm} can be promising targets for KM3NeT, Baikal-GVD, P-ONE and TRIDENT. We note that, even among x-ray bright Seyfert galaxies, neutrino brightness may not be determined by $L_X$ alone. In the coronal framework, the Eddington ratio may regulate the coronal compactness, the spectral hardness, photon densities, and resulting interaction efficiencies, and thus they control the relative importance of $pp$ interactions, photomeson production, and Bethe-Heitler pair production. Consequently, significant source-to-source variations in the neutrino flux are expected at fixed $L_X$, and only a subset of nearby Seyferts may be neutrino active at a detectable level. This dispersion has direct implications for search strategies. Catalog-based stacking and cross-correlation analyses can enhance detectability, but they should ideally be performed with prespecified, physically motivated rankings that incorporate not only x-ray luminosity but also proxies for $\lambda_{\rm Edd}$ and $\mathcal R$. In addition, the predicted correlation between neutrino emission and MeV-band gamma-ray output provides an independent ranking axis: sources exhibiting the characteristic MeV cascade features are particularly compelling neutrino targets. Therefore, the coordinated use of x-ray monitoring, MeV gamma-ray observations, together with high-energy neutrino data will be essential to robustly identify and characterize the neutrino-active galaxies. We leave detailed investigations into this for future work.

\section{Summary and Implications}\label{sec:summary}
The IceCube Collaboration has established a large all-sky flux of astrophysical neutrinos in the medium-energy range, and the existence of hidden neutrino sources has been supported by recent indications of neutrino signals from nearby Seyfert galaxies. Moreover, recent analyses indicate that the all-sky spectrum is convex with a spectral turnover around $\sim 3-30~\mathrm{TeV}$. Such a spectral structure is highly informative because it disfavors simple, unbroken power-law source models, and strengthens the case for acceleration of cosmic rays in dense environments involving efficient energy losses.  
In this work, we revisited the magnetically powered corona model of jet-quiet (radio-quiet) AGNs, building on the framework of MKM20, and demonstrated that a population of such coronae can reproduce IceCube's convex all-flavor spectrum while remaining consistent with x-ray and gamma-ray background measurements and with the multimessenger data of the nearby Seyfert galaxy NGC~1068 within observational and modeling uncertainties. If jet-quiet AGNs are established as neutrino sources, the essential physical picture we may obtain is that compact, magnetized x-ray coronae surrounding SMBHs host turbulence and magnetic reconnections that can accelerate nonthermal protons in dense environments, which will bring us big impacts on our understanding of extreme phenomena in the vicinity of black holes.  

We showed that the existence of a spectral curvature is likely but a break around $\sim10$~TeV energies can be useful for constraining the models. Confirming a spectral turnover, below which the spectral index is $s_\nu\lesssim2$, may support the magnetically powered corona model. We also showed that the contribution from NGC~1068-like objects is generally subdominant in the diffuse neutrino background, which is especially the case if one considers the fact that NGC~1068 has a large value of the Eddington ratio. More compact coronae may better explain the multimessenger data of NGC~1068. Although current uncertainties are large, 
we emphasize that the Eddington ratio can induce significant source-to-source variations in neutrino and gamma-ray outputs among nearby AGNs even if microphysics parameters of particle acceleration are nearly universal. Our results imply the importance of understanding the dependence of coronal properties (e.g., ${\mathcal R}$, $\alpha$ and $\beta$) as a function of the Eddington ratio. 

Considering a wide range of systematics uncertainty in the corona model, we showed that the predicted scaling relation between the neutrino (gamma-ray) and x-ray luminosities is consistent with $L_\nu \propto L_X$ ($L_\gamma \propto L_X$), as inferred by earlier works~\cite{Murase:2016gly,Murase:2019vdl}, despite complications due to the Bethe-Heitler pair production process. The corona model has a predictive power, and with its theory prior we demonstrate how multimessenger observations with upcoming neutrino detectors such as IceCube-Gen2 and KM3NeT, as well as future MeV gamma-ray missions, can be used for testing the hypothesis that jet-quiet AGNs give the major contribution to the all-sky neutrino flux at medium energies. We encourage systematic investigations with prespecified, physically ranked catalogs for stacking and cross-correlation searches targeting the most promising jet-quiet AGNs.



\medskip
\begin{acknowledgments}
We thank Vedant Basu for providing us with the medium-energy starting neutrino data. K.M. also thanks Paolo Coppi for useful discussions. The work of K.M. was supported by the NSF Grants No.~AST-2108466, No.~AST-2108467, and No.~2308021, and KAKENHI No.~20H01901 and No.~20H05852.
S.S.K. also acknowledges the support by KAKENHI No.~22K14028, No.~21H04487, No.~23H04899, and the Tohoku Initiative for Fostering Global Researchers for Interdisciplinary Sciences (TI-FRIS) of MEXT's Strategic Professional Development Program for Young Researchers. 
M.M. acknowledges support from the NSF Grant No.~AST-2108466, the Fermi Forward Discovery Group, LLC under Contract No.~89243024CSC000002 with the U.S. Department of Energy, Office of Science, Office of High Energy Physics. 
M.B. acknowledges support from the Eberly Research Fellowship at the Pennsylvania State University and the Simons Collaboration on Extreme Electrodynamics of Compact Sources (SCEECS) Postdoctoral Fellowship at the Wisconsin IceCube Particle Astrophysics Center (WIPAC), University of Wisconsin-Madison. 
K.M. thanks the 2023 Topical Workshop: NGC~1068 as cosmic laboratory sponsored by SFB1258 and Cluster of Excellence ORIGINS and the 2024 Nordita Scientific Program: Extragalactic and Galactic Neutrino Astronomy. 
The results of this work were presented at the 2024 workshop, ``Cosmic Rays and Neutrinos in the Multi-Messenger Era'' and the 22nd AAS HEAD meeting in 2025. When this work was being completed, we became aware of the related, independent work~\cite{Wang:2025aov}. K.M. also acknowledges Mamoru Yanagisawa for his generous donation and continuous support throughout this project, despite a three-year delay.  
\end{acknowledgments}


\bibliography{kmurase.bib}

\hyphenation{Post-Script Springer}
\begin{thebibliography}{130}%
\makeatletter
\providecommand \@ifxundefined [1]{%
 \@ifx{#1\undefined}
}%
\providecommand \@ifnum [1]{%
 \ifnum #1\expandafter \@firstoftwo
 \else \expandafter \@secondoftwo
 \fi
}%
\providecommand \@ifx [1]{%
 \ifx #1\expandafter \@firstoftwo
 \else \expandafter \@secondoftwo
 \fi
}%
\providecommand \natexlab [1]{#1}%
\providecommand \enquote  [1]{``#1''}%
\providecommand \bibnamefont  [1]{#1}%
\providecommand \bibfnamefont [1]{#1}%
\providecommand \citenamefont [1]{#1}%
\providecommand \href@noop [0]{\@secondoftwo}%
\providecommand \href [0]{\begingroup \@sanitize@url \@href}%
\providecommand \@href[1]{\@@startlink{#1}\@@href}%
\providecommand \@@href[1]{\endgroup#1\@@endlink}%
\providecommand \@sanitize@url [0]{\catcode `\\12\catcode `\$12\catcode
  `\&12\catcode `\#12\catcode `\^12\catcode `\_12\catcode `\%12\relax}%
\providecommand \@@startlink[1]{}%
\providecommand \@@endlink[0]{}%
\providecommand \url  [0]{\begingroup\@sanitize@url \@url }%
\providecommand \@url [1]{\endgroup\@href {#1}{\urlprefix }}%
\providecommand \urlprefix  [0]{URL }%
\providecommand \Eprint [0]{\href }%
\providecommand \doibase [0]{http://dx.doi.org/}%
\providecommand \selectlanguage [0]{\@gobble}%
\providecommand \bibinfo  [0]{\@secondoftwo}%
\providecommand \bibfield  [0]{\@secondoftwo}%
\providecommand \translation [1]{[#1]}%
\providecommand \BibitemOpen [0]{}%
\providecommand \bibitemStop [0]{}%
\providecommand \bibitemNoStop [0]{.\EOS\space}%
\providecommand \EOS [0]{\spacefactor3000\relax}%
\providecommand \BibitemShut  [1]{\csname bibitem#1\endcsname}%
\let\auto@bib@innerbib\@empty
\bibitem [{\citenamefont {Aartsen}\ \emph {et~al.}(2020)\citenamefont {Aartsen}
  \emph {et~al.}}]{IceCube:2020acn}%
  \BibitemOpen
  \bibfield  {author} {\bibinfo {author} {\bibfnamefont {M.~G.}\ \bibnamefont
  {Aartsen}} \emph {et~al.} (\bibinfo {collaboration} {IceCube
  Collaboration}),\ }\bibfield  {title} {\enquote {\bibinfo {title}
  {{Characteristics of the diffuse astrophysical electron and tau neutrino flux
  with six years of IceCube high energy cascade data}},}\ }\href {\doibase
  10.1103/PhysRevLett.125.121104} {\bibfield  {journal} {\bibinfo  {journal}
  {Phys. Rev. Lett.}\ }\textbf {\bibinfo {volume} {125}},\ \bibinfo {pages}
  {121104} (\bibinfo {year} {2020})},\ \Eprint
  {http://arxiv.org/abs/2001.09520} {arXiv:2001.09520 [astro-ph.HE]}
  \BibitemShut {NoStop}%
\bibitem [{\citenamefont {Abbasi}\ \emph {et~al.}(2024)\citenamefont {Abbasi}
  \emph {et~al.}}]{IceCube:2024fxo}%
  \BibitemOpen
  \bibfield  {author} {\bibinfo {author} {\bibfnamefont {R.}~\bibnamefont
  {Abbasi}} \emph {et~al.} (\bibinfo {collaboration} {IceCube Collaboration}),\
  }\bibfield  {title} {\enquote {\bibinfo {title} {{Characterization of the
  astrophysical diffuse neutrino flux using starting track events in
  IceCube}},}\ }\href {\doibase 10.1103/PhysRevD.110.022001} {\bibfield
  {journal} {\bibinfo  {journal} {Phys. Rev. D}\ }\textbf {\bibinfo {volume}
  {110}},\ \bibinfo {pages} {022001} (\bibinfo {year} {2024})},\ \Eprint
  {http://arxiv.org/abs/2402.18026} {arXiv:2402.18026 [astro-ph.HE]}
  \BibitemShut {NoStop}%
\bibitem [{\citenamefont {Abbasi}\ \emph
  {et~al.}(2025{\natexlab{a}})\citenamefont {Abbasi} \emph
  {et~al.}}]{IceCube:2025tgp}%
  \BibitemOpen
  \bibfield  {author} {\bibinfo {author} {\bibfnamefont {R.}~\bibnamefont
  {Abbasi}} \emph {et~al.} (\bibinfo {collaboration} {IceCube Collaboration}),\
  }\bibfield  {title} {\enquote {\bibinfo {title} {{Evidence for a Spectral
  Break or Curvature in the Spectrum of Astrophysical Neutrinos from 5 TeV--10
  PeV}},}\ }\href@noop {} {\  (\bibinfo {year} {2025}{\natexlab{a}})},\ \Eprint
  {http://arxiv.org/abs/2507.22233} {arXiv:2507.22233 [astro-ph.HE]}
  \BibitemShut {NoStop}%
\bibitem [{\citenamefont {Ackermann}\ \emph {et~al.}(2016)\citenamefont
  {Ackermann} \emph {et~al.}}]{TheFermi-LAT:2015ykq}%
  \BibitemOpen
  \bibfield  {author} {\bibinfo {author} {\bibfnamefont {M.}~\bibnamefont
  {Ackermann}} \emph {et~al.} (\bibinfo {collaboration} {Fermi LAT
  Collaboration}),\ }\bibfield  {title} {\enquote {\bibinfo {title} {{Resolving
  the Extragalactic $\gamma$-Ray Background above 50 GeV with the Fermi Large
  Area Telescope}},}\ }\href {\doibase 10.1103/PhysRevLett.116.151105}
  {\bibfield  {journal} {\bibinfo  {journal} {Phys. Rev. Lett.}\ }\textbf
  {\bibinfo {volume} {116}},\ \bibinfo {pages} {151105} (\bibinfo {year}
  {2016})},\ \Eprint {http://arxiv.org/abs/1511.00693} {arXiv:1511.00693
  [astro-ph.CO]} \BibitemShut {NoStop}%
\bibitem [{\citenamefont {Murase}\ \emph {et~al.}(2016)\citenamefont {Murase},
  \citenamefont {Guetta},\ and\ \citenamefont {Ahlers}}]{Murase:2015xka}%
  \BibitemOpen
  \bibfield  {author} {\bibinfo {author} {\bibfnamefont {Kohta}\ \bibnamefont
  {Murase}}, \bibinfo {author} {\bibfnamefont {Dafne}\ \bibnamefont {Guetta}},
  \ and\ \bibinfo {author} {\bibfnamefont {Markus}\ \bibnamefont {Ahlers}},\
  }\bibfield  {title} {\enquote {\bibinfo {title} {{Hidden Cosmic-Ray
  Accelerators as an Origin of TeV-PeV Cosmic Neutrinos}},}\ }\href {\doibase
  10.1103/PhysRevLett.116.071101} {\bibfield  {journal} {\bibinfo  {journal}
  {Phys. Rev. Lett.}\ }\textbf {\bibinfo {volume} {116}},\ \bibinfo {pages}
  {071101} (\bibinfo {year} {2016})},\ \Eprint
  {http://arxiv.org/abs/1509.00805} {arXiv:1509.00805 [astro-ph.HE]}
  \BibitemShut {NoStop}%
\bibitem [{\citenamefont {Capanema}\ \emph {et~al.}(2020)\citenamefont
  {Capanema}, \citenamefont {Esmaili},\ and\ \citenamefont
  {Murase}}]{Capanema:2020rjj}%
  \BibitemOpen
  \bibfield  {author} {\bibinfo {author} {\bibfnamefont {Antonio}\ \bibnamefont
  {Capanema}}, \bibinfo {author} {\bibfnamefont {Arman}\ \bibnamefont
  {Esmaili}}, \ and\ \bibinfo {author} {\bibfnamefont {Kohta}\ \bibnamefont
  {Murase}},\ }\bibfield  {title} {\enquote {\bibinfo {title} {{New constraints
  on the origin of medium-energy neutrinos observed by IceCube}},}\ }\href
  {\doibase 10.1103/PhysRevD.101.103012} {\bibfield  {journal} {\bibinfo
  {journal} {Phys. Rev. D}\ }\textbf {\bibinfo {volume} {101}},\ \bibinfo
  {pages} {103012} (\bibinfo {year} {2020})},\ \Eprint
  {http://arxiv.org/abs/2002.07192} {arXiv:2002.07192 [hep-ph]} \BibitemShut
  {NoStop}%
\bibitem [{\citenamefont {Fang}\ \emph {et~al.}(2022)\citenamefont {Fang},
  \citenamefont {Gallagher},\ and\ \citenamefont {Halzen}}]{Fang:2022trf}%
  \BibitemOpen
  \bibfield  {author} {\bibinfo {author} {\bibfnamefont {Ke}~\bibnamefont
  {Fang}}, \bibinfo {author} {\bibfnamefont {John~S.}\ \bibnamefont
  {Gallagher}}, \ and\ \bibinfo {author} {\bibfnamefont {Francis}\ \bibnamefont
  {Halzen}},\ }\bibfield  {title} {\enquote {\bibinfo {title} {{The TeV Diffuse
  Cosmic Neutrino Spectrum and the Nature of Astrophysical Neutrino
  Sources}},}\ }\href {\doibase 10.3847/1538-4357/ac7649} {\bibfield  {journal}
  {\bibinfo  {journal} {Astrophys. J.}\ }\textbf {\bibinfo {volume} {933}},\
  \bibinfo {pages} {190} (\bibinfo {year} {2022})},\ \Eprint
  {http://arxiv.org/abs/2205.03740} {arXiv:2205.03740 [astro-ph.HE]}
  \BibitemShut {NoStop}%
\bibitem [{\citenamefont {Abbasi}\ \emph {et~al.}(2023)\citenamefont {Abbasi}
  \emph {et~al.}}]{IceCube:2023ame}%
  \BibitemOpen
  \bibfield  {author} {\bibinfo {author} {\bibfnamefont {R.}~\bibnamefont
  {Abbasi}} \emph {et~al.} (\bibinfo {collaboration} {IceCube}),\ }\bibfield
  {title} {\enquote {\bibinfo {title} {{Observation of high-energy neutrinos
  from the Galactic plane}},}\ }\href {\doibase 10.1126/science.adc9818}
  {\bibfield  {journal} {\bibinfo  {journal} {Science}\ }\textbf {\bibinfo
  {volume} {380}},\ \bibinfo {pages} {adc9818} (\bibinfo {year} {2023})},\
  \Eprint {http://arxiv.org/abs/2307.04427} {arXiv:2307.04427 [astro-ph.HE]}
  \BibitemShut {NoStop}%
\bibitem [{\citenamefont {Aartsen}\ \emph
  {et~al.}(2013{\natexlab{a}})\citenamefont {Aartsen} \emph
  {et~al.}}]{Aartsen:2013bka}%
  \BibitemOpen
  \bibfield  {author} {\bibinfo {author} {\bibfnamefont {M.G.}\ \bibnamefont
  {Aartsen}} \emph {et~al.} (\bibinfo {collaboration} {IceCube
  Collaboration}),\ }\bibfield  {title} {\enquote {\bibinfo {title} {{First
  observation of PeV-energy neutrinos with IceCube}},}\ }\href {\doibase
  10.1103/PhysRevLett.111.021103} {\bibfield  {journal} {\bibinfo  {journal}
  {Phys.Rev.Lett.}\ }\textbf {\bibinfo {volume} {111}},\ \bibinfo {pages}
  {021103} (\bibinfo {year} {2013}{\natexlab{a}})},\ \Eprint
  {http://arxiv.org/abs/1304.5356} {arXiv:1304.5356 [astro-ph.HE]} \BibitemShut
  {NoStop}%
\bibitem [{\citenamefont {Aartsen}\ \emph
  {et~al.}(2013{\natexlab{b}})\citenamefont {Aartsen} \emph
  {et~al.}}]{Aartsen:2013jdh}%
  \BibitemOpen
  \bibfield  {author} {\bibinfo {author} {\bibfnamefont {M.G.}\ \bibnamefont
  {Aartsen}} \emph {et~al.} (\bibinfo {collaboration} {IceCube
  Collaboration}),\ }\bibfield  {title} {\enquote {\bibinfo {title} {{Evidence
  for High-Energy Extraterrestrial Neutrinos at the IceCube Detector}},}\
  }\href {\doibase 10.1126/science.1242856} {\bibfield  {journal} {\bibinfo
  {journal} {Science}\ }\textbf {\bibinfo {volume} {342}},\ \bibinfo {pages}
  {1242856} (\bibinfo {year} {2013}{\natexlab{b}})},\ \Eprint
  {http://arxiv.org/abs/1311.5238} {arXiv:1311.5238 [astro-ph.HE]} \BibitemShut
  {NoStop}%
\bibitem [{\citenamefont {Waxman}\ and\ \citenamefont
  {Bahcall}(1998)}]{Waxman:1998yy}%
  \BibitemOpen
  \bibfield  {author} {\bibinfo {author} {\bibfnamefont {Eli}\ \bibnamefont
  {Waxman}}\ and\ \bibinfo {author} {\bibfnamefont {John~N.}\ \bibnamefont
  {Bahcall}},\ }\bibfield  {title} {\enquote {\bibinfo {title} {{High-energy
  neutrinos from astrophysical sources: An Upper bound}},}\ }\href {\doibase
  10.1103/PhysRevD.59.023002} {\bibfield  {journal} {\bibinfo  {journal}
  {Phys.Rev.}\ }\textbf {\bibinfo {volume} {D59}},\ \bibinfo {pages} {023002}
  (\bibinfo {year} {1998})},\ \Eprint {http://arxiv.org/abs/hep-ph/9807282}
  {arXiv:hep-ph/9807282 [hep-ph]} \BibitemShut {NoStop}%
\bibitem [{\citenamefont {Murase}\ and\ \citenamefont
  {Beacom}(2010)}]{Murase:2010gj}%
  \BibitemOpen
  \bibfield  {author} {\bibinfo {author} {\bibfnamefont {Kohta}\ \bibnamefont
  {Murase}}\ and\ \bibinfo {author} {\bibfnamefont {John~F.}\ \bibnamefont
  {Beacom}},\ }\bibfield  {title} {\enquote {\bibinfo {title} {{Neutrino
  Background Flux from Sources of Ultrahigh-Energy Cosmic-Ray Nuclei}},}\
  }\href {\doibase 10.1103/PhysRevD.81.123001} {\bibfield  {journal} {\bibinfo
  {journal} {Phys.Rev.}\ }\textbf {\bibinfo {volume} {D81}},\ \bibinfo {pages}
  {123001} (\bibinfo {year} {2010})},\ \Eprint {http://arxiv.org/abs/1003.4959}
  {arXiv:1003.4959 [astro-ph.HE]} \BibitemShut {NoStop}%
\bibitem [{\citenamefont {Murase}\ and\ \citenamefont
  {Fukugita}(2019)}]{Murase:2018utn}%
  \BibitemOpen
  \bibfield  {author} {\bibinfo {author} {\bibfnamefont {Kohta}\ \bibnamefont
  {Murase}}\ and\ \bibinfo {author} {\bibfnamefont {Masataka}\ \bibnamefont
  {Fukugita}},\ }\bibfield  {title} {\enquote {\bibinfo {title} {{Energetics of
  High-Energy Cosmic Radiations}},}\ }\href {\doibase
  10.1103/PhysRevD.99.063012} {\bibfield  {journal} {\bibinfo  {journal} {Phys.
  Rev. D}\ }\textbf {\bibinfo {volume} {99}},\ \bibinfo {pages} {063012}
  (\bibinfo {year} {2019})},\ \Eprint {http://arxiv.org/abs/1806.04194}
  {arXiv:1806.04194 [astro-ph.HE]} \BibitemShut {NoStop}%
\bibitem [{\citenamefont {Murase}\ \emph
  {et~al.}(2020{\natexlab{a}})\citenamefont {Murase}, \citenamefont {Kimura},\
  and\ \citenamefont {M\'esz\'aros}}]{Murase:2019vdl}%
  \BibitemOpen
  \bibfield  {author} {\bibinfo {author} {\bibfnamefont {Kohta}\ \bibnamefont
  {Murase}}, \bibinfo {author} {\bibfnamefont {Shigeo~S.}\ \bibnamefont
  {Kimura}}, \ and\ \bibinfo {author} {\bibfnamefont {Peter}\ \bibnamefont
  {M\'esz\'aros}},\ }\bibfield  {title} {\enquote {\bibinfo {title} {{Hidden
  Cores of Active Galactic Nuclei as the Origin of Medium-Energy Neutrinos:
  Critical Tests with the MeV Gamma-Ray Connection}},}\ }\href {\doibase
  10.1103/PhysRevLett.125.011101} {\bibfield  {journal} {\bibinfo  {journal}
  {Phys. Rev. Lett.}\ }\textbf {\bibinfo {volume} {125}},\ \bibinfo {pages}
  {011101} (\bibinfo {year} {2020}{\natexlab{a}})},\ \Eprint
  {http://arxiv.org/abs/1904.04226} {arXiv:1904.04226 [astro-ph.HE]}
  \BibitemShut {NoStop}%
\bibitem [{\citenamefont {Abbasi}\ \emph {et~al.}(2022)\citenamefont {Abbasi}
  \emph {et~al.}}]{IceCube2022NGC1068}%
  \BibitemOpen
  \bibfield  {author} {\bibinfo {author} {\bibfnamefont {R.}~\bibnamefont
  {Abbasi}} \emph {et~al.} (\bibinfo {collaboration} {IceCube Collaboration}),\
  }\bibfield  {title} {\enquote {\bibinfo {title} {{Evidence for neutrino
  emission from the nearby active galaxy NGC 1068}},}\ }\href {\doibase
  10.1126/science.abg3395} {\bibfield  {journal} {\bibinfo  {journal}
  {Science}\ }\textbf {\bibinfo {volume} {378}},\ \bibinfo {pages} {538--543}
  (\bibinfo {year} {2022})}\BibitemShut {NoStop}%
\bibitem [{\citenamefont {Murase}(2022{\natexlab{a}})}]{Murase:2022azo}%
  \BibitemOpen
  \bibfield  {author} {\bibinfo {author} {\bibfnamefont {Kohta}\ \bibnamefont
  {Murase}},\ }\bibfield  {title} {\enquote {\bibinfo {title} {{Neutrinos
  unveil hidden galactic activities}},}\ }\href {\doibase
  10.1126/science.ade4190} {\bibfield  {journal} {\bibinfo  {journal}
  {Science}\ }\textbf {\bibinfo {volume} {378}},\ \bibinfo {pages} {474--475}
  (\bibinfo {year} {2022}{\natexlab{a}})}\BibitemShut {NoStop}%
\bibitem [{\citenamefont {Ajello}\ \emph {et~al.}(2023)\citenamefont {Ajello},
  \citenamefont {Murase},\ and\ \citenamefont {McDaniel}}]{Ajello:2023hkh}%
  \BibitemOpen
  \bibfield  {author} {\bibinfo {author} {\bibfnamefont {Marco}\ \bibnamefont
  {Ajello}}, \bibinfo {author} {\bibfnamefont {Kohta}\ \bibnamefont {Murase}},
  \ and\ \bibinfo {author} {\bibfnamefont {Alex}\ \bibnamefont {McDaniel}},\
  }\bibfield  {title} {\enquote {\bibinfo {title} {{Disentangling the Hadronic
  Components in NGC 1068}},}\ }\href {\doibase 10.3847/2041-8213/acf296}
  {\bibfield  {journal} {\bibinfo  {journal} {Astrophys. J. Lett.}\ }\textbf
  {\bibinfo {volume} {954}},\ \bibinfo {pages} {L49} (\bibinfo {year}
  {2023})},\ \Eprint {http://arxiv.org/abs/2307.02333} {arXiv:2307.02333
  [astro-ph.HE]} \BibitemShut {NoStop}%
\bibitem [{\citenamefont {Murase}(2022{\natexlab{b}})}]{Murase:2022dog}%
  \BibitemOpen
  \bibfield  {author} {\bibinfo {author} {\bibfnamefont {Kohta}\ \bibnamefont
  {Murase}},\ }\bibfield  {title} {\enquote {\bibinfo {title} {{Hidden Hearts
  of Neutrino Active Galaxies}},}\ }\href {\doibase 10.3847/2041-8213/aca53c}
  {\bibfield  {journal} {\bibinfo  {journal} {Astrophys. J. Lett.}\ }\textbf
  {\bibinfo {volume} {941}},\ \bibinfo {pages} {L17} (\bibinfo {year}
  {2022}{\natexlab{b}})},\ \Eprint {http://arxiv.org/abs/2211.04460}
  {arXiv:2211.04460 [astro-ph.HE]} \BibitemShut {NoStop}%
\bibitem [{\citenamefont {Abbasi}\ \emph
  {et~al.}(2025{\natexlab{b}})\citenamefont {Abbasi} \emph
  {et~al.}}]{IceCube:2024dou}%
  \BibitemOpen
  \bibfield  {author} {\bibinfo {author} {\bibfnamefont {R.}~\bibnamefont
  {Abbasi}} \emph {et~al.} (\bibinfo {collaboration} {IceCube Collaboration}),\
  }\bibfield  {title} {\enquote {\bibinfo {title} {{IceCube Search for Neutrino
  Emission from X-Ray Bright Seyfert Galaxies}},}\ }\href {\doibase
  10.3847/1538-4357/addd05} {\bibfield  {journal} {\bibinfo  {journal}
  {Astrophys. J.}\ }\textbf {\bibinfo {volume} {988}},\ \bibinfo {pages} {141}
  (\bibinfo {year} {2025}{\natexlab{b}})},\ \Eprint
  {http://arxiv.org/abs/2406.07601} {arXiv:2406.07601 [astro-ph.HE]}
  \BibitemShut {NoStop}%
\bibitem [{\citenamefont {Abbasi}\ \emph
  {et~al.}(2025{\natexlab{c}})\citenamefont {Abbasi} \emph
  {et~al.}}]{IceCube:2024ayt}%
  \BibitemOpen
  \bibfield  {author} {\bibinfo {author} {\bibfnamefont {R.}~\bibnamefont
  {Abbasi}} \emph {et~al.} (\bibinfo {collaboration} {IceCube Collaboration}),\
  }\bibfield  {title} {\enquote {\bibinfo {title} {{Search for Neutrino
  Emission from Hard X-Ray AGN with IceCube}},}\ }\href {\doibase
  10.3847/1538-4357/ada94b} {\bibfield  {journal} {\bibinfo  {journal}
  {Astrophys. J.}\ }\textbf {\bibinfo {volume} {981}},\ \bibinfo {pages} {131}
  (\bibinfo {year} {2025}{\natexlab{c}})},\ \Eprint
  {http://arxiv.org/abs/2406.06684} {arXiv:2406.06684 [astro-ph.HE]}
  \BibitemShut {NoStop}%
\bibitem [{\citenamefont {Abbasi}\ \emph
  {et~al.}(2025{\natexlab{d}})\citenamefont {Abbasi} \emph
  {et~al.}}]{IceCube:2025lev}%
  \BibitemOpen
  \bibfield  {author} {\bibinfo {author} {\bibfnamefont {R.}~\bibnamefont
  {Abbasi}} \emph {et~al.} (\bibinfo {collaboration} {IceCube Collaboration}),\
  }\bibfield  {title} {\enquote {\bibinfo {title} {{All-sky Neutrino
  Point-source Search with IceCube Combined Track and Cascade Data}},}\ }\href
  {\doibase 10.3847/1538-4357/ae113f} {\bibfield  {journal} {\bibinfo
  {journal} {Astrophys. J.}\ }\textbf {\bibinfo {volume} {995}},\ \bibinfo
  {pages} {11} (\bibinfo {year} {2025}{\natexlab{d}})},\ \Eprint
  {http://arxiv.org/abs/2507.07275} {arXiv:2507.07275 [astro-ph.HE]}
  \BibitemShut {NoStop}%
\bibitem [{\citenamefont {Abbasi}\ \emph
  {et~al.}(2025{\natexlab{e}})\citenamefont {Abbasi} \emph
  {et~al.}}]{IceCube:2025ssu}%
  \BibitemOpen
  \bibfield  {author} {\bibinfo {author} {\bibfnamefont {R.}~\bibnamefont
  {Abbasi}} \emph {et~al.} (\bibinfo {collaboration} {IceCube Collaboration}),\
  }\bibfield  {title} {\enquote {\bibinfo {title} {{Evidence of Neutrino
  Emission from X-ray Bright Seyfert Galaxies with IceCube}},}\ }\href
  {\doibase 10.22323/1.501.1219} {\bibfield  {journal} {\bibinfo  {journal}
  {PoS}\ }\textbf {\bibinfo {volume} {ICRC2025}},\ \bibinfo {pages} {1219}
  (\bibinfo {year} {2025}{\natexlab{e}})}\BibitemShut {NoStop}%
\bibitem [{\citenamefont {Abbasi}\ \emph
  {et~al.}(2025{\natexlab{f}})\citenamefont {Abbasi} \emph
  {et~al.}}]{Abbasi:2025tas}%
  \BibitemOpen
  \bibfield  {author} {\bibinfo {author} {\bibfnamefont {R.}~\bibnamefont
  {Abbasi}} \emph {et~al.},\ }\bibfield  {title} {\enquote {\bibinfo {title}
  {{Evidence for Neutrino Emission from X-ray Bright Active Galactic Nuclei
  with IceCube}},}\ }\href@noop {} {\  (\bibinfo {year}
  {2025}{\natexlab{f}})},\ \Eprint {http://arxiv.org/abs/2510.13403}
  {arXiv:2510.13403 [astro-ph.HE]} \BibitemShut {NoStop}%
\bibitem [{\citenamefont {Abbasi}\ \emph {et~al.}(2026)\citenamefont {Abbasi}
  \emph {et~al.}}]{IceCube:2026hzq}%
  \BibitemOpen
  \bibfield  {author} {\bibinfo {author} {\bibfnamefont {R.}~\bibnamefont
  {Abbasi}} \emph {et~al.} (\bibinfo {collaboration} {IceCube Collaboration}),\
  }\bibfield  {title} {\enquote {\bibinfo {title} {{Evidence for neutrino
  emission from X-ray Bright Seyfert Galaxies in the Southern Hemisphere using
  Enhanced Starting Track Events with IceCube}},}\ }\href@noop {} {\  (\bibinfo
  {year} {2026})},\ \Eprint {http://arxiv.org/abs/2602.10208} {arXiv:2602.10208
  [astro-ph.HE]} \BibitemShut {NoStop}%
\bibitem [{\citenamefont {Neronov}\ \emph {et~al.}(2024)\citenamefont
  {Neronov}, \citenamefont {Savchenko},\ and\ \citenamefont
  {Semikoz}}]{Neronov:2023aks}%
  \BibitemOpen
  \bibfield  {author} {\bibinfo {author} {\bibfnamefont {A.}~\bibnamefont
  {Neronov}}, \bibinfo {author} {\bibfnamefont {D.}~\bibnamefont {Savchenko}},
  \ and\ \bibinfo {author} {\bibfnamefont {D.~V.}\ \bibnamefont {Semikoz}},\
  }\bibfield  {title} {\enquote {\bibinfo {title} {{Neutrino Signal from a
  Population of Seyfert Galaxies}},}\ }\href {\doibase
  10.1103/PhysRevLett.132.101002} {\bibfield  {journal} {\bibinfo  {journal}
  {Phys. Rev. Lett.}\ }\textbf {\bibinfo {volume} {132}},\ \bibinfo {pages}
  {101002} (\bibinfo {year} {2024})},\ \Eprint
  {http://arxiv.org/abs/2306.09018} {arXiv:2306.09018 [astro-ph.HE]}
  \BibitemShut {NoStop}%
\bibitem [{\citenamefont {Murase}\ and\ \citenamefont
  {Stecker}(2023)}]{Murase:2022feu}%
  \BibitemOpen
  \bibfield  {author} {\bibinfo {author} {\bibfnamefont {Kohta}\ \bibnamefont
  {Murase}}\ and\ \bibinfo {author} {\bibfnamefont {Floyd~W.}\ \bibnamefont
  {Stecker}},\ }\bibfield  {title} {\enquote {\bibinfo {title} {{Chapter 10:
  High-Energy Neutrinos from Active Galactic Nuclei}},}\ }\href {\doibase
  10.1142/9789811282645_0010} {\ ,\ \bibinfo {pages} {483--540} (\bibinfo
  {year} {2023})},\ \Eprint {http://arxiv.org/abs/2202.03381} {arXiv:2202.03381
  [astro-ph.HE]} \BibitemShut {NoStop}%
\bibitem [{\citenamefont {Padovani}\ \emph
  {et~al.}(2024{\natexlab{a}})\citenamefont {Padovani} \emph
  {et~al.}}]{Padovani:2024ibi}%
  \BibitemOpen
  \bibfield  {author} {\bibinfo {author} {\bibfnamefont {P.}~\bibnamefont
  {Padovani}} \emph {et~al.},\ }\bibfield  {title} {\enquote {\bibinfo {title}
  {{High-energy neutrinos from the vicinity of the supermassive black hole in
  NGC{\,}1068}},}\ }\href {\doibase 10.1038/s41550-024-02339-z} {\bibfield
  {journal} {\bibinfo  {journal} {Nature Astron.}\ }\textbf {\bibinfo {volume}
  {8}},\ \bibinfo {pages} {1077--1087} (\bibinfo {year}
  {2024}{\natexlab{a}})},\ \Eprint {http://arxiv.org/abs/2405.20146}
  {arXiv:2405.20146 [astro-ph.HE]} \BibitemShut {NoStop}%
\bibitem [{\citenamefont {Murase}\ \emph {et~al.}(2024)\citenamefont {Murase},
  \citenamefont {Karwin}, \citenamefont {Kimura}, \citenamefont {Ajello},\ and\
  \citenamefont {Buson}}]{Murase:2023ccp}%
  \BibitemOpen
  \bibfield  {author} {\bibinfo {author} {\bibfnamefont {Kohta}\ \bibnamefont
  {Murase}}, \bibinfo {author} {\bibfnamefont {Christopher~M.}\ \bibnamefont
  {Karwin}}, \bibinfo {author} {\bibfnamefont {Shigeo~S.}\ \bibnamefont
  {Kimura}}, \bibinfo {author} {\bibfnamefont {Marco}\ \bibnamefont {Ajello}},
  \ and\ \bibinfo {author} {\bibfnamefont {Sara}\ \bibnamefont {Buson}},\
  }\bibfield  {title} {\enquote {\bibinfo {title} {{Sub-GeV Gamma Rays from
  Nearby Seyfert Galaxies and Implications for Coronal Neutrino Emission}},}\
  }\href {\doibase 10.3847/2041-8213/ad19c5} {\bibfield  {journal} {\bibinfo
  {journal} {Astrophys. J. Lett.}\ }\textbf {\bibinfo {volume} {961}},\
  \bibinfo {pages} {L34} (\bibinfo {year} {2024})},\ \Eprint
  {http://arxiv.org/abs/2312.16089} {arXiv:2312.16089 [astro-ph.HE]}
  \BibitemShut {NoStop}%
\bibitem [{\citenamefont {Inoue}\ \emph {et~al.}(2019)\citenamefont {Inoue},
  \citenamefont {Khangulyan}, \citenamefont {Inoue},\ and\ \citenamefont
  {Doi}}]{Inoue:2019fil}%
  \BibitemOpen
  \bibfield  {author} {\bibinfo {author} {\bibfnamefont {Yoshiyuki}\
  \bibnamefont {Inoue}}, \bibinfo {author} {\bibfnamefont {Dmitry}\
  \bibnamefont {Khangulyan}}, \bibinfo {author} {\bibfnamefont {Susumu}\
  \bibnamefont {Inoue}}, \ and\ \bibinfo {author} {\bibfnamefont {Akihiro}\
  \bibnamefont {Doi}},\ }\bibfield  {title} {\enquote {\bibinfo {title} {{On
  high-energy particles in accretion disk coronae of supermassive black holes:
  implications for MeV gamma rays and high-energy neutrinos from AGN cores}},}\
  }\href {\doibase 10.3847/1538-4357/ab2715} {\bibfield  {journal} {\bibinfo
  {journal} {Astrophys. J.}\ }\textbf {\bibinfo {volume} {880}},\ \bibinfo
  {pages} {40} (\bibinfo {year} {2019})},\ \Eprint
  {http://arxiv.org/abs/1904.00554} {arXiv:1904.00554 [astro-ph.HE]}
  \BibitemShut {NoStop}%
\bibitem [{\citenamefont {Inoue}\ \emph {et~al.}(2022)\citenamefont {Inoue},
  \citenamefont {Cerruti}, \citenamefont {Murase},\ and\ \citenamefont
  {Liu}}]{Inoue:2022yak}%
  \BibitemOpen
  \bibfield  {author} {\bibinfo {author} {\bibfnamefont {Susumu}\ \bibnamefont
  {Inoue}}, \bibinfo {author} {\bibfnamefont {Matteo}\ \bibnamefont {Cerruti}},
  \bibinfo {author} {\bibfnamefont {Kohta}\ \bibnamefont {Murase}}, \ and\
  \bibinfo {author} {\bibfnamefont {Ruo-Yu}\ \bibnamefont {Liu}},\ }\bibfield
  {title} {\enquote {\bibinfo {title} {{High-energy neutrinos and gamma rays
  from winds and tori in active galactic nuclei}},}\ }\href@noop {} {\
  (\bibinfo {year} {2022})},\ \Eprint {http://arxiv.org/abs/2207.02097}
  {arXiv:2207.02097 [astro-ph.HE]} \BibitemShut {NoStop}%
\bibitem [{\citenamefont {Ajello}\ \emph {et~al.}(2008)\citenamefont {Ajello}
  \emph {et~al.}}]{Ajello:2008xb}%
  \BibitemOpen
  \bibfield  {author} {\bibinfo {author} {\bibfnamefont {M.}~\bibnamefont
  {Ajello}} \emph {et~al.},\ }\bibfield  {title} {\enquote {\bibinfo {title}
  {{Cosmic X-ray background and Earth albedo Spectra with Swift/BAT}},}\ }\href
  {\doibase 10.1086/592595} {\bibfield  {journal} {\bibinfo  {journal}
  {Astrophys. J.}\ }\textbf {\bibinfo {volume} {689}},\ \bibinfo {pages} {666}
  (\bibinfo {year} {2008})},\ \Eprint {http://arxiv.org/abs/0808.3377}
  {arXiv:0808.3377 [astro-ph]} \BibitemShut {NoStop}%
\bibitem [{\citenamefont {{Fukada}}\ \emph {et~al.}(1975)\citenamefont
  {{Fukada}}, \citenamefont {{Hayakawa}}, \citenamefont {{Kasahara}},
  \citenamefont {{Makino}}, \citenamefont {{Tanaka}},\ and\ \citenamefont
  {{Sreekantan}}}]{1975Natur.254..398F}%
  \BibitemOpen
  \bibfield  {author} {\bibinfo {author} {\bibfnamefont {Y.}~\bibnamefont
  {{Fukada}}}, \bibinfo {author} {\bibfnamefont {S.}~\bibnamefont
  {{Hayakawa}}}, \bibinfo {author} {\bibfnamefont {I.}~\bibnamefont
  {{Kasahara}}}, \bibinfo {author} {\bibfnamefont {F.}~\bibnamefont
  {{Makino}}}, \bibinfo {author} {\bibfnamefont {Y.}~\bibnamefont {{Tanaka}}},
  \ and\ \bibinfo {author} {\bibfnamefont {B.~V.}\ \bibnamefont
  {{Sreekantan}}},\ }\bibfield  {title} {\enquote {\bibinfo {title} {{Energy
  spectrum of diffuse component of cosmic soft gamma rays}},}\ }\href {\doibase
  10.1038/254398a0} {\bibfield  {journal} {\bibinfo  {journal} {Nature}\
  }\textbf {\bibinfo {volume} {254}},\ \bibinfo {pages} {398} (\bibinfo {year}
  {1975})}\BibitemShut {NoStop}%
\bibitem [{\citenamefont {{Watanabe}}\ \emph {et~al.}(1997)\citenamefont
  {{Watanabe}}, \citenamefont {{Hartmann}}, \citenamefont {{Leising}},
  \citenamefont {{The}}, \citenamefont {{Share}},\ and\ \citenamefont
  {{Kinzer}}}]{1997AIPC..410.1223W}%
  \BibitemOpen
  \bibfield  {author} {\bibinfo {author} {\bibfnamefont {K.}~\bibnamefont
  {{Watanabe}}}, \bibinfo {author} {\bibfnamefont {D.~H.}\ \bibnamefont
  {{Hartmann}}}, \bibinfo {author} {\bibfnamefont {M.~D.}\ \bibnamefont
  {{Leising}}}, \bibinfo {author} {\bibfnamefont {L.~S.}\ \bibnamefont
  {{The}}}, \bibinfo {author} {\bibfnamefont {G.~H.}\ \bibnamefont {{Share}}},
  \ and\ \bibinfo {author} {\bibfnamefont {R.~L.}\ \bibnamefont {{Kinzer}}},\
  }\bibfield  {title} {\enquote {\bibinfo {title} {{The Cosmic
  {\ensuremath{\gamma}}-ray Background from supernovae}},}\ }in\ \href
  {\doibase 10.1063/1.53933} {\emph {\bibinfo {booktitle} {Proceedings of the
  Fourth Compton Symposium}}},\ \bibinfo {series} {American Institute of
  Physics Conference Series}, Vol.\ \bibinfo {volume} {410},\ \bibinfo {editor}
  {edited by\ \bibinfo {editor} {\bibfnamefont {Charles~D.}\ \bibnamefont
  {{Dermer}}}, \bibinfo {editor} {\bibfnamefont {Mark~S.}\ \bibnamefont
  {{Strickman}}}, \ and\ \bibinfo {editor} {\bibfnamefont {James~D.}\
  \bibnamefont {{Kurfess}}}}\ (\bibinfo {year} {1997})\ pp.\ \bibinfo {pages}
  {1223--1227}\BibitemShut {NoStop}%
\bibitem [{\citenamefont {{Weidenspointner}}\ \emph {et~al.}(2000)\citenamefont
  {{Weidenspointner}}, \citenamefont {{Varendorff}}, \citenamefont
  {{Kappadath}}, \citenamefont {{Bennett}}, \citenamefont {{Bloemen}},
  \citenamefont {{Diehl}}, \citenamefont {{Hermsen}}, \citenamefont {{Lichti}},
  \citenamefont {{Ryan}},\ and\ \citenamefont
  {{Sch{\"o}nfelder}}}]{2000AIPC..510..467W}%
  \BibitemOpen
  \bibfield  {author} {\bibinfo {author} {\bibfnamefont {G.}~\bibnamefont
  {{Weidenspointner}}}, \bibinfo {author} {\bibfnamefont {M.}~\bibnamefont
  {{Varendorff}}}, \bibinfo {author} {\bibfnamefont {S.~C.}\ \bibnamefont
  {{Kappadath}}}, \bibinfo {author} {\bibfnamefont {K.}~\bibnamefont
  {{Bennett}}}, \bibinfo {author} {\bibfnamefont {H.}~\bibnamefont
  {{Bloemen}}}, \bibinfo {author} {\bibfnamefont {R.}~\bibnamefont {{Diehl}}},
  \bibinfo {author} {\bibfnamefont {W.}~\bibnamefont {{Hermsen}}}, \bibinfo
  {author} {\bibfnamefont {G.~G.}\ \bibnamefont {{Lichti}}}, \bibinfo {author}
  {\bibfnamefont {J.}~\bibnamefont {{Ryan}}}, \ and\ \bibinfo {author}
  {\bibfnamefont {V.}~\bibnamefont {{Sch{\"o}nfelder}}},\ }\bibfield  {title}
  {\enquote {\bibinfo {title} {{The cosmic diffuse gamma-ray background
  measured with COMPTEL}},}\ }in\ \href {\doibase 10.1063/1.1307028} {\emph
  {\bibinfo {booktitle} {American Institute of Physics Conference Series}}},\
  Vol.\ \bibinfo {volume} {510},\ \bibinfo {editor} {edited by\ \bibinfo
  {editor} {\bibfnamefont {Mark~L.}\ \bibnamefont {{McConnell}}}\ and\ \bibinfo
  {editor} {\bibfnamefont {James~M.}\ \bibnamefont {{Ryan}}}}\ (\bibinfo {year}
  {2000})\ pp.\ \bibinfo {pages} {467--470}\BibitemShut {NoStop}%
\bibitem [{\citenamefont {Ackermann}\ \emph {et~al.}(2015)\citenamefont
  {Ackermann} \emph {et~al.}}]{Ackermann:2014usa}%
  \BibitemOpen
  \bibfield  {author} {\bibinfo {author} {\bibfnamefont {M.}~\bibnamefont
  {Ackermann}} \emph {et~al.} (\bibinfo {collaboration} {Fermi LAT
  collaboration}),\ }\bibfield  {title} {\enquote {\bibinfo {title} {{The
  spectrum of isotropic diffuse gamma-ray emission between 100 MeV and 820
  GeV}},}\ }\href {\doibase 10.1088/0004-637X/799/1/86} {\bibfield  {journal}
  {\bibinfo  {journal} {Astrophys.J.}\ }\textbf {\bibinfo {volume} {799}},\
  \bibinfo {pages} {86} (\bibinfo {year} {2015})},\ \Eprint
  {http://arxiv.org/abs/1410.3696} {arXiv:1410.3696 [astro-ph.HE]} \BibitemShut
  {NoStop}%
\bibitem [{\citenamefont {Kimura}\ \emph {et~al.}(2021)\citenamefont {Kimura},
  \citenamefont {Murase},\ and\ \citenamefont {M\'esz\'aros}}]{Kimura:2020thg}%
  \BibitemOpen
  \bibfield  {author} {\bibinfo {author} {\bibfnamefont {Shigeo~S.}\
  \bibnamefont {Kimura}}, \bibinfo {author} {\bibfnamefont {Kohta}\
  \bibnamefont {Murase}}, \ and\ \bibinfo {author} {\bibfnamefont {Peter}\
  \bibnamefont {M\'esz\'aros}},\ }\bibfield  {title} {\enquote {\bibinfo
  {title} {{Soft gamma rays from low accreting supermassive black holes and
  connection to energetic neutrinos}},}\ }\href {\doibase
  10.1038/s41467-021-25111-7} {\bibfield  {journal} {\bibinfo  {journal}
  {Nature Commun.}\ }\textbf {\bibinfo {volume} {12}},\ \bibinfo {pages} {5615}
  (\bibinfo {year} {2021})},\ \Eprint {http://arxiv.org/abs/2005.01934}
  {arXiv:2005.01934 [astro-ph.HE]} \BibitemShut {NoStop}%
\bibitem [{\citenamefont {Fang}\ and\ \citenamefont
  {Murase}(2018)}]{Fang:2017zjf}%
  \BibitemOpen
  \bibfield  {author} {\bibinfo {author} {\bibfnamefont {Ke}~\bibnamefont
  {Fang}}\ and\ \bibinfo {author} {\bibfnamefont {Kohta}\ \bibnamefont
  {Murase}},\ }\bibfield  {title} {\enquote {\bibinfo {title} {{Linking
  High-Energy Cosmic Particles by Black Hole Jets Embedded in Large-Scale
  Structures}},}\ }\href {\doibase 10.1038/s41567-017-0025-4} {\bibfield
  {journal} {\bibinfo  {journal} {Nature Phys.}\ }\textbf {\bibinfo {volume}
  {14}},\ \bibinfo {pages} {396--398} (\bibinfo {year} {2018})},\ \Eprint
  {http://arxiv.org/abs/1704.00015} {arXiv:1704.00015 [astro-ph.HE]}
  \BibitemShut {NoStop}%
\bibitem [{\citenamefont {Ajello}\ \emph {et~al.}(2014)\citenamefont {Ajello},
  \citenamefont {Romani}, \citenamefont {Gasparrini}, \citenamefont {Shaw},
  \citenamefont {Bolmer} \emph {et~al.}}]{Ajello:2013lka}%
  \BibitemOpen
  \bibfield  {author} {\bibinfo {author} {\bibfnamefont {M.}~\bibnamefont
  {Ajello}}, \bibinfo {author} {\bibfnamefont {R.W.}\ \bibnamefont {Romani}},
  \bibinfo {author} {\bibfnamefont {D.}~\bibnamefont {Gasparrini}}, \bibinfo
  {author} {\bibfnamefont {M.S.}\ \bibnamefont {Shaw}}, \bibinfo {author}
  {\bibfnamefont {J.}~\bibnamefont {Bolmer}},  \emph {et~al.},\ }\bibfield
  {title} {\enquote {\bibinfo {title} {{The Cosmic Evolution of Fermi BL
  Lacertae Objects}},}\ }\href {\doibase 10.1088/0004-637X/780/1/73} {\bibfield
   {journal} {\bibinfo  {journal} {Astrophys.J.}\ }\textbf {\bibinfo {volume}
  {780}},\ \bibinfo {pages} {73} (\bibinfo {year} {2014})},\ \Eprint
  {http://arxiv.org/abs/1310.0006} {arXiv:1310.0006 [astro-ph.CO]} \BibitemShut
  {NoStop}%
\bibitem [{\citenamefont {{Schlickeiser}}(1989)}]{Schlickeiser1989}%
  \BibitemOpen
  \bibfield  {author} {\bibinfo {author} {\bibfnamefont {Reinhard}\
  \bibnamefont {{Schlickeiser}}},\ }\bibfield  {title} {\enquote {\bibinfo
  {title} {{Cosmic-Ray Transport and Acceleration. I. Derivation of the Kinetic
  Equation and Application to Cosmic Rays in Static Cold Media}},}\ }\href
  {\doibase 10.1086/167009} {\bibfield  {journal} {\bibinfo  {journal} {\apj}\
  }\textbf {\bibinfo {volume} {336}},\ \bibinfo {pages} {243} (\bibinfo {year}
  {1989})}\BibitemShut {NoStop}%
\bibitem [{\citenamefont {{Park}}\ and\ \citenamefont
  {{Petrosian}}(1995)}]{ParkPetrosian1995}%
  \BibitemOpen
  \bibfield  {author} {\bibinfo {author} {\bibfnamefont {Brian~T.}\
  \bibnamefont {{Park}}}\ and\ \bibinfo {author} {\bibfnamefont {Vahe}\
  \bibnamefont {{Petrosian}}},\ }\bibfield  {title} {\enquote {\bibinfo {title}
  {{Fokker-Planck Equations of Stochastic Acceleration: Green's Functions and
  Boundary Conditions}},}\ }\href {\doibase 10.1086/175828} {\bibfield
  {journal} {\bibinfo  {journal} {\apj}\ }\textbf {\bibinfo {volume} {446}},\
  \bibinfo {pages} {699} (\bibinfo {year} {1995})}\BibitemShut {NoStop}%
\bibitem [{\citenamefont {Becker}\ \emph {et~al.}(2006)\citenamefont {Becker},
  \citenamefont {Le},\ and\ \citenamefont {Dermer}}]{Becker:2006nz}%
  \BibitemOpen
  \bibfield  {author} {\bibinfo {author} {\bibfnamefont {P.~A.}\ \bibnamefont
  {Becker}}, \bibinfo {author} {\bibfnamefont {Truong}\ \bibnamefont {Le}}, \
  and\ \bibinfo {author} {\bibfnamefont {Charles~D.}\ \bibnamefont {Dermer}},\
  }\bibfield  {title} {\enquote {\bibinfo {title} {{Time-Dependent Stochastic
  Particle Acceleration in Astrophysical Plasmas: Exact Solutions Including
  Momentum-Dependent Escape}},}\ }\href {\doibase 10.1086/505319} {\bibfield
  {journal} {\bibinfo  {journal} {Astrophys. J.}\ }\textbf {\bibinfo {volume}
  {647}},\ \bibinfo {pages} {539--551} (\bibinfo {year} {2006})},\ \Eprint
  {http://arxiv.org/abs/astro-ph/0604504} {arXiv:astro-ph/0604504} \BibitemShut
  {NoStop}%
\bibitem [{\citenamefont {Stawarz}\ and\ \citenamefont
  {Petrosian}(2008)}]{Stawarz:2008sp}%
  \BibitemOpen
  \bibfield  {author} {\bibinfo {author} {\bibfnamefont {Lukasz}\ \bibnamefont
  {Stawarz}}\ and\ \bibinfo {author} {\bibfnamefont {Vahe}\ \bibnamefont
  {Petrosian}},\ }\bibfield  {title} {\enquote {\bibinfo {title} {{On the
  Momentum Diffusion of Radiating Ultrarelativistic Electrons in a Turbulent
  Magnetic Field}},}\ }\href {\doibase 10.1086/588813} {\bibfield  {journal}
  {\bibinfo  {journal} {Astrophys. J.}\ }\textbf {\bibinfo {volume} {681}},\
  \bibinfo {pages} {1725--1744} (\bibinfo {year} {2008})},\ \Eprint
  {http://arxiv.org/abs/0803.0989} {arXiv:0803.0989 [astro-ph]} \BibitemShut
  {NoStop}%
\bibitem [{\citenamefont {Murase}\ \emph
  {et~al.}(2012{\natexlab{a}})\citenamefont {Murase}, \citenamefont {Asano},
  \citenamefont {Terasawa},\ and\ \citenamefont {Meszaros}}]{Murase:2011cx}%
  \BibitemOpen
  \bibfield  {author} {\bibinfo {author} {\bibfnamefont {Kohta}\ \bibnamefont
  {Murase}}, \bibinfo {author} {\bibfnamefont {Katsuaki}\ \bibnamefont
  {Asano}}, \bibinfo {author} {\bibfnamefont {Toshio}\ \bibnamefont
  {Terasawa}}, \ and\ \bibinfo {author} {\bibfnamefont {Peter}\ \bibnamefont
  {Meszaros}},\ }\bibfield  {title} {\enquote {\bibinfo {title} {{The Role of
  Stochastic Acceleration in the Prompt Emission of Gamma-Ray Bursts:
  Application to Hadronic Injection}},}\ }\href {\doibase
  10.1088/0004-637X/746/2/164} {\bibfield  {journal} {\bibinfo  {journal}
  {Astrophys. J.}\ }\textbf {\bibinfo {volume} {746}},\ \bibinfo {pages} {164}
  (\bibinfo {year} {2012}{\natexlab{a}})},\ \Eprint
  {http://arxiv.org/abs/1107.5575} {arXiv:1107.5575 [astro-ph.HE]} \BibitemShut
  {NoStop}%
\bibitem [{\citenamefont {Brunetti}\ and\ \citenamefont
  {Lazarian}(2016)}]{Brunetti:2016oaw}%
  \BibitemOpen
  \bibfield  {author} {\bibinfo {author} {\bibfnamefont {G.}~\bibnamefont
  {Brunetti}}\ and\ \bibinfo {author} {\bibfnamefont {A.}~\bibnamefont
  {Lazarian}},\ }\bibfield  {title} {\enquote {\bibinfo {title} {{Stochastic
  reacceleration of relativistic electrons by turbulent reconnection: a
  mechanism for cluster-scale radio emission?}}}\ }\href {\doibase
  10.1093/mnras/stw496} {\bibfield  {journal} {\bibinfo  {journal} {Mon. Not.
  Roy. Astron. Soc.}\ }\textbf {\bibinfo {volume} {458}},\ \bibinfo {pages}
  {2584--2595} (\bibinfo {year} {2016})},\ \Eprint
  {http://arxiv.org/abs/1603.00458} {arXiv:1603.00458 [astro-ph.HE]}
  \BibitemShut {NoStop}%
\bibitem [{\citenamefont {Nishiwaki}\ \emph {et~al.}(2021)\citenamefont
  {Nishiwaki}, \citenamefont {Asano},\ and\ \citenamefont
  {Murase}}]{Nishiwaki:2021axb}%
  \BibitemOpen
  \bibfield  {author} {\bibinfo {author} {\bibfnamefont {Kosuke}\ \bibnamefont
  {Nishiwaki}}, \bibinfo {author} {\bibfnamefont {Katsuaki}\ \bibnamefont
  {Asano}}, \ and\ \bibinfo {author} {\bibfnamefont {Kohta}\ \bibnamefont
  {Murase}},\ }\bibfield  {title} {\enquote {\bibinfo {title} {{Particle
  Reacceleration by Turbulence and Radio Constraints on Multimessenger
  High-energy Emission from the Coma Cluster}},}\ }\href {\doibase
  10.3847/1538-4357/ac1cdb} {\bibfield  {journal} {\bibinfo  {journal}
  {Astrophys. J.}\ }\textbf {\bibinfo {volume} {922}},\ \bibinfo {pages} {190}
  (\bibinfo {year} {2021})},\ \Eprint {http://arxiv.org/abs/2105.04541}
  {arXiv:2105.04541 [astro-ph.HE]} \BibitemShut {NoStop}%
\bibitem [{\citenamefont {Berezinsky}(1977)}]{Berezinsky1977}%
  \BibitemOpen
  \bibfield  {author} {\bibinfo {author} {\bibfnamefont {V.~S.}\ \bibnamefont
  {Berezinsky}},\ }\bibfield  {title} {\enquote {\bibinfo {title} {---},}\ }in\
  \href@noop {} {\emph {\bibinfo {booktitle} {Proc. 7th Int. Conf.
  \textit{Neutrino-77}}}},\ Vol.~\bibinfo {volume} {1}\ (\bibinfo {address}
  {USSR},\ \bibinfo {year} {1977})\ p.\ \bibinfo {pages} {177}\BibitemShut
  {NoStop}%
\bibitem [{\citenamefont {Stecker}\ \emph {et~al.}(1991)\citenamefont
  {Stecker}, \citenamefont {Done}, \citenamefont {Salamon},\ and\ \citenamefont
  {Sommers}}]{Stecker:1991vm}%
  \BibitemOpen
  \bibfield  {author} {\bibinfo {author} {\bibfnamefont {Floyd~W.}\
  \bibnamefont {Stecker}}, \bibinfo {author} {\bibfnamefont {C.}~\bibnamefont
  {Done}}, \bibinfo {author} {\bibfnamefont {Michael~H.}\ \bibnamefont
  {Salamon}}, \ and\ \bibinfo {author} {\bibfnamefont {P.}~\bibnamefont
  {Sommers}},\ }\bibfield  {title} {\enquote {\bibinfo {title} {{High-energy
  neutrinos from active galactic nuclei}},}\ }\href {\doibase
  10.1103/PhysRevLett.66.2697} {\bibfield  {journal} {\bibinfo  {journal}
  {Phys.Rev.Lett.}\ }\textbf {\bibinfo {volume} {66}},\ \bibinfo {pages}
  {2697--2700} (\bibinfo {year} {1991})}\BibitemShut {NoStop}%
\bibitem [{\citenamefont {Hopkins}\ \emph {et~al.}(2007)\citenamefont
  {Hopkins}, \citenamefont {Richards},\ and\ \citenamefont
  {Hernquist}}]{Hopkins:2006fq}%
  \BibitemOpen
  \bibfield  {author} {\bibinfo {author} {\bibfnamefont {Philip~F.}\
  \bibnamefont {Hopkins}}, \bibinfo {author} {\bibfnamefont {Gordon~T.}\
  \bibnamefont {Richards}}, \ and\ \bibinfo {author} {\bibfnamefont {Lars}\
  \bibnamefont {Hernquist}},\ }\bibfield  {title} {\enquote {\bibinfo {title}
  {{An Observational Determination of the Bolometric Quasar Luminosity
  Function}},}\ }\href {\doibase 10.1086/509629} {\bibfield  {journal}
  {\bibinfo  {journal} {Astrophys. J.}\ }\textbf {\bibinfo {volume} {654}},\
  \bibinfo {pages} {731--753} (\bibinfo {year} {2007})},\ \Eprint
  {http://arxiv.org/abs/astro-ph/0605678} {arXiv:astro-ph/0605678} \BibitemShut
  {NoStop}%
\bibitem [{\citenamefont {Kimura}\ \emph {et~al.}(2015)\citenamefont {Kimura},
  \citenamefont {Murase},\ and\ \citenamefont {Toma}}]{Kimura:2014jba}%
  \BibitemOpen
  \bibfield  {author} {\bibinfo {author} {\bibfnamefont {Shigeo~S.}\
  \bibnamefont {Kimura}}, \bibinfo {author} {\bibfnamefont {Kohta}\
  \bibnamefont {Murase}}, \ and\ \bibinfo {author} {\bibfnamefont {Kenji}\
  \bibnamefont {Toma}},\ }\bibfield  {title} {\enquote {\bibinfo {title}
  {{Neutrino and Cosmic-Ray Emission and Cumulative Background from Radiatively
  Inefficient Accretion Flows in Low-Luminosity Active Galactic Nuclei}},}\
  }\href {\doibase 10.1088/0004-637X/806/2/159} {\bibfield  {journal} {\bibinfo
   {journal} {Astrophys.J.}\ }\textbf {\bibinfo {volume} {806}},\ \bibinfo
  {pages} {159} (\bibinfo {year} {2015})},\ \Eprint
  {http://arxiv.org/abs/1411.3588} {arXiv:1411.3588 [astro-ph.HE]} \BibitemShut
  {NoStop}%
\bibitem [{\citenamefont {Kimura}\ \emph
  {et~al.}(2019{\natexlab{a}})\citenamefont {Kimura}, \citenamefont {Murase},\
  and\ \citenamefont {M\'esz\'aros}}]{Kimura:2019yjo}%
  \BibitemOpen
  \bibfield  {author} {\bibinfo {author} {\bibfnamefont {Shigeo~S.}\
  \bibnamefont {Kimura}}, \bibinfo {author} {\bibfnamefont {Kohta}\
  \bibnamefont {Murase}}, \ and\ \bibinfo {author} {\bibfnamefont {Peter}\
  \bibnamefont {M\'esz\'aros}},\ }\bibfield  {title} {\enquote {\bibinfo
  {title} {{Multimessenger tests of cosmic-ray acceleration in radiatively
  inefficient accretion flows}},}\ }\href {\doibase
  10.1103/PhysRevD.100.083014} {\bibfield  {journal} {\bibinfo  {journal}
  {Phys. Rev. D}\ }\textbf {\bibinfo {volume} {100}},\ \bibinfo {pages}
  {083014} (\bibinfo {year} {2019}{\natexlab{a}})},\ \Eprint
  {http://arxiv.org/abs/1908.08421} {arXiv:1908.08421 [astro-ph.HE]}
  \BibitemShut {NoStop}%
\bibitem [{\citenamefont {Ricci}\ \emph {et~al.}(2017)\citenamefont {Ricci}
  \emph {et~al.}}]{Ricci:2017dhj}%
  \BibitemOpen
  \bibfield  {author} {\bibinfo {author} {\bibfnamefont {Claudio}\ \bibnamefont
  {Ricci}} \emph {et~al.},\ }\bibfield  {title} {\enquote {\bibinfo {title}
  {{BAT AGN Spectroscopic Survey - V. X-ray properties of the Swift/BAT
  70-month AGN catalog}},}\ }\href {\doibase 10.3847/1538-4365/aa96ad}
  {\bibfield  {journal} {\bibinfo  {journal} {Astrophys. J. Suppl.}\ }\textbf
  {\bibinfo {volume} {233}},\ \bibinfo {pages} {17} (\bibinfo {year} {2017})},\
  \Eprint {http://arxiv.org/abs/1709.03989} {arXiv:1709.03989 [astro-ph.HE]}
  \BibitemShut {NoStop}%
\bibitem [{\citenamefont {Ho}(2008)}]{Ho:2008rf}%
  \BibitemOpen
  \bibfield  {author} {\bibinfo {author} {\bibfnamefont {Luis~C.}\ \bibnamefont
  {Ho}},\ }\bibfield  {title} {\enquote {\bibinfo {title} {{Nuclear Activity in
  Nearby Galaxies}},}\ }\href {\doibase 10.1146/annurev.astro.45.051806.110546}
  {\bibfield  {journal} {\bibinfo  {journal} {Ann. Rev. Astron. Astrophys.}\
  }\textbf {\bibinfo {volume} {46}},\ \bibinfo {pages} {475--539} (\bibinfo
  {year} {2008})},\ \Eprint {http://arxiv.org/abs/0803.2268} {arXiv:0803.2268
  [astro-ph]} \BibitemShut {NoStop}%
\bibitem [{\citenamefont {Das}\ \emph {et~al.}(2024)\citenamefont {Das},
  \citenamefont {Zhang},\ and\ \citenamefont {Murase}}]{Das:2024vug}%
  \BibitemOpen
  \bibfield  {author} {\bibinfo {author} {\bibfnamefont {Abhishek}\
  \bibnamefont {Das}}, \bibinfo {author} {\bibfnamefont {B.~Theodore}\
  \bibnamefont {Zhang}}, \ and\ \bibinfo {author} {\bibfnamefont {Kohta}\
  \bibnamefont {Murase}},\ }\bibfield  {title} {\enquote {\bibinfo {title}
  {{Revealing the Production Mechanism of High-energy Neutrinos from NGC
  1068}},}\ }\href {\doibase 10.3847/1538-4357/ad5a04} {\bibfield  {journal}
  {\bibinfo  {journal} {Astrophys. J.}\ }\textbf {\bibinfo {volume} {972}},\
  \bibinfo {pages} {44} (\bibinfo {year} {2024})},\ \Eprint
  {http://arxiv.org/abs/2405.09332} {arXiv:2405.09332 [astro-ph.HE]}
  \BibitemShut {NoStop}%
\bibitem [{\citenamefont {Nemmen}\ \emph {et~al.}(2024)\citenamefont {Nemmen},
  \citenamefont {Vemado}, \citenamefont {Almeida}, \citenamefont {Garcia},\
  and\ \citenamefont {Motta}}]{Nemmen:2023kpa}%
  \BibitemOpen
  \bibfield  {author} {\bibinfo {author} {\bibfnamefont {Rodrigo}\ \bibnamefont
  {Nemmen}}, \bibinfo {author} {\bibfnamefont {Artur}\ \bibnamefont {Vemado}},
  \bibinfo {author} {\bibfnamefont {Ivan}\ \bibnamefont {Almeida}}, \bibinfo
  {author} {\bibfnamefont {Javier}\ \bibnamefont {Garcia}}, \ and\ \bibinfo
  {author} {\bibfnamefont {Pedro~N.}\ \bibnamefont {Motta}},\ }\bibfield
  {title} {\enquote {\bibinfo {title} {{Emergence of hot corona and truncated
  disc in simulations of accreting stellar mass black holes}},}\ }\href
  {\doibase 10.1093/mnras/stae1133} {\bibfield  {journal} {\bibinfo  {journal}
  {Mon. Not. Roy. Astron. Soc.}\ }\textbf {\bibinfo {volume} {531}},\ \bibinfo
  {pages} {805--814} (\bibinfo {year} {2024})},\ \Eprint
  {http://arxiv.org/abs/2305.11429} {arXiv:2305.11429 [astro-ph.HE]}
  \BibitemShut {NoStop}%
\bibitem [{\citenamefont {Hagen}\ \emph {et~al.}(2024)\citenamefont {Hagen},
  \citenamefont {Done}, \citenamefont {Silverman}, \citenamefont {Li},
  \citenamefont {Liu}, \citenamefont {Ren}, \citenamefont {Buchner},
  \citenamefont {Merloni}, \citenamefont {Nagao},\ and\ \citenamefont
  {Salvato}}]{Hagen:2024lyg}%
  \BibitemOpen
  \bibfield  {author} {\bibinfo {author} {\bibfnamefont {Scott}\ \bibnamefont
  {Hagen}}, \bibinfo {author} {\bibfnamefont {Chris}\ \bibnamefont {Done}},
  \bibinfo {author} {\bibfnamefont {John~D.}\ \bibnamefont {Silverman}},
  \bibinfo {author} {\bibfnamefont {Junyao}\ \bibnamefont {Li}}, \bibinfo
  {author} {\bibfnamefont {Teng}\ \bibnamefont {Liu}}, \bibinfo {author}
  {\bibfnamefont {Wenke}\ \bibnamefont {Ren}}, \bibinfo {author} {\bibfnamefont
  {Johannes}\ \bibnamefont {Buchner}}, \bibinfo {author} {\bibfnamefont
  {Andrea}\ \bibnamefont {Merloni}}, \bibinfo {author} {\bibfnamefont {Tohru}\
  \bibnamefont {Nagao}}, \ and\ \bibinfo {author} {\bibfnamefont {Mara}\
  \bibnamefont {Salvato}},\ }\bibfield  {title} {\enquote {\bibinfo {title}
  {{Systematic collapse of the accretion disc across the supermassive black
  hole population}},}\ }\href {\doibase 10.1093/mnras/stae2272} {\bibfield
  {journal} {\bibinfo  {journal} {Mon. Not. Roy. Astron. Soc.}\ }\textbf
  {\bibinfo {volume} {534}},\ \bibinfo {pages} {2803--2818} (\bibinfo {year}
  {2024})},\ \Eprint {http://arxiv.org/abs/2406.06674} {arXiv:2406.06674
  [astro-ph.HE]} \BibitemShut {NoStop}%
\bibitem [{\citenamefont {Kang}\ \emph {et~al.}(2025)\citenamefont {Kang},
  \citenamefont {Done}, \citenamefont {Hagen}, \citenamefont {Temple},
  \citenamefont {Silverman}, \citenamefont {Li},\ and\ \citenamefont
  {Liu}}]{Kang:2024coc}%
  \BibitemOpen
  \bibfield  {author} {\bibinfo {author} {\bibfnamefont {Jia-Lai}\ \bibnamefont
  {Kang}}, \bibinfo {author} {\bibfnamefont {Chris}\ \bibnamefont {Done}},
  \bibinfo {author} {\bibfnamefont {Scott}\ \bibnamefont {Hagen}}, \bibinfo
  {author} {\bibfnamefont {Matthew~J.}\ \bibnamefont {Temple}}, \bibinfo
  {author} {\bibfnamefont {John~D.}\ \bibnamefont {Silverman}}, \bibinfo
  {author} {\bibfnamefont {Junyao}\ \bibnamefont {Li}}, \ and\ \bibinfo
  {author} {\bibfnamefont {Teng}\ \bibnamefont {Liu}},\ }\bibfield  {title}
  {\enquote {\bibinfo {title} {{Systematic collapse of the accretion disc in
  AGN confirmed by UV photometry and broad-line spectra}},}\ }\href {\doibase
  10.1093/mnras/staf145} {\bibfield  {journal} {\bibinfo  {journal} {Mon. Not.
  Roy. Astron. Soc.}\ }\textbf {\bibinfo {volume} {538}},\ \bibinfo {pages}
  {121--131} (\bibinfo {year} {2025})},\ \Eprint
  {http://arxiv.org/abs/2410.06730} {arXiv:2410.06730 [astro-ph.HE]}
  \BibitemShut {NoStop}%
\bibitem [{\citenamefont {Groselj}\ \emph {et~al.}(2026)\citenamefont
  {Groselj}, \citenamefont {Philippov}, \citenamefont {Beloborodov},\ and\
  \citenamefont {Mushotzky}}]{Groselj:2026nix}%
  \BibitemOpen
  \bibfield  {author} {\bibinfo {author} {\bibfnamefont {Daniel}\ \bibnamefont
  {Groselj}}, \bibinfo {author} {\bibfnamefont {Alexander}\ \bibnamefont
  {Philippov}}, \bibinfo {author} {\bibfnamefont {Andrei~M.}\ \bibnamefont
  {Beloborodov}}, \ and\ \bibinfo {author} {\bibfnamefont {Richard}\
  \bibnamefont {Mushotzky}},\ }\bibfield  {title} {\enquote {\bibinfo {title}
  {{High-energy Emission from Turbulent Electron-ion Coronae of Accreting Black
  Holes}},}\ }\href@noop {} {\  (\bibinfo {year} {2026})},\ \Eprint
  {http://arxiv.org/abs/2601.00518} {arXiv:2601.00518 [astro-ph.HE]}
  \BibitemShut {NoStop}%
\bibitem [{\citenamefont {Fiorillo}\ \emph
  {et~al.}(2024{\natexlab{a}})\citenamefont {Fiorillo}, \citenamefont
  {Petropoulou}, \citenamefont {Comisso}, \citenamefont {Peretti},\ and\
  \citenamefont {Sironi}}]{Fiorillo:2023dts}%
  \BibitemOpen
  \bibfield  {author} {\bibinfo {author} {\bibfnamefont {Damiano F.~G.}\
  \bibnamefont {Fiorillo}}, \bibinfo {author} {\bibfnamefont {Maria}\
  \bibnamefont {Petropoulou}}, \bibinfo {author} {\bibfnamefont {Luca}\
  \bibnamefont {Comisso}}, \bibinfo {author} {\bibfnamefont {Enrico}\
  \bibnamefont {Peretti}}, \ and\ \bibinfo {author} {\bibfnamefont {Lorenzo}\
  \bibnamefont {Sironi}},\ }\bibfield  {title} {\enquote {\bibinfo {title}
  {{TeV Neutrinos and Hard X-Rays from Relativistic Reconnection in the Corona
  of NGC 1068}},}\ }\href {\doibase 10.3847/2041-8213/ad192b} {\bibfield
  {journal} {\bibinfo  {journal} {Astrophys. J.}\ }\textbf {\bibinfo {volume}
  {961}},\ \bibinfo {pages} {L14} (\bibinfo {year} {2024}{\natexlab{a}})},\
  \Eprint {http://arxiv.org/abs/2310.18254} {arXiv:2310.18254 [astro-ph.HE]}
  \BibitemShut {NoStop}%
\bibitem [{\citenamefont {Mbarek}\ \emph {et~al.}(2024)\citenamefont {Mbarek},
  \citenamefont {Philippov}, \citenamefont {Chernoglazov}, \citenamefont
  {Levinson},\ and\ \citenamefont {Mushotzky}}]{Mbarek:2023yeq}%
  \BibitemOpen
  \bibfield  {author} {\bibinfo {author} {\bibfnamefont {Rostom}\ \bibnamefont
  {Mbarek}}, \bibinfo {author} {\bibfnamefont {Alexander}\ \bibnamefont
  {Philippov}}, \bibinfo {author} {\bibfnamefont {Alexander}\ \bibnamefont
  {Chernoglazov}}, \bibinfo {author} {\bibfnamefont {Amir}\ \bibnamefont
  {Levinson}}, \ and\ \bibinfo {author} {\bibfnamefont {Richard}\ \bibnamefont
  {Mushotzky}},\ }\bibfield  {title} {\enquote {\bibinfo {title} {{Interplay
  between accelerated protons, x rays and neutrinos in the corona of NGC 1068:
  Constraints from kinetic plasma simulations}},}\ }\href {\doibase
  10.1103/PhysRevD.109.L101306} {\bibfield  {journal} {\bibinfo  {journal}
  {Phys. Rev. D}\ }\textbf {\bibinfo {volume} {109}},\ \bibinfo {pages}
  {L101306} (\bibinfo {year} {2024})},\ \Eprint
  {http://arxiv.org/abs/2310.15222} {arXiv:2310.15222 [astro-ph.HE]}
  \BibitemShut {NoStop}%
\bibitem [{\citenamefont {Kheirandish}\ \emph {et~al.}(2021)\citenamefont
  {Kheirandish}, \citenamefont {Murase},\ and\ \citenamefont
  {Kimura}}]{Kheirandish:2021wkm}%
  \BibitemOpen
  \bibfield  {author} {\bibinfo {author} {\bibfnamefont {Ali}\ \bibnamefont
  {Kheirandish}}, \bibinfo {author} {\bibfnamefont {Kohta}\ \bibnamefont
  {Murase}}, \ and\ \bibinfo {author} {\bibfnamefont {Shigeo~S.}\ \bibnamefont
  {Kimura}},\ }\bibfield  {title} {\enquote {\bibinfo {title} {{High-energy
  Neutrinos from Magnetized Coronae of Active Galactic Nuclei and Prospects for
  Identification of Seyfert Galaxies and Quasars in Neutrino Telescopes}},}\
  }\href {\doibase 10.3847/1538-4357/ac1c77} {\bibfield  {journal} {\bibinfo
  {journal} {Astrophys. J.}\ }\textbf {\bibinfo {volume} {922}},\ \bibinfo
  {pages} {45} (\bibinfo {year} {2021})},\ \Eprint
  {http://arxiv.org/abs/2102.04475} {arXiv:2102.04475 [astro-ph.HE]}
  \BibitemShut {NoStop}%
\bibitem [{\citenamefont {Mayers}\ \emph {et~al.}(2018)\citenamefont {Mayers}
  \emph {et~al.}}]{Mayers:2018hau}%
  \BibitemOpen
  \bibfield  {author} {\bibinfo {author} {\bibfnamefont {Julian~A.}\
  \bibnamefont {Mayers}} \emph {et~al.},\ }\bibfield  {title} {\enquote
  {\bibinfo {title} {{Correlations between X-ray properties and Black Hole Mass
  in AGN: towards a new method to estimate black hole mass from short exposure
  X-ray observations}},}\ }\href@noop {} {\  (\bibinfo {year} {2018})},\
  \Eprint {http://arxiv.org/abs/1803.06891} {arXiv:1803.06891 [astro-ph.GA]}
  \BibitemShut {NoStop}%
\bibitem [{\citenamefont {Ueda}\ \emph {et~al.}(2014)\citenamefont {Ueda},
  \citenamefont {Akiyama}, \citenamefont {Hasinger}, \citenamefont {Miyaji},\
  and\ \citenamefont {Watson}}]{Ueda:2014tma}%
  \BibitemOpen
  \bibfield  {author} {\bibinfo {author} {\bibfnamefont {Yoshihiro}\
  \bibnamefont {Ueda}}, \bibinfo {author} {\bibfnamefont {Masayuki}\
  \bibnamefont {Akiyama}}, \bibinfo {author} {\bibfnamefont {Günther}\
  \bibnamefont {Hasinger}}, \bibinfo {author} {\bibfnamefont {Takamitsu}\
  \bibnamefont {Miyaji}}, \ and\ \bibinfo {author} {\bibfnamefont {Michael~G.}\
  \bibnamefont {Watson}},\ }\bibfield  {title} {\enquote {\bibinfo {title}
  {{Toward the Standard Population Synthesis Model of the X-Ray Background:
  Evolution of X-Ray Luminosity and Absorption Functions of Active Galactic
  Nuclei Including Compton-Thick Populations}},}\ }\href {\doibase
  10.1088/0004-637X/786/2/104} {\bibfield  {journal} {\bibinfo  {journal}
  {Astrophys.J.}\ }\textbf {\bibinfo {volume} {786}},\ \bibinfo {pages} {104}
  (\bibinfo {year} {2014})},\ \Eprint {http://arxiv.org/abs/1402.1836}
  {arXiv:1402.1836 [astro-ph.CO]} \BibitemShut {NoStop}%
\bibitem [{\citenamefont {Shen}\ \emph {et~al.}(2020)\citenamefont {Shen},
  \citenamefont {Hopkins}, \citenamefont {Faucher-Gigu{\`e}re}, \citenamefont
  {Alexander}, \citenamefont {Richards}, \citenamefont {Ross},\ and\
  \citenamefont {Hickox}}]{Shen:2020obl}%
  \BibitemOpen
  \bibfield  {author} {\bibinfo {author} {\bibfnamefont {Xuejian}\ \bibnamefont
  {Shen}}, \bibinfo {author} {\bibfnamefont {Philip~F.}\ \bibnamefont
  {Hopkins}}, \bibinfo {author} {\bibfnamefont {Claude-Andr{\'e}}\ \bibnamefont
  {Faucher-Gigu{\`e}re}}, \bibinfo {author} {\bibfnamefont {D.~M.}\
  \bibnamefont {Alexander}}, \bibinfo {author} {\bibfnamefont {Gordon~T.}\
  \bibnamefont {Richards}}, \bibinfo {author} {\bibfnamefont {Nicholas~P.}\
  \bibnamefont {Ross}}, \ and\ \bibinfo {author} {\bibfnamefont {R.~C.}\
  \bibnamefont {Hickox}},\ }\bibfield  {title} {\enquote {\bibinfo {title}
  {{The bolometric quasar luminosity function at z = 0{\textendash}7}},}\
  }\href {\doibase 10.1093/mnras/staa1381} {\bibfield  {journal} {\bibinfo
  {journal} {Mon. Not. Roy. Astron. Soc.}\ }\textbf {\bibinfo {volume} {495}},\
  \bibinfo {pages} {3252--3275} (\bibinfo {year} {2020})},\ \Eprint
  {http://arxiv.org/abs/2001.02696} {arXiv:2001.02696 [astro-ph.GA]}
  \BibitemShut {NoStop}%
\bibitem [{\citenamefont {Gilli}\ \emph {et~al.}(2007)\citenamefont {Gilli},
  \citenamefont {Comastri},\ and\ \citenamefont {Hasinger}}]{Gilli:2006zi}%
  \BibitemOpen
  \bibfield  {author} {\bibinfo {author} {\bibfnamefont {Roberto}\ \bibnamefont
  {Gilli}}, \bibinfo {author} {\bibfnamefont {Andrea}\ \bibnamefont
  {Comastri}}, \ and\ \bibinfo {author} {\bibfnamefont {Guenther}\ \bibnamefont
  {Hasinger}},\ }\bibfield  {title} {\enquote {\bibinfo {title} {{The synthesis
  of the cosmic X-ray background in the Chandra and XMM-Newton era}},}\ }\href
  {\doibase 10.1051/0004-6361:20066334} {\bibfield  {journal} {\bibinfo
  {journal} {Astron. Astrophys.}\ }\textbf {\bibinfo {volume} {463}},\ \bibinfo
  {pages} {79} (\bibinfo {year} {2007})},\ \Eprint
  {http://arxiv.org/abs/astro-ph/0610939} {arXiv:astro-ph/0610939} \BibitemShut
  {NoStop}%
\bibitem [{\citenamefont {Balbus}\ and\ \citenamefont
  {Hawley}(1991)}]{Balbus:1991ay}%
  \BibitemOpen
  \bibfield  {author} {\bibinfo {author} {\bibfnamefont {Steven~A.}\
  \bibnamefont {Balbus}}\ and\ \bibinfo {author} {\bibfnamefont {John~F.}\
  \bibnamefont {Hawley}},\ }\bibfield  {title} {\enquote {\bibinfo {title} {{A
  powerful local shear instability in weakly magnetized disks. 1. Linear
  analysis. 2. Nonlinear evolution}},}\ }\href {\doibase 10.1086/170270}
  {\bibfield  {journal} {\bibinfo  {journal} {Astrophys. J.}\ }\textbf
  {\bibinfo {volume} {376}},\ \bibinfo {pages} {214--233} (\bibinfo {year}
  {1991})}\BibitemShut {NoStop}%
\bibitem [{\citenamefont {Balbus}\ and\ \citenamefont
  {Hawley}(1998)}]{Balbus:1998ja}%
  \BibitemOpen
  \bibfield  {author} {\bibinfo {author} {\bibfnamefont {Steven~A.}\
  \bibnamefont {Balbus}}\ and\ \bibinfo {author} {\bibfnamefont {John~F.}\
  \bibnamefont {Hawley}},\ }\bibfield  {title} {\enquote {\bibinfo {title}
  {{Instability, turbulence, and enhanced transport in accretion disks}},}\
  }\href {\doibase 10.1103/RevModPhys.70.1} {\bibfield  {journal} {\bibinfo
  {journal} {Rev. Mod. Phys.}\ }\textbf {\bibinfo {volume} {70}},\ \bibinfo
  {pages} {1--53} (\bibinfo {year} {1998})}\BibitemShut {NoStop}%
\bibitem [{\citenamefont {Galeev}\ \emph {et~al.}(1979)\citenamefont {Galeev},
  \citenamefont {Rosner},\ and\ \citenamefont {Vaiana}}]{Galeev:1979td}%
  \BibitemOpen
  \bibfield  {author} {\bibinfo {author} {\bibfnamefont {A.~A.}\ \bibnamefont
  {Galeev}}, \bibinfo {author} {\bibfnamefont {R.}~\bibnamefont {Rosner}}, \
  and\ \bibinfo {author} {\bibfnamefont {G.~S.}\ \bibnamefont {Vaiana}},\
  }\bibfield  {title} {\enquote {\bibinfo {title} {{Structured coronae of
  accretion disks}},}\ }\href {\doibase 10.1086/156957} {\bibfield  {journal}
  {\bibinfo  {journal} {Astrophys. J.}\ }\textbf {\bibinfo {volume} {229}},\
  \bibinfo {pages} {318--326} (\bibinfo {year} {1979})}\BibitemShut {NoStop}%
\bibitem [{\citenamefont {Miller}\ and\ \citenamefont
  {Stone}(2000)}]{Miller:1999ix}%
  \BibitemOpen
  \bibfield  {author} {\bibinfo {author} {\bibfnamefont {K.~A.}\ \bibnamefont
  {Miller}}\ and\ \bibinfo {author} {\bibfnamefont {J.~M.}\ \bibnamefont
  {Stone}},\ }\bibfield  {title} {\enquote {\bibinfo {title} {{The Formation
  and structure of a strongly magnetized corona above weakly magnetized
  accretion disks}},}\ }\href {\doibase 10.1086/308736} {\bibfield  {journal}
  {\bibinfo  {journal} {Astrophys. J.}\ }\textbf {\bibinfo {volume} {534}},\
  \bibinfo {pages} {398--419} (\bibinfo {year} {2000})},\ \Eprint
  {http://arxiv.org/abs/astro-ph/9912135} {arXiv:astro-ph/9912135} \BibitemShut
  {NoStop}%
\bibitem [{\citenamefont {Fromang}\ and\ \citenamefont
  {Nelson}(2006)}]{Fromang:2006vd}%
  \BibitemOpen
  \bibfield  {author} {\bibinfo {author} {\bibfnamefont {Sebastien}\
  \bibnamefont {Fromang}}\ and\ \bibinfo {author} {\bibfnamefont {Richard~P.}\
  \bibnamefont {Nelson}},\ }\bibfield  {title} {\enquote {\bibinfo {title}
  {{Global MHD simulations of stratified and turbulent protoplanetary discs. 1.
  Model properties}},}\ }\href {\doibase 10.1051/0004-6361:20065643} {\bibfield
   {journal} {\bibinfo  {journal} {Astron. Astrophys.}\ }\textbf {\bibinfo
  {volume} {457}},\ \bibinfo {pages} {343} (\bibinfo {year} {2006})},\ \Eprint
  {http://arxiv.org/abs/astro-ph/0606729} {arXiv:astro-ph/0606729} \BibitemShut
  {NoStop}%
\bibitem [{\citenamefont {{Tout}}\ and\ \citenamefont
  {{Pringle}}(1992)}]{1992MNRAS.259..604T}%
  \BibitemOpen
  \bibfield  {author} {\bibinfo {author} {\bibfnamefont {C.~A.}\ \bibnamefont
  {{Tout}}}\ and\ \bibinfo {author} {\bibfnamefont {J.~E.}\ \bibnamefont
  {{Pringle}}},\ }\bibfield  {title} {\enquote {\bibinfo {title} {{Accretion
  disc viscosity: a simple model for a magnetic dynamo.}}}\ }\href {\doibase
  10.1093/mnras/259.4.604} {\bibfield  {journal} {\bibinfo  {journal} {\mnras}\
  }\textbf {\bibinfo {volume} {259}},\ \bibinfo {pages} {604--612} (\bibinfo
  {year} {1992})}\BibitemShut {NoStop}%
\bibitem [{\citenamefont {Johansen}\ and\ \citenamefont
  {Levin}(2008)}]{Johansen:2008ed}%
  \BibitemOpen
  \bibfield  {author} {\bibinfo {author} {\bibfnamefont {Anders}\ \bibnamefont
  {Johansen}}\ and\ \bibinfo {author} {\bibfnamefont {Yuri}\ \bibnamefont
  {Levin}},\ }\bibfield  {title} {\enquote {\bibinfo {title} {{High accretion
  rates in magnetised Keplerian discs mediated by a Parker instability driven
  dynamo}},}\ }\href {\doibase 10.1051/0004-6361:200810385} {\bibfield
  {journal} {\bibinfo  {journal} {Astron. Astrophys.}\ }\textbf {\bibinfo
  {volume} {490}},\ \bibinfo {pages} {501} (\bibinfo {year} {2008})},\ \Eprint
  {http://arxiv.org/abs/0808.3579} {arXiv:0808.3579 [astro-ph]} \BibitemShut
  {NoStop}%
\bibitem [{\citenamefont {Uzdensky}\ and\ \citenamefont
  {Goodman}(2008)}]{Uzdensky:2008ce}%
  \BibitemOpen
  \bibfield  {author} {\bibinfo {author} {\bibfnamefont {Dmitri}\ \bibnamefont
  {Uzdensky}}\ and\ \bibinfo {author} {\bibfnamefont {Jeremy}\ \bibnamefont
  {Goodman}},\ }\bibfield  {title} {\enquote {\bibinfo {title} {{Statistical
  Description of a Magnetized Corona above a Turbulent Accretion Disk}},}\
  }\href {\doibase 10.1086/588812} {\bibfield  {journal} {\bibinfo  {journal}
  {Astrophys. J.}\ }\textbf {\bibinfo {volume} {682}},\ \bibinfo {pages}
  {608--629} (\bibinfo {year} {2008})},\ \Eprint
  {http://arxiv.org/abs/0803.0337} {arXiv:0803.0337 [astro-ph]} \BibitemShut
  {NoStop}%
\bibitem [{\citenamefont {Lazarian}\ and\ \citenamefont
  {Vishniac}(1999)}]{Lazarian:1998wd}%
  \BibitemOpen
  \bibfield  {author} {\bibinfo {author} {\bibfnamefont {A.}~\bibnamefont
  {Lazarian}}\ and\ \bibinfo {author} {\bibfnamefont {E.~T.}\ \bibnamefont
  {Vishniac}},\ }\bibfield  {title} {\enquote {\bibinfo {title} {{Reconnection
  in a weakly stochastic field}},}\ }\href {\doibase 10.1086/307233} {\bibfield
   {journal} {\bibinfo  {journal} {Astrophys. J.}\ }\textbf {\bibinfo {volume}
  {517}},\ \bibinfo {pages} {700--718} (\bibinfo {year} {1999})},\ \Eprint
  {http://arxiv.org/abs/astro-ph/9811037} {arXiv:astro-ph/9811037} \BibitemShut
  {NoStop}%
\bibitem [{\citenamefont {Uzdensky}\ and\ \citenamefont
  {Loureiro}(2016)}]{Uzdensky:2014uda}%
  \BibitemOpen
  \bibfield  {author} {\bibinfo {author} {\bibfnamefont {Dmitri~A.}\
  \bibnamefont {Uzdensky}}\ and\ \bibinfo {author} {\bibfnamefont {Nuno~F.}\
  \bibnamefont {Loureiro}},\ }\bibfield  {title} {\enquote {\bibinfo {title}
  {{Magnetic Reconnection Onset via Disruption of a Forming Current Sheet by
  the Tearing Instability}},}\ }\href {\doibase 10.1103/PhysRevLett.116.105003}
  {\bibfield  {journal} {\bibinfo  {journal} {Phys. Rev. Lett.}\ }\textbf
  {\bibinfo {volume} {116}},\ \bibinfo {pages} {105003} (\bibinfo {year}
  {2016})},\ \Eprint {http://arxiv.org/abs/1411.4295} {arXiv:1411.4295
  [astro-ph.SR]} \BibitemShut {NoStop}%
\bibitem [{\citenamefont {Zhdankin}\ \emph {et~al.}(2019)\citenamefont
  {Zhdankin}, \citenamefont {Uzdensky}, \citenamefont {Werner},\ and\
  \citenamefont {Begelman}}]{Zhdankin:2018lhq}%
  \BibitemOpen
  \bibfield  {author} {\bibinfo {author} {\bibfnamefont {Vladimir}\
  \bibnamefont {Zhdankin}}, \bibinfo {author} {\bibfnamefont {Dmitri~A.}\
  \bibnamefont {Uzdensky}}, \bibinfo {author} {\bibfnamefont {Gregory~R.}\
  \bibnamefont {Werner}}, \ and\ \bibinfo {author} {\bibfnamefont
  {Mitchell~C.}\ \bibnamefont {Begelman}},\ }\bibfield  {title} {\enquote
  {\bibinfo {title} {{Electron and ion energization in relativistic plasma
  turbulence}},}\ }\href {\doibase 10.1103/PhysRevLett.122.055101} {\bibfield
  {journal} {\bibinfo  {journal} {Phys. Rev. Lett.}\ }\textbf {\bibinfo
  {volume} {122}},\ \bibinfo {pages} {055101} (\bibinfo {year} {2019})},\
  \Eprint {http://arxiv.org/abs/1809.01966} {arXiv:1809.01966 [astro-ph.HE]}
  \BibitemShut {NoStop}%
\bibitem [{\citenamefont {Comisso}\ and\ \citenamefont
  {Sironi}(2019)}]{Comisso:2019frj}%
  \BibitemOpen
  \bibfield  {author} {\bibinfo {author} {\bibfnamefont {Luca}\ \bibnamefont
  {Comisso}}\ and\ \bibinfo {author} {\bibfnamefont {Lorenzo}\ \bibnamefont
  {Sironi}},\ }\bibfield  {title} {\enquote {\bibinfo {title} {{The interplay
  of magnetically-dominated turbulence and magnetic reconnection in producing
  nonthermal particles}},}\ }\href {\doibase 10.3847/1538-4357/ab4c33}
  {\bibfield  {journal} {\bibinfo  {journal} {Astrophys. J.}\ }\textbf
  {\bibinfo {volume} {886}},\ \bibinfo {pages} {122} (\bibinfo {year}
  {2019})},\ \Eprint {http://arxiv.org/abs/1909.01420} {arXiv:1909.01420
  [astro-ph.HE]} \BibitemShut {NoStop}%
\bibitem [{\citenamefont {Lynn}\ \emph {et~al.}(2014)\citenamefont {Lynn},
  \citenamefont {Quataert}, \citenamefont {Chandran},\ and\ \citenamefont
  {Parrish}}]{Lynn:2014dya}%
  \BibitemOpen
  \bibfield  {author} {\bibinfo {author} {\bibfnamefont {Jacob~W.}\
  \bibnamefont {Lynn}}, \bibinfo {author} {\bibfnamefont {Eliot}\ \bibnamefont
  {Quataert}}, \bibinfo {author} {\bibfnamefont {Benjamin D.~G.}\ \bibnamefont
  {Chandran}}, \ and\ \bibinfo {author} {\bibfnamefont {Ian~J.}\ \bibnamefont
  {Parrish}},\ }\bibfield  {title} {\enquote {\bibinfo {title} {{Acceleration
  of Relativistic Electrons by Magnetohydrodynamic Turbulence: Implications for
  Non-thermal Emission from Black Hole Accretion Disks}},}\ }\href {\doibase
  10.1088/0004-637X/791/1/71} {\bibfield  {journal} {\bibinfo  {journal}
  {Astrophys. J.}\ }\textbf {\bibinfo {volume} {791}},\ \bibinfo {pages} {71}
  (\bibinfo {year} {2014})},\ \Eprint {http://arxiv.org/abs/1403.3123}
  {arXiv:1403.3123 [astro-ph.HE]} \BibitemShut {NoStop}%
\bibitem [{\citenamefont {Kimura}\ \emph
  {et~al.}(2019{\natexlab{b}})\citenamefont {Kimura}, \citenamefont {Tomida},\
  and\ \citenamefont {Murase}}]{Kimura:2018clk}%
  \BibitemOpen
  \bibfield  {author} {\bibinfo {author} {\bibfnamefont {Shigeo~S.}\
  \bibnamefont {Kimura}}, \bibinfo {author} {\bibfnamefont {Kengo}\
  \bibnamefont {Tomida}}, \ and\ \bibinfo {author} {\bibfnamefont {Kohta}\
  \bibnamefont {Murase}},\ }\bibfield  {title} {\enquote {\bibinfo {title}
  {{Acceleration and Escape Processes of High-energy Particles in Turbulence
  inside Hot Accretion Flows}},}\ }\href {\doibase 10.1093/mnras/stz329}
  {\bibfield  {journal} {\bibinfo  {journal} {Mon. Not. Roy. Astron. Soc.}\
  }\textbf {\bibinfo {volume} {485}},\ \bibinfo {pages} {163--178} (\bibinfo
  {year} {2019}{\natexlab{b}})},\ \Eprint {http://arxiv.org/abs/1812.03901}
  {arXiv:1812.03901 [astro-ph.HE]} \BibitemShut {NoStop}%
\bibitem [{\citenamefont {Sun}\ and\ \citenamefont {Bai}(2021)}]{Sun:2021ods}%
  \BibitemOpen
  \bibfield  {author} {\bibinfo {author} {\bibfnamefont {Xiaochen}\
  \bibnamefont {Sun}}\ and\ \bibinfo {author} {\bibfnamefont {Xue-Ning}\
  \bibnamefont {Bai}},\ }\bibfield  {title} {\enquote {\bibinfo {title}
  {{Particle diffusion and acceleration in magnetorotational instability
  turbulence}},}\ }\href {\doibase 10.1093/mnras/stab1643} {\bibfield
  {journal} {\bibinfo  {journal} {Mon. Not. Roy. Astron. Soc.}\ }\textbf
  {\bibinfo {volume} {506}},\ \bibinfo {pages} {1128--1147} (\bibinfo {year}
  {2021})},\ \Eprint {http://arxiv.org/abs/2106.03098} {arXiv:2106.03098
  [astro-ph.HE]} \BibitemShut {NoStop}%
\bibitem [{\citenamefont {Lemoine}(2019)}]{Lemoine:2019ofp}%
  \BibitemOpen
  \bibfield  {author} {\bibinfo {author} {\bibfnamefont {Martin}\ \bibnamefont
  {Lemoine}},\ }\bibfield  {title} {\enquote {\bibinfo {title} {{Generalized
  Fermi acceleration}},}\ }\href {\doibase 10.1103/PhysRevD.99.083006}
  {\bibfield  {journal} {\bibinfo  {journal} {Phys. Rev. D}\ }\textbf {\bibinfo
  {volume} {99}},\ \bibinfo {pages} {083006} (\bibinfo {year} {2019})},\
  \Eprint {http://arxiv.org/abs/1903.05917} {arXiv:1903.05917 [astro-ph.HE]}
  \BibitemShut {NoStop}%
\bibitem [{\citenamefont {Lemoine}(2022)}]{Lemoine:2022rpj}%
  \BibitemOpen
  \bibfield  {author} {\bibinfo {author} {\bibfnamefont {Martin}\ \bibnamefont
  {Lemoine}},\ }\bibfield  {title} {\enquote {\bibinfo {title}
  {{First-Principles Fermi Acceleration in Magnetized Turbulence}},}\ }\href
  {\doibase 10.1103/PhysRevLett.129.215101} {\bibfield  {journal} {\bibinfo
  {journal} {Phys. Rev. Lett.}\ }\textbf {\bibinfo {volume} {129}},\ \bibinfo
  {pages} {215101} (\bibinfo {year} {2022})},\ \Eprint
  {http://arxiv.org/abs/2210.01038} {arXiv:2210.01038 [astro-ph.HE]}
  \BibitemShut {NoStop}%
\bibitem [{\citenamefont {Lemoine}\ \emph {et~al.}(2024)\citenamefont
  {Lemoine}, \citenamefont {Murase},\ and\ \citenamefont
  {Rieger}}]{Lemoine:2023wsw}%
  \BibitemOpen
  \bibfield  {author} {\bibinfo {author} {\bibfnamefont {Martin}\ \bibnamefont
  {Lemoine}}, \bibinfo {author} {\bibfnamefont {Kohta}\ \bibnamefont {Murase}},
  \ and\ \bibinfo {author} {\bibfnamefont {Frank}\ \bibnamefont {Rieger}},\
  }\bibfield  {title} {\enquote {\bibinfo {title} {{Nonlinear aspects of
  stochastic particle acceleration}},}\ }\href {\doibase
  10.1103/PhysRevD.109.063006} {\bibfield  {journal} {\bibinfo  {journal}
  {Phys. Rev. D}\ }\textbf {\bibinfo {volume} {109}},\ \bibinfo {pages}
  {063006} (\bibinfo {year} {2024})},\ \Eprint
  {http://arxiv.org/abs/2312.04443} {arXiv:2312.04443 [astro-ph.HE]}
  \BibitemShut {NoStop}%
\bibitem [{\citenamefont {Lemoine}\ and\ \citenamefont
  {Rieger}(2025)}]{Lemoine:2024roa}%
  \BibitemOpen
  \bibfield  {author} {\bibinfo {author} {\bibfnamefont {Martin}\ \bibnamefont
  {Lemoine}}\ and\ \bibinfo {author} {\bibfnamefont {Frank}\ \bibnamefont
  {Rieger}},\ }\bibfield  {title} {\enquote {\bibinfo {title} {{Neutrinos from
  stochastic acceleration in black hole environments}},}\ }\href {\doibase
  10.1051/0004-6361/202453296} {\bibfield  {journal} {\bibinfo  {journal}
  {Astron. Astrophys.}\ }\textbf {\bibinfo {volume} {697}},\ \bibinfo {pages}
  {A124} (\bibinfo {year} {2025})},\ \Eprint {http://arxiv.org/abs/2412.01457}
  {arXiv:2412.01457 [astro-ph.HE]} \BibitemShut {NoStop}%
\bibitem [{\citenamefont {Fiorillo}\ \emph
  {et~al.}(2024{\natexlab{b}})\citenamefont {Fiorillo}, \citenamefont
  {Comisso}, \citenamefont {Peretti}, \citenamefont {Petropoulou},\ and\
  \citenamefont {Sironi}}]{Fiorillo:2024akm}%
  \BibitemOpen
  \bibfield  {author} {\bibinfo {author} {\bibfnamefont {Damiano F.~G.}\
  \bibnamefont {Fiorillo}}, \bibinfo {author} {\bibfnamefont {Luca}\
  \bibnamefont {Comisso}}, \bibinfo {author} {\bibfnamefont {Enrico}\
  \bibnamefont {Peretti}}, \bibinfo {author} {\bibfnamefont {Maria}\
  \bibnamefont {Petropoulou}}, \ and\ \bibinfo {author} {\bibfnamefont
  {Lorenzo}\ \bibnamefont {Sironi}},\ }\bibfield  {title} {\enquote {\bibinfo
  {title} {{A Magnetized Strongly Turbulent Corona as the Source of Neutrinos
  from NGC 1068}},}\ }\href {\doibase 10.3847/1538-4357/ad7021} {\bibfield
  {journal} {\bibinfo  {journal} {Astrophys. J.}\ }\textbf {\bibinfo {volume}
  {974}},\ \bibinfo {pages} {75} (\bibinfo {year} {2024}{\natexlab{b}})},\
  \Eprint {http://arxiv.org/abs/2407.01678} {arXiv:2407.01678 [astro-ph.HE]}
  \BibitemShut {NoStop}%
\bibitem [{\citenamefont {Kawashima}\ and\ \citenamefont
  {Asano}(2025)}]{Kawashima:2025wzq}%
  \BibitemOpen
  \bibfield  {author} {\bibinfo {author} {\bibfnamefont {Tomohisa}\
  \bibnamefont {Kawashima}}\ and\ \bibinfo {author} {\bibfnamefont {Katsuaki}\
  \bibnamefont {Asano}},\ }\bibfield  {title} {\enquote {\bibinfo {title}
  {{High-energy Neutrino Emission from a Radiatively Inefficient Accretion Flow
  with a Three-dimensional General Relativistic Magnetohydrodynamic
  Simulation}},}\ }\href {\doibase 10.3847/1538-4357/adee96} {\bibfield
  {journal} {\bibinfo  {journal} {Astrophys. J.}\ }\textbf {\bibinfo {volume}
  {989}},\ \bibinfo {pages} {155} (\bibinfo {year} {2025})},\ \Eprint
  {http://arxiv.org/abs/2503.22891} {arXiv:2503.22891 [astro-ph.HE]}
  \BibitemShut {NoStop}%
\bibitem [{\citenamefont {Takamoto}\ and\ \citenamefont
  {Lazarian}(2017)}]{Takamoto:2017vhf}%
  \BibitemOpen
  \bibfield  {author} {\bibinfo {author} {\bibfnamefont {Makoto}\ \bibnamefont
  {Takamoto}}\ and\ \bibinfo {author} {\bibfnamefont {Alexandre}\ \bibnamefont
  {Lazarian}},\ }\bibfield  {title} {\enquote {\bibinfo {title} {{Strong
  coupling of Alfv{\'e}n and fast modes in compressible relativistic
  magnetohydrodynamic turbulence in magnetically dominated plasmas}},}\ }\href
  {\doibase 10.1093/mnras/stx2292} {\bibfield  {journal} {\bibinfo  {journal}
  {Mon. Not. Roy. Astron. Soc.}\ }\textbf {\bibinfo {volume} {472}},\ \bibinfo
  {pages} {4542--4550} (\bibinfo {year} {2017})},\ \Eprint
  {http://arxiv.org/abs/1709.00785} {arXiv:1709.00785 [astro-ph.HE]}
  \BibitemShut {NoStop}%
\bibitem [{\citenamefont {Yan}\ and\ \citenamefont
  {Lazarian}(2002)}]{Yan:2002qm}%
  \BibitemOpen
  \bibfield  {author} {\bibinfo {author} {\bibfnamefont {Hui-rong}\
  \bibnamefont {Yan}}\ and\ \bibinfo {author} {\bibfnamefont {A.}~\bibnamefont
  {Lazarian}},\ }\bibfield  {title} {\enquote {\bibinfo {title} {{Scattering of
  cosmic rays by magnetohydrodynamic interstellar turbulence}},}\ }\href
  {\doibase 10.1103/PhysRevLett.89.281102} {\bibfield  {journal} {\bibinfo
  {journal} {Phys. Rev. Lett.}\ }\textbf {\bibinfo {volume} {89}},\ \bibinfo
  {pages} {281102} (\bibinfo {year} {2002})},\ \Eprint
  {http://arxiv.org/abs/astro-ph/0205285} {arXiv:astro-ph/0205285} \BibitemShut
  {NoStop}%
\bibitem [{\citenamefont {Cho}\ and\ \citenamefont
  {Lazarian}(2006)}]{Cho:2005mb}%
  \BibitemOpen
  \bibfield  {author} {\bibinfo {author} {\bibfnamefont {Jungyeon}\
  \bibnamefont {Cho}}\ and\ \bibinfo {author} {\bibfnamefont {A.}~\bibnamefont
  {Lazarian}},\ }\bibfield  {title} {\enquote {\bibinfo {title} {{Particle
  acceleration by mhd turbulence}},}\ }\href {\doibase 10.1086/498967}
  {\bibfield  {journal} {\bibinfo  {journal} {Astrophys. J.}\ }\textbf
  {\bibinfo {volume} {638}},\ \bibinfo {pages} {811--826} (\bibinfo {year}
  {2006})},\ \Eprint {http://arxiv.org/abs/astro-ph/0509385}
  {arXiv:astro-ph/0509385} \BibitemShut {NoStop}%
\bibitem [{\citenamefont {Teraki}\ and\ \citenamefont
  {Asano}(2019)}]{Teraki:2019qam}%
  \BibitemOpen
  \bibfield  {author} {\bibinfo {author} {\bibfnamefont {Yuto}\ \bibnamefont
  {Teraki}}\ and\ \bibinfo {author} {\bibfnamefont {Katsuaki}\ \bibnamefont
  {Asano}},\ }\bibfield  {title} {\enquote {\bibinfo {title} {{Particle Energy
  Diffusion in Linear Magnetohydrodynamic Waves}},}\ }\href {\doibase
  10.3847/1538-4357/ab1b13} {\bibfield  {journal} {\bibinfo  {journal}
  {Astrophys. J.}\ }\textbf {\bibinfo {volume} {877}},\ \bibinfo {pages} {71}
  (\bibinfo {year} {2019})},\ \Eprint {http://arxiv.org/abs/1904.08579}
  {arXiv:1904.08579 [astro-ph.HE]} \BibitemShut {NoStop}%
\bibitem [{\citenamefont {Demidem}\ \emph {et~al.}(2020)\citenamefont
  {Demidem}, \citenamefont {Lemoine},\ and\ \citenamefont
  {Casse}}]{Demidem:2019jzn}%
  \BibitemOpen
  \bibfield  {author} {\bibinfo {author} {\bibfnamefont {Camilia}\ \bibnamefont
  {Demidem}}, \bibinfo {author} {\bibfnamefont {Martin}\ \bibnamefont
  {Lemoine}}, \ and\ \bibinfo {author} {\bibfnamefont {Fabien}\ \bibnamefont
  {Casse}},\ }\bibfield  {title} {\enquote {\bibinfo {title} {{Particle
  acceleration in relativistic turbulence: a theoretical appraisal}},}\ }\href
  {\doibase 10.1103/PhysRevD.102.023003} {\bibfield  {journal} {\bibinfo
  {journal} {Phys. Rev. D}\ }\textbf {\bibinfo {volume} {102}},\ \bibinfo
  {pages} {023003} (\bibinfo {year} {2020})},\ \Eprint
  {http://arxiv.org/abs/1909.12885} {arXiv:1909.12885 [astro-ph.HE]}
  \BibitemShut {NoStop}%
\bibitem [{\citenamefont {{Yang}}\ \emph {et~al.}(2015)\citenamefont {{Yang}},
  \citenamefont {{Zhang}}, \citenamefont {{He}}, \citenamefont {{Peter}},
  \citenamefont {{Tu}}, \citenamefont {{Wang}}, \citenamefont {{Zhang}},\ and\
  \citenamefont {{Feng}}}]{2015ApJ...800..111Y}%
  \BibitemOpen
  \bibfield  {author} {\bibinfo {author} {\bibfnamefont {Liping}\ \bibnamefont
  {{Yang}}}, \bibinfo {author} {\bibfnamefont {Lei}\ \bibnamefont {{Zhang}}},
  \bibinfo {author} {\bibfnamefont {Jiansen}\ \bibnamefont {{He}}}, \bibinfo
  {author} {\bibfnamefont {Hardi}\ \bibnamefont {{Peter}}}, \bibinfo {author}
  {\bibfnamefont {Chuanyi}\ \bibnamefont {{Tu}}}, \bibinfo {author}
  {\bibfnamefont {Linghua}\ \bibnamefont {{Wang}}}, \bibinfo {author}
  {\bibfnamefont {Shaohua}\ \bibnamefont {{Zhang}}}, \ and\ \bibinfo {author}
  {\bibfnamefont {Xueshang}\ \bibnamefont {{Feng}}},\ }\bibfield  {title}
  {\enquote {\bibinfo {title} {{Numerical Simulation of Fast-mode Magnetosonic
  Waves Excited by Plasmoid Ejections in the Solar Corona}},}\ }\href {\doibase
  10.1088/0004-637X/800/2/111} {\bibfield  {journal} {\bibinfo  {journal}
  {\apj}\ }\textbf {\bibinfo {volume} {800}},\ \bibinfo {eid} {111} (\bibinfo
  {year} {2015})}\BibitemShut {NoStop}%
\bibitem [{\citenamefont {{Mondal}}\ \emph {et~al.}(2024)\citenamefont
  {{Mondal}}, \citenamefont {{Srivastava}}, \citenamefont {{Pontin}},
  \citenamefont {{Priest}}, \citenamefont {{Kwon}},\ and\ \citenamefont
  {{Yuan}}}]{2024ApJ...977..235M}%
  \BibitemOpen
  \bibfield  {author} {\bibinfo {author} {\bibfnamefont {Sripan}\ \bibnamefont
  {{Mondal}}}, \bibinfo {author} {\bibfnamefont {Abhishekh~Kumar}\ \bibnamefont
  {{Srivastava}}}, \bibinfo {author} {\bibfnamefont {David~I.}\ \bibnamefont
  {{Pontin}}}, \bibinfo {author} {\bibfnamefont {Eric~R.}\ \bibnamefont
  {{Priest}}}, \bibinfo {author} {\bibfnamefont {R.~Y.}\ \bibnamefont
  {{Kwon}}}, \ and\ \bibinfo {author} {\bibfnamefont {Ding}\ \bibnamefont
  {{Yuan}}},\ }\bibfield  {title} {\enquote {\bibinfo {title} {{Generation of
  Fast Magnetoacoustic Waves in the Corona by Impulsive Bursty
  Reconnection}},}\ }\href {\doibase 10.3847/1538-4357/ad9022} {\bibfield
  {journal} {\bibinfo  {journal} {\apj}\ }\textbf {\bibinfo {volume} {977}},\
  \bibinfo {eid} {235} (\bibinfo {year} {2024})},\ \Eprint
  {http://arxiv.org/abs/2411.02180} {arXiv:2411.02180 [astro-ph.SR]}
  \BibitemShut {NoStop}%
\bibitem [{\citenamefont {Kimura}\ \emph {et~al.}(2026)\citenamefont {Kimura}
  \emph {et~al.}}]{Kimura:2026}%
  \BibitemOpen
  \bibfield  {author} {\bibinfo {author} {\bibfnamefont {S.~S.}\ \bibnamefont
  {Kimura}} \emph {et~al.},\ }\href@noop {} {\bibfield  {journal} {\bibinfo
  {journal} {In preparation}\ } (\bibinfo {year} {2026})}\BibitemShut {NoStop}%
\bibitem [{\citenamefont {Kawazura}\ and\ \citenamefont
  {Kimura}(2024)}]{Kawazura:2024usv}%
  \BibitemOpen
  \bibfield  {author} {\bibinfo {author} {\bibfnamefont {Yohei}\ \bibnamefont
  {Kawazura}}\ and\ \bibinfo {author} {\bibfnamefont {Shigeo~S.}\ \bibnamefont
  {Kimura}},\ }\bibfield  {title} {\enquote {\bibinfo {title} {{Inertial range
  of magnetorotational turbulence}},}\ }\href {\doibase 10.1126/sciadv.adp4965}
  {\bibfield  {journal} {\bibinfo  {journal} {Sci. Adv.}\ }\textbf {\bibinfo
  {volume} {10}},\ \bibinfo {pages} {adp4965} (\bibinfo {year} {2024})},\
  \Eprint {http://arxiv.org/abs/2404.09252} {arXiv:2404.09252
  [physics.plasm-ph]} \BibitemShut {NoStop}%
\bibitem [{\citenamefont {Murase}\ \emph
  {et~al.}(2020{\natexlab{b}})\citenamefont {Murase}, \citenamefont {Kimura},
  \citenamefont {Zhang}, \citenamefont {Oikonomou},\ and\ \citenamefont
  {Petropoulou}}]{Murase:2020lnu}%
  \BibitemOpen
  \bibfield  {author} {\bibinfo {author} {\bibfnamefont {Kohta}\ \bibnamefont
  {Murase}}, \bibinfo {author} {\bibfnamefont {Shigeo~S.}\ \bibnamefont
  {Kimura}}, \bibinfo {author} {\bibfnamefont {B.~Theodore}\ \bibnamefont
  {Zhang}}, \bibinfo {author} {\bibfnamefont {Foteini}\ \bibnamefont
  {Oikonomou}}, \ and\ \bibinfo {author} {\bibfnamefont {Maria}\ \bibnamefont
  {Petropoulou}},\ }\bibfield  {title} {\enquote {\bibinfo {title}
  {{High-Energy Neutrino and Gamma-Ray Emission from Tidal Disruption
  Events}},}\ }\href {\doibase 10.3847/1538-4357/abb3c0} {\bibfield  {journal}
  {\bibinfo  {journal} {Astrophys. J.}\ }\textbf {\bibinfo {volume} {902}},\
  \bibinfo {pages} {108} (\bibinfo {year} {2020}{\natexlab{b}})},\ \Eprint
  {http://arxiv.org/abs/2005.08937} {arXiv:2005.08937 [astro-ph.HE]}
  \BibitemShut {NoStop}%
\bibitem [{\citenamefont {Murase}\ \emph {et~al.}(2013)\citenamefont {Murase},
  \citenamefont {Ahlers},\ and\ \citenamefont {Lacki}}]{Murase:2013rfa}%
  \BibitemOpen
  \bibfield  {author} {\bibinfo {author} {\bibfnamefont {Kohta}\ \bibnamefont
  {Murase}}, \bibinfo {author} {\bibfnamefont {Markus}\ \bibnamefont {Ahlers}},
  \ and\ \bibinfo {author} {\bibfnamefont {Brian~C.}\ \bibnamefont {Lacki}},\
  }\bibfield  {title} {\enquote {\bibinfo {title} {{Testing the Hadronuclear
  Origin of PeV Neutrinos Observed with IceCube}},}\ }\href {\doibase
  10.1103/PhysRevD.88.121301} {\bibfield  {journal} {\bibinfo  {journal}
  {Phys.Rev.}\ }\textbf {\bibinfo {volume} {D88}},\ \bibinfo {pages} {121301}
  (\bibinfo {year} {2013})},\ \Eprint {http://arxiv.org/abs/1306.3417}
  {arXiv:1306.3417 [astro-ph.HE]} \BibitemShut {NoStop}%
\bibitem [{\citenamefont {Zhang}\ and\ \citenamefont
  {Murase}(2023)}]{Zhang:2023ewt}%
  \BibitemOpen
  \bibfield  {author} {\bibinfo {author} {\bibfnamefont {B.~Theodore}\
  \bibnamefont {Zhang}}\ and\ \bibinfo {author} {\bibfnamefont {Kohta}\
  \bibnamefont {Murase}},\ }\bibfield  {title} {\enquote {\bibinfo {title}
  {{Nuclear and electromagnetic cascades induced by ultra-high-energy cosmic
  rays in radio galaxies: implications for Centaurus A}},}\ }\href {\doibase
  10.1093/mnras/stad1829} {\bibfield  {journal} {\bibinfo  {journal} {Mon. Not.
  Roy. Astron. Soc.}\ }\textbf {\bibinfo {volume} {524}},\ \bibinfo {pages}
  {76--89} (\bibinfo {year} {2023})},\ \Eprint
  {http://arxiv.org/abs/2302.14048} {arXiv:2302.14048 [astro-ph.HE]}
  \BibitemShut {NoStop}%
\bibitem [{\citenamefont {Murase}(2024)}]{Murase:2023chr}%
  \BibitemOpen
  \bibfield  {author} {\bibinfo {author} {\bibfnamefont {Kohta}\ \bibnamefont
  {Murase}},\ }\bibfield  {title} {\enquote {\bibinfo {title} {{Interacting
  supernovae as high-energy multimessenger transients}},}\ }\href {\doibase
  10.1103/PhysRevD.109.103020} {\bibfield  {journal} {\bibinfo  {journal}
  {Phys. Rev. D}\ }\textbf {\bibinfo {volume} {109}},\ \bibinfo {pages}
  {103020} (\bibinfo {year} {2024})},\ \Eprint
  {http://arxiv.org/abs/2312.17239} {arXiv:2312.17239 [astro-ph.HE]}
  \BibitemShut {NoStop}%
\bibitem [{\citenamefont {Hasinger}\ \emph {et~al.}(2005)\citenamefont
  {Hasinger}, \citenamefont {Miyaji},\ and\ \citenamefont
  {Schmidt}}]{Hasinger:2005sb}%
  \BibitemOpen
  \bibfield  {author} {\bibinfo {author} {\bibfnamefont {Gunther}\ \bibnamefont
  {Hasinger}}, \bibinfo {author} {\bibfnamefont {Takamitsu}\ \bibnamefont
  {Miyaji}}, \ and\ \bibinfo {author} {\bibfnamefont {Maarten}\ \bibnamefont
  {Schmidt}},\ }\bibfield  {title} {\enquote {\bibinfo {title}
  {{Luminosity-dependent evolution of soft x-ray selected AGN: New Chandra and
  XMM-Newton surveys}},}\ }\href {\doibase 10.1051/0004-6361:20042134}
  {\bibfield  {journal} {\bibinfo  {journal} {Astron. Astrophys.}\ }\textbf
  {\bibinfo {volume} {441}},\ \bibinfo {pages} {417--434} (\bibinfo {year}
  {2005})},\ \Eprint {http://arxiv.org/abs/astro-ph/0506118}
  {arXiv:astro-ph/0506118} \BibitemShut {NoStop}%
\bibitem [{\citenamefont {Murase}\ and\ \citenamefont
  {Waxman}(2016)}]{Murase:2016gly}%
  \BibitemOpen
  \bibfield  {author} {\bibinfo {author} {\bibfnamefont {Kohta}\ \bibnamefont
  {Murase}}\ and\ \bibinfo {author} {\bibfnamefont {Eli}\ \bibnamefont
  {Waxman}},\ }\bibfield  {title} {\enquote {\bibinfo {title} {{Constraining
  High-Energy Cosmic Neutrino Sources: Implications and Prospects}},}\ }\href
  {\doibase 10.1103/PhysRevD.94.103006} {\bibfield  {journal} {\bibinfo
  {journal} {Phys. Rev.}\ }\textbf {\bibinfo {volume} {D94}},\ \bibinfo {pages}
  {103006} (\bibinfo {year} {2016})},\ \Eprint
  {http://arxiv.org/abs/1607.01601} {arXiv:1607.01601 [astro-ph.HE]}
  \BibitemShut {NoStop}%
\bibitem [{\citenamefont {{Zou}}\ \emph {et~al.}(2024)\citenamefont {{Zou}},
  \citenamefont {{Yu}}, \citenamefont {{Brandt}}, \citenamefont {{Tak}},
  \citenamefont {{Yang}},\ and\ \citenamefont {{Ni}}}]{Zou+2024}%
  \BibitemOpen
  \bibfield  {author} {\bibinfo {author} {\bibfnamefont {Fan}\ \bibnamefont
  {{Zou}}}, \bibinfo {author} {\bibfnamefont {Zhibo}\ \bibnamefont {{Yu}}},
  \bibinfo {author} {\bibfnamefont {W.~N.}\ \bibnamefont {{Brandt}}}, \bibinfo
  {author} {\bibfnamefont {Hyungsuk}\ \bibnamefont {{Tak}}}, \bibinfo {author}
  {\bibfnamefont {Guang}\ \bibnamefont {{Yang}}}, \ and\ \bibinfo {author}
  {\bibfnamefont {Qingling}\ \bibnamefont {{Ni}}},\ }\bibfield  {title}
  {\enquote {\bibinfo {title} {{Mapping the Growth of Supermassive Black Holes
  as a Function of Galaxy Stellar Mass and Redshift}},}\ }\href {\doibase
  10.3847/1538-4357/ad27cc} {\bibfield  {journal} {\bibinfo  {journal} {\apj}\
  }\textbf {\bibinfo {volume} {964}},\ \bibinfo {eid} {183} (\bibinfo {year}
  {2024})},\ \Eprint {http://arxiv.org/abs/2404.00097} {arXiv:2404.00097
  [astro-ph.GA]} \BibitemShut {NoStop}%
\bibitem [{\citenamefont {Murase}(2008)}]{Murase:2008sp}%
  \BibitemOpen
  \bibfield  {author} {\bibinfo {author} {\bibfnamefont {Kohta}\ \bibnamefont
  {Murase}},\ }\bibfield  {title} {\enquote {\bibinfo {title} {{Prompt
  High-Energy Neutrinos from Gamma-Ray Bursts in the Photospheric and
  Synchrotron Self-Compton Scenarios}},}\ }\href {\doibase
  10.1103/PhysRevD.78.101302} {\bibfield  {journal} {\bibinfo  {journal}
  {Phys.Rev.}\ }\textbf {\bibinfo {volume} {D78}},\ \bibinfo {pages} {101302}
  (\bibinfo {year} {2008})},\ \Eprint {http://arxiv.org/abs/0807.0919}
  {arXiv:0807.0919 [astro-ph]} \BibitemShut {NoStop}%
\bibitem [{\citenamefont {Abbasi}\ \emph {et~al.}(2021)\citenamefont {Abbasi}
  \emph {et~al.}}]{IceCube:2020wum}%
  \BibitemOpen
  \bibfield  {author} {\bibinfo {author} {\bibfnamefont {R.}~\bibnamefont
  {Abbasi}} \emph {et~al.} (\bibinfo {collaboration} {IceCube Collaboration}),\
  }\bibfield  {title} {\enquote {\bibinfo {title} {{The IceCube high-energy
  starting event sample: Description and flux characterization with 7.5 years
  of data}},}\ }\href {\doibase 10.1103/PhysRevD.104.022002} {\bibfield
  {journal} {\bibinfo  {journal} {Phys. Rev. D}\ }\textbf {\bibinfo {volume}
  {104}},\ \bibinfo {pages} {022002} (\bibinfo {year} {2021})},\ \Eprint
  {http://arxiv.org/abs/2011.03545} {arXiv:2011.03545 [astro-ph.HE]}
  \BibitemShut {NoStop}%
\bibitem [{\citenamefont {Murase}\ \emph {et~al.}(2008)\citenamefont {Murase},
  \citenamefont {Inoue},\ and\ \citenamefont {Nagataki}}]{Murase:2008yt}%
  \BibitemOpen
  \bibfield  {author} {\bibinfo {author} {\bibfnamefont {Kohta}\ \bibnamefont
  {Murase}}, \bibinfo {author} {\bibfnamefont {Susumu}\ \bibnamefont {Inoue}},
  \ and\ \bibinfo {author} {\bibfnamefont {Shigehiro}\ \bibnamefont
  {Nagataki}},\ }\bibfield  {title} {\enquote {\bibinfo {title} {{Cosmic Rays
  Above the Second Knee from Clusters of Galaxies and Associated High-Energy
  Neutrino Emission}},}\ }\href {\doibase 10.1086/595882} {\bibfield  {journal}
  {\bibinfo  {journal} {Astrophys.J.}\ }\textbf {\bibinfo {volume} {689}},\
  \bibinfo {pages} {L105} (\bibinfo {year} {2008})},\ \Eprint
  {http://arxiv.org/abs/0805.0104} {arXiv:0805.0104 [astro-ph]} \BibitemShut
  {NoStop}%
\bibitem [{\citenamefont {Kotera}\ \emph {et~al.}(2009)\citenamefont {Kotera},
  \citenamefont {Allard}, \citenamefont {Murase}, \citenamefont {Aoi},
  \citenamefont {Dubois} \emph {et~al.}}]{Kotera:2009ms}%
  \BibitemOpen
  \bibfield  {author} {\bibinfo {author} {\bibfnamefont {K.}~\bibnamefont
  {Kotera}}, \bibinfo {author} {\bibfnamefont {D.}~\bibnamefont {Allard}},
  \bibinfo {author} {\bibfnamefont {K.}~\bibnamefont {Murase}}, \bibinfo
  {author} {\bibfnamefont {J.}~\bibnamefont {Aoi}}, \bibinfo {author}
  {\bibfnamefont {Y.}~\bibnamefont {Dubois}},  \emph {et~al.},\ }\bibfield
  {title} {\enquote {\bibinfo {title} {{Propagation of ultrahigh energy nuclei
  in clusters of galaxies: resulting composition and secondary emissions}},}\
  }\href {\doibase 10.1088/0004-637X/707/1/370} {\bibfield  {journal} {\bibinfo
   {journal} {Astrophys.J.}\ }\textbf {\bibinfo {volume} {707}},\ \bibinfo
  {pages} {370--386} (\bibinfo {year} {2009})},\ \Eprint
  {http://arxiv.org/abs/0907.2433} {arXiv:0907.2433 [astro-ph.HE]} \BibitemShut
  {NoStop}%
\bibitem [{\citenamefont {Anchordoqui}\ \emph {et~al.}(2021)\citenamefont
  {Anchordoqui}, \citenamefont {Krizmanic},\ and\ \citenamefont
  {Stecker}}]{Anchordoqui:2021vms}%
  \BibitemOpen
  \bibfield  {author} {\bibinfo {author} {\bibfnamefont {Luis~A.}\ \bibnamefont
  {Anchordoqui}}, \bibinfo {author} {\bibfnamefont {John~F.}\ \bibnamefont
  {Krizmanic}}, \ and\ \bibinfo {author} {\bibfnamefont {Floyd~W.}\
  \bibnamefont {Stecker}},\ }\bibfield  {title} {\enquote {\bibinfo {title}
  {{High-Energy Neutrinos from NGC 1068}},}\ }\href {\doibase
  10.22323/1.395.0993} {\bibfield  {journal} {\bibinfo  {journal} {PoS}\
  }\textbf {\bibinfo {volume} {ICRC2021}},\ \bibinfo {pages} {993} (\bibinfo
  {year} {2021})},\ \Eprint {http://arxiv.org/abs/2102.12409} {arXiv:2102.12409
  [astro-ph.HE]} \BibitemShut {NoStop}%
\bibitem [{\citenamefont {Eichmann}\ \emph {et~al.}(2022)\citenamefont
  {Eichmann}, \citenamefont {Oikonomou}, \citenamefont {Salvatore},
  \citenamefont {Dettmar},\ and\ \citenamefont
  {Becker~Tjus}}]{Eichmann:2022lxh}%
  \BibitemOpen
  \bibfield  {author} {\bibinfo {author} {\bibfnamefont {Bj\"orn}\ \bibnamefont
  {Eichmann}}, \bibinfo {author} {\bibfnamefont {Foteini}\ \bibnamefont
  {Oikonomou}}, \bibinfo {author} {\bibfnamefont {Silvia}\ \bibnamefont
  {Salvatore}}, \bibinfo {author} {\bibfnamefont {Ralf-J\"urgen}\ \bibnamefont
  {Dettmar}}, \ and\ \bibinfo {author} {\bibfnamefont {Julia}\ \bibnamefont
  {Becker~Tjus}},\ }\bibfield  {title} {\enquote {\bibinfo {title} {{Solving
  the Multimessenger Puzzle of the AGN-starburst Composite Galaxy NGC 1068}},}\
  }\href {\doibase 10.3847/1538-4357/ac9588} {\bibfield  {journal} {\bibinfo
  {journal} {Astrophys. J.}\ }\textbf {\bibinfo {volume} {939}},\ \bibinfo
  {pages} {43} (\bibinfo {year} {2022})},\ \Eprint
  {http://arxiv.org/abs/2207.00102} {arXiv:2207.00102 [astro-ph.HE]}
  \BibitemShut {NoStop}%
\bibitem [{\citenamefont {Saurenhaus}\ \emph {et~al.}(2026)\citenamefont
  {Saurenhaus}, \citenamefont {Capel}, \citenamefont {Oikonomou},\ and\
  \citenamefont {Buchner}}]{Saurenhaus:2025ysu}%
  \BibitemOpen
  \bibfield  {author} {\bibinfo {author} {\bibfnamefont {Lena}\ \bibnamefont
  {Saurenhaus}}, \bibinfo {author} {\bibfnamefont {Francesca}\ \bibnamefont
  {Capel}}, \bibinfo {author} {\bibfnamefont {Foteini}\ \bibnamefont
  {Oikonomou}}, \ and\ \bibinfo {author} {\bibfnamefont {Johannes}\
  \bibnamefont {Buchner}},\ }\bibfield  {title} {\enquote {\bibinfo {title}
  {{Constraining the contribution of Seyfert galaxies to the diffuse neutrino
  flux in light of point source observations}},}\ }\href {\doibase
  10.1103/f66p-k6z9} {\bibfield  {journal} {\bibinfo  {journal} {Phys. Rev. D}\
  }\textbf {\bibinfo {volume} {113}},\ \bibinfo {pages} {023019} (\bibinfo
  {year} {2026})},\ \Eprint {http://arxiv.org/abs/2507.06110} {arXiv:2507.06110
  [astro-ph.HE]} \BibitemShut {NoStop}%
\bibitem [{\citenamefont {Carpio}\ \emph {et~al.}(2026)\citenamefont {Carpio},
  \citenamefont {Kheirandish},\ and\ \citenamefont {Murase}}]{Carpio:2026}%
  \BibitemOpen
  \bibfield  {author} {\bibinfo {author} {\bibfnamefont {Jose~A.}\ \bibnamefont
  {Carpio}}, \bibinfo {author} {\bibfnamefont {Ali}\ \bibnamefont
  {Kheirandish}}, \ and\ \bibinfo {author} {\bibfnamefont {Kohta}\ \bibnamefont
  {Murase}},\ }\href@noop {} {\bibfield  {journal} {\bibinfo  {journal}
  {Astrophysical Journal, submitted}\ } (\bibinfo {year} {2026})}\BibitemShut
  {NoStop}%
\bibitem [{\citenamefont {Blanco}\ \emph {et~al.}(2025)\citenamefont {Blanco},
  \citenamefont {Hooper}, \citenamefont {Linden},\ and\ \citenamefont
  {Pinetti}}]{Blanco:2023dfp}%
  \BibitemOpen
  \bibfield  {author} {\bibinfo {author} {\bibfnamefont {Carlos}\ \bibnamefont
  {Blanco}}, \bibinfo {author} {\bibfnamefont {Dan}\ \bibnamefont {Hooper}},
  \bibinfo {author} {\bibfnamefont {Tim}\ \bibnamefont {Linden}}, \ and\
  \bibinfo {author} {\bibfnamefont {Elena}\ \bibnamefont {Pinetti}},\
  }\bibfield  {title} {\enquote {\bibinfo {title} {{Neutrino and gamma-ray
  emissions from NGC 1068}},}\ }\href {\doibase 10.1103/wnjh-7nwp} {\bibfield
  {journal} {\bibinfo  {journal} {Phys. Rev. D}\ }\textbf {\bibinfo {volume}
  {112}},\ \bibinfo {pages} {123016} (\bibinfo {year} {2025})},\ \Eprint
  {http://arxiv.org/abs/2307.03259} {arXiv:2307.03259 [astro-ph.HE]}
  \BibitemShut {NoStop}%
\bibitem [{\citenamefont {Padovani}\ \emph
  {et~al.}(2024{\natexlab{b}})\citenamefont {Padovani}, \citenamefont {Gilli},
  \citenamefont {Resconi}, \citenamefont {Bellenghi},\ and\ \citenamefont
  {Henningsen}}]{Padovani:2024tgx}%
  \BibitemOpen
  \bibfield  {author} {\bibinfo {author} {\bibfnamefont {P.}~\bibnamefont
  {Padovani}}, \bibinfo {author} {\bibfnamefont {R.}~\bibnamefont {Gilli}},
  \bibinfo {author} {\bibfnamefont {E.}~\bibnamefont {Resconi}}, \bibinfo
  {author} {\bibfnamefont {C.}~\bibnamefont {Bellenghi}}, \ and\ \bibinfo
  {author} {\bibfnamefont {F.}~\bibnamefont {Henningsen}},\ }\bibfield  {title}
  {\enquote {\bibinfo {title} {{The neutrino background from non-jetted active
  galactic nuclei}},}\ }\href {\doibase 10.1051/0004-6361/202450025} {\bibfield
   {journal} {\bibinfo  {journal} {Astron. Astrophys.}\ }\textbf {\bibinfo
  {volume} {684}},\ \bibinfo {pages} {L21} (\bibinfo {year}
  {2024}{\natexlab{b}})},\ \Eprint {http://arxiv.org/abs/2404.05690}
  {arXiv:2404.05690 [astro-ph.HE]} \BibitemShut {NoStop}%
\bibitem [{\citenamefont {Fiorillo}\ \emph {et~al.}(2025)\citenamefont
  {Fiorillo}, \citenamefont {Comisso}, \citenamefont {Peretti}, \citenamefont
  {Petropoulou},\ and\ \citenamefont {Sironi}}]{Fiorillo:2025ehn}%
  \BibitemOpen
  \bibfield  {author} {\bibinfo {author} {\bibfnamefont {Damiano F.~G.}\
  \bibnamefont {Fiorillo}}, \bibinfo {author} {\bibfnamefont {Luca}\
  \bibnamefont {Comisso}}, \bibinfo {author} {\bibfnamefont {Enrico}\
  \bibnamefont {Peretti}}, \bibinfo {author} {\bibfnamefont {Maria}\
  \bibnamefont {Petropoulou}}, \ and\ \bibinfo {author} {\bibfnamefont
  {Lorenzo}\ \bibnamefont {Sironi}},\ }\bibfield  {title} {\enquote {\bibinfo
  {title} {{The Contribution of Turbulent Active Galactic Nucleus Coronae to
  the Diffuse Neutrino Flux}},}\ }\href {\doibase 10.3847/1538-4357/adec9c}
  {\bibfield  {journal} {\bibinfo  {journal} {Astrophys. J.}\ }\textbf
  {\bibinfo {volume} {989}},\ \bibinfo {pages} {215} (\bibinfo {year}
  {2025})},\ \Eprint {http://arxiv.org/abs/2504.06336} {arXiv:2504.06336
  [astro-ph.HE]} \BibitemShut {NoStop}%
\bibitem [{\citenamefont {Caputo}\ \emph {et~al.}(2022)\citenamefont {Caputo}
  \emph {et~al.}}]{Caputo:2022xpx}%
  \BibitemOpen
  \bibfield  {author} {\bibinfo {author} {\bibfnamefont {Regina}\ \bibnamefont
  {Caputo}} \emph {et~al.},\ }\bibfield  {title} {\enquote {\bibinfo {title}
  {{All-sky Medium Energy Gamma-ray Observatory eXplorer mission concept}},}\
  }\href {\doibase 10.1117/1.JATIS.8.4.044003} {\bibfield  {journal} {\bibinfo
  {journal} {J. Astron. Telesc. Instrum. Syst.}\ }\textbf {\bibinfo {volume}
  {8}},\ \bibinfo {pages} {044003} (\bibinfo {year} {2022})},\ \Eprint
  {http://arxiv.org/abs/2208.04990} {arXiv:2208.04990 [astro-ph.IM]}
  \BibitemShut {NoStop}%
\bibitem [{\citenamefont {De~Angelis}\ \emph {et~al.}(2017)\citenamefont
  {De~Angelis} \emph {et~al.}}]{e-ASTROGAM:2016bph}%
  \BibitemOpen
  \bibfield  {author} {\bibinfo {author} {\bibfnamefont {A.}~\bibnamefont
  {De~Angelis}} \emph {et~al.} (\bibinfo {collaboration} {e-ASTROGAM}),\
  }\bibfield  {title} {\enquote {\bibinfo {title} {{The e-ASTROGAM mission}},}\
  }\href {\doibase 10.1007/s10686-017-9533-6} {\bibfield  {journal} {\bibinfo
  {journal} {Exper. Astron.}\ }\textbf {\bibinfo {volume} {44}},\ \bibinfo
  {pages} {25--82} (\bibinfo {year} {2017})},\ \Eprint
  {http://arxiv.org/abs/1611.02232} {arXiv:1611.02232 [astro-ph.HE]}
  \BibitemShut {NoStop}%
\bibitem [{\citenamefont {Aramaki}\ \emph {et~al.}(2020)\citenamefont
  {Aramaki}, \citenamefont {Hansson~Adrian}, \citenamefont {Karagiorgi},\ and\
  \citenamefont {Odaka}}]{Aramaki:2019bpi}%
  \BibitemOpen
  \bibfield  {author} {\bibinfo {author} {\bibfnamefont {Tsuguo}\ \bibnamefont
  {Aramaki}}, \bibinfo {author} {\bibfnamefont {Per}\ \bibnamefont
  {Hansson~Adrian}}, \bibinfo {author} {\bibfnamefont {Georgia}\ \bibnamefont
  {Karagiorgi}}, \ and\ \bibinfo {author} {\bibfnamefont {Hirokazu}\
  \bibnamefont {Odaka}},\ }\bibfield  {title} {\enquote {\bibinfo {title}
  {{Dual MeV Gamma-Ray and Dark Matter Observatory - GRAMS Project}},}\ }\href
  {\doibase 10.1016/j.astropartphys.2019.07.002} {\bibfield  {journal}
  {\bibinfo  {journal} {Astropart. Phys.}\ }\textbf {\bibinfo {volume} {114}},\
  \bibinfo {pages} {107--114} (\bibinfo {year} {2020})},\ \Eprint
  {http://arxiv.org/abs/1901.03430} {arXiv:1901.03430 [astro-ph.HE]}
  \BibitemShut {NoStop}%
\bibitem [{\citenamefont {Murase}\ \emph
  {et~al.}(2012{\natexlab{b}})\citenamefont {Murase}, \citenamefont {Beacom},\
  and\ \citenamefont {Takami}}]{Murase:2012df}%
  \BibitemOpen
  \bibfield  {author} {\bibinfo {author} {\bibfnamefont {Kohta}\ \bibnamefont
  {Murase}}, \bibinfo {author} {\bibfnamefont {John~F.}\ \bibnamefont
  {Beacom}}, \ and\ \bibinfo {author} {\bibfnamefont {Hajime}\ \bibnamefont
  {Takami}},\ }\bibfield  {title} {\enquote {\bibinfo {title} {{Gamma-Ray and
  Neutrino Backgrounds as Probes of the High-Energy Universe: Hints of
  Cascades, General Constraints, and Implications for TeV Searches}},}\ }\href
  {\doibase 10.1088/1475-7516/2012/08/030} {\bibfield  {journal} {\bibinfo
  {journal} {JCAP}\ }\textbf {\bibinfo {volume} {1208}},\ \bibinfo {pages}
  {030} (\bibinfo {year} {2012}{\natexlab{b}})},\ \Eprint
  {http://arxiv.org/abs/1205.5755} {arXiv:1205.5755 [astro-ph.HE]} \BibitemShut
  {NoStop}%
\bibitem [{\citenamefont {Inoue}\ \emph {et~al.}(2008)\citenamefont {Inoue},
  \citenamefont {Totani},\ and\ \citenamefont {Ueda}}]{Inoue:2007tn}%
  \BibitemOpen
  \bibfield  {author} {\bibinfo {author} {\bibfnamefont {Yoshiyuki}\
  \bibnamefont {Inoue}}, \bibinfo {author} {\bibfnamefont {Tomonori}\
  \bibnamefont {Totani}}, \ and\ \bibinfo {author} {\bibfnamefont {Yoshihiro}\
  \bibnamefont {Ueda}},\ }\bibfield  {title} {\enquote {\bibinfo {title} {{The
  Cosmic MeV Gamma-ray Background and Hard X-ray Spectra of Active Galactic
  Nuclei: Implications for the Origin of Hot AGN Coronae}},}\ }\href {\doibase
  10.1086/525848} {\bibfield  {journal} {\bibinfo  {journal} {Astrophys. J.
  Lett.}\ }\textbf {\bibinfo {volume} {672}},\ \bibinfo {pages} {L5} (\bibinfo
  {year} {2008})},\ \Eprint {http://arxiv.org/abs/0709.3877} {arXiv:0709.3877
  [astro-ph]} \BibitemShut {NoStop}%
\bibitem [{\citenamefont {Marinucci}\ \emph {et~al.}(2016)\citenamefont
  {Marinucci} \emph {et~al.}}]{Marinucci:2015fqo}%
  \BibitemOpen
  \bibfield  {author} {\bibinfo {author} {\bibfnamefont {A.}~\bibnamefont
  {Marinucci}} \emph {et~al.},\ }\bibfield  {title} {\enquote {\bibinfo {title}
  {{NuSTAR catches the unveiling nucleus of NGC 1068}},}\ }\href {\doibase
  10.1093/mnrasl/slv178} {\bibfield  {journal} {\bibinfo  {journal} {Mon. Not.
  Roy. Astron. Soc.}\ }\textbf {\bibinfo {volume} {456}},\ \bibinfo {pages}
  {L94--L98} (\bibinfo {year} {2016})},\ \Eprint
  {http://arxiv.org/abs/1511.03503} {arXiv:1511.03503 [astro-ph.HE]}
  \BibitemShut {NoStop}%
\bibitem [{\citenamefont {Lodato}\ and\ \citenamefont
  {Bertin}(2003)}]{Lodato:2002cv}%
  \BibitemOpen
  \bibfield  {author} {\bibinfo {author} {\bibfnamefont {Giuseppe}\
  \bibnamefont {Lodato}}\ and\ \bibinfo {author} {\bibfnamefont
  {G.}~\bibnamefont {Bertin}},\ }\bibfield  {title} {\enquote {\bibinfo {title}
  {{Non-Keplerian rotation in the nucleus of NGC 1068: Evidence for a massive
  accretion disk?}}}\ }\href {\doibase 10.1051/0004-6361:20021672} {\bibfield
  {journal} {\bibinfo  {journal} {Astron. Astrophys.}\ }\textbf {\bibinfo
  {volume} {398}},\ \bibinfo {pages} {517--524} (\bibinfo {year} {2003})},\
  \Eprint {http://arxiv.org/abs/astro-ph/0211113} {arXiv:astro-ph/0211113}
  \BibitemShut {NoStop}%
\bibitem [{\citenamefont {Yang}\ \emph {et~al.}(2001)\citenamefont {Yang},
  \citenamefont {Wilson},\ and\ \citenamefont {Ferruit}}]{Yang:2001ce}%
  \BibitemOpen
  \bibfield  {author} {\bibinfo {author} {\bibfnamefont {Y.}~\bibnamefont
  {Yang}}, \bibinfo {author} {\bibfnamefont {A.~S.}\ \bibnamefont {Wilson}}, \
  and\ \bibinfo {author} {\bibfnamefont {P.}~\bibnamefont {Ferruit}},\
  }\bibfield  {title} {\enquote {\bibinfo {title} {{Chandra x-ray observation
  of ngc 4151}},}\ }\href {\doibase 10.1086/323693} {\bibfield  {journal}
  {\bibinfo  {journal} {Astrophys. J.}\ }\textbf {\bibinfo {volume} {563}},\
  \bibinfo {pages} {124} (\bibinfo {year} {2001})},\ \Eprint
  {http://arxiv.org/abs/astro-ph/0108166} {arXiv:astro-ph/0108166} \BibitemShut
  {NoStop}%
\bibitem [{\citenamefont {Gianolli}\ \emph {et~al.}(2023)\citenamefont
  {Gianolli} \emph {et~al.}}]{Gianolli:2023zji}%
  \BibitemOpen
  \bibfield  {author} {\bibinfo {author} {\bibfnamefont {V.~E.}\ \bibnamefont
  {Gianolli}} \emph {et~al.},\ }\bibfield  {title} {\enquote {\bibinfo {title}
  {{Uncovering the geometry of the hot X-ray corona in the Seyfert galaxy NGC
  4151 with IXPE}},}\ }\href {\doibase 10.1093/mnras/stad1697} {\bibfield
  {journal} {\bibinfo  {journal} {Mon. Not. Roy. Astron. Soc.}\ }\textbf
  {\bibinfo {volume} {523}},\ \bibinfo {pages} {4468--4476} (\bibinfo {year}
  {2023})},\ \Eprint {http://arxiv.org/abs/2303.12541} {arXiv:2303.12541
  [astro-ph.GA]} \BibitemShut {NoStop}%
\bibitem [{\citenamefont {{Bentz}}\ \emph {et~al.}(2022)\citenamefont
  {{Bentz}}, \citenamefont {{Williams}},\ and\ \citenamefont
  {{Treu}}}]{Bentz+2022}%
  \BibitemOpen
  \bibfield  {author} {\bibinfo {author} {\bibfnamefont {Misty~C.}\
  \bibnamefont {{Bentz}}}, \bibinfo {author} {\bibfnamefont {Peter~R.}\
  \bibnamefont {{Williams}}}, \ and\ \bibinfo {author} {\bibfnamefont
  {Tommaso}\ \bibnamefont {{Treu}}},\ }\bibfield  {title} {\enquote {\bibinfo
  {title} {{The Broad Line Region and Black Hole Mass of NGC 4151}},}\ }\href
  {\doibase 10.3847/1538-4357/ac7c0a} {\bibfield  {journal} {\bibinfo
  {journal} {\apj}\ }\textbf {\bibinfo {volume} {934}},\ \bibinfo {eid} {168}
  (\bibinfo {year} {2022})},\ \Eprint {http://arxiv.org/abs/2206.03513}
  {arXiv:2206.03513 [astro-ph.GA]} \BibitemShut {NoStop}%
\bibitem [{\citenamefont {Kormendy}\ and\ \citenamefont
  {Ho}(2013)}]{Kormendy:2013dxa}%
  \BibitemOpen
  \bibfield  {author} {\bibinfo {author} {\bibfnamefont {John}\ \bibnamefont
  {Kormendy}}\ and\ \bibinfo {author} {\bibfnamefont {Luis~C.}\ \bibnamefont
  {Ho}},\ }\bibfield  {title} {\enquote {\bibinfo {title} {{Coevolution (Or
  Not) of Supermassive Black Holes and Host Galaxies}},}\ }\href {\doibase
  10.1146/annurev-astro-082708-101811} {\bibfield  {journal} {\bibinfo
  {journal} {Ann. Rev. Astron. Astrophys.}\ }\textbf {\bibinfo {volume} {51}},\
  \bibinfo {pages} {511--653} (\bibinfo {year} {2013})},\ \Eprint
  {http://arxiv.org/abs/1304.7762} {arXiv:1304.7762 [astro-ph.CO]} \BibitemShut
  {NoStop}%
\bibitem [{\citenamefont {{Oknyanskij}}\ and\ \citenamefont
  {{Lyuty}}(2007)}]{2007OAP....20..160O}%
  \BibitemOpen
  \bibfield  {author} {\bibinfo {author} {\bibfnamefont {V.~L.}\ \bibnamefont
  {{Oknyanskij}}}\ and\ \bibinfo {author} {\bibfnamefont {V.~M.}\ \bibnamefont
  {{Lyuty}}},\ }\bibfield  {title} {\enquote {\bibinfo {title} {{Optical
  Variability of NGC 4151 during 100 Years}},}\ }\href@noop {} {\bibfield
  {journal} {\bibinfo  {journal} {Odessa Astronomical Publications}\ }\textbf
  {\bibinfo {volume} {20}},\ \bibinfo {pages} {160} (\bibinfo {year}
  {2007})}\BibitemShut {NoStop}%
\bibitem [{\citenamefont {Hankla}\ \emph {et~al.}(2026)\citenamefont {Hankla},
  \citenamefont {Philippov}, \citenamefont {Mbarek}, \citenamefont {Mushotzky},
  \citenamefont {Musoke}, \citenamefont {Gro{\v{s}}elj},\ and\ \citenamefont
  {Liska}}]{Hankla:2025bqc}%
  \BibitemOpen
  \bibfield  {author} {\bibinfo {author} {\bibfnamefont {Amelia~M.}\
  \bibnamefont {Hankla}}, \bibinfo {author} {\bibfnamefont {Alexander}\
  \bibnamefont {Philippov}}, \bibinfo {author} {\bibfnamefont {Rostom}\
  \bibnamefont {Mbarek}}, \bibinfo {author} {\bibfnamefont {Richard~F.}\
  \bibnamefont {Mushotzky}}, \bibinfo {author} {\bibfnamefont {G.}~\bibnamefont
  {Musoke}}, \bibinfo {author} {\bibfnamefont {Daniel}\ \bibnamefont
  {Gro{\v{s}}elj}}, \ and\ \bibinfo {author} {\bibfnamefont {Matthew}\
  \bibnamefont {Liska}},\ }\bibfield  {title} {\enquote {\bibinfo {title} {{An
  Outflow from the X-Ray Corona as the Origin of Millimeter Emission from
  Radio-quiet AGNs}},}\ }\href {\doibase 10.3847/1538-4357/ae2478} {\bibfield
  {journal} {\bibinfo  {journal} {Astrophys. J.}\ }\textbf {\bibinfo {volume}
  {997}},\ \bibinfo {pages} {224} (\bibinfo {year} {2026})},\ \Eprint
  {http://arxiv.org/abs/2512.01662} {arXiv:2512.01662 [astro-ph.HE]}
  \BibitemShut {NoStop}%
\bibitem [{\citenamefont {Kun}\ \emph {et~al.}(2024)\citenamefont {Kun},
  \citenamefont {Bartos}, \citenamefont {Becker~Tjus}, \citenamefont
  {Biermann}, \citenamefont {Franckowiak}, \citenamefont {Halzen},
  \citenamefont {del Palacio},\ and\ \citenamefont {Woo}}]{Kun:2024meq}%
  \BibitemOpen
  \bibfield  {author} {\bibinfo {author} {\bibfnamefont {Emma}\ \bibnamefont
  {Kun}}, \bibinfo {author} {\bibfnamefont {Imre}\ \bibnamefont {Bartos}},
  \bibinfo {author} {\bibfnamefont {Julia}\ \bibnamefont {Becker~Tjus}},
  \bibinfo {author} {\bibfnamefont {Peter~L.}\ \bibnamefont {Biermann}},
  \bibinfo {author} {\bibfnamefont {Anna}\ \bibnamefont {Franckowiak}},
  \bibinfo {author} {\bibfnamefont {Francis}\ \bibnamefont {Halzen}}, \bibinfo
  {author} {\bibfnamefont {Santiago}\ \bibnamefont {del Palacio}}, \ and\
  \bibinfo {author} {\bibfnamefont {Jooyun}\ \bibnamefont {Woo}},\ }\bibfield
  {title} {\enquote {\bibinfo {title} {{Possible correlation between unabsorbed
  hard x rays and neutrinos in radio-loud and radio-quiet active galactic
  nuclei}},}\ }\href {\doibase 10.1103/PhysRevD.110.123014} {\bibfield
  {journal} {\bibinfo  {journal} {Phys. Rev. D}\ }\textbf {\bibinfo {volume}
  {110}},\ \bibinfo {pages} {123014} (\bibinfo {year} {2024})},\ \Eprint
  {http://arxiv.org/abs/2404.06867} {arXiv:2404.06867 [astro-ph.HE]}
  \BibitemShut {NoStop}%
\bibitem [{\citenamefont {Ambrosone}(2024)}]{Ambrosone:2024zrf}%
  \BibitemOpen
  \bibfield  {author} {\bibinfo {author} {\bibfnamefont {Antonio}\ \bibnamefont
  {Ambrosone}},\ }\bibfield  {title} {\enquote {\bibinfo {title} {{Berezinsky
  hidden sources: an emergent tension in the high-energy neutrino sky?}}}\
  }\href {\doibase 10.1088/1475-7516/2024/09/075} {\bibfield  {journal}
  {\bibinfo  {journal} {JCAP}\ }\textbf {\bibinfo {volume} {09}},\ \bibinfo
  {pages} {075} (\bibinfo {year} {2024})},\ \Eprint
  {http://arxiv.org/abs/2406.13336} {arXiv:2406.13336 [astro-ph.HE]}
  \BibitemShut {NoStop}%
\bibitem [{\citenamefont {Aiello}\ \emph {et~al.}(2019)\citenamefont {Aiello}
  \emph {et~al.}}]{KM3NeT:2018wnd}%
  \BibitemOpen
  \bibfield  {author} {\bibinfo {author} {\bibfnamefont {S.}~\bibnamefont
  {Aiello}} \emph {et~al.} (\bibinfo {collaboration} {KM3NeT}),\ }\bibfield
  {title} {\enquote {\bibinfo {title} {{Sensitivity of the KM3NeT/ARCA neutrino
  telescope to point-like neutrino sources}},}\ }\href {\doibase
  10.1016/j.astropartphys.2019.04.002} {\bibfield  {journal} {\bibinfo
  {journal} {Astropart. Phys.}\ }\textbf {\bibinfo {volume} {111}},\ \bibinfo
  {pages} {100--110} (\bibinfo {year} {2019})},\ \Eprint
  {http://arxiv.org/abs/1810.08499} {arXiv:1810.08499 [astro-ph.HE]}
  \BibitemShut {NoStop}%
\bibitem [{BAS(BASS Project)}]{BASS}%
  \BibitemOpen
  \href@noop {} {\bibfield  {journal} {\bibinfo  {journal}
  {https://www.bass-survey.com}\ } (\bibinfo {year} {BASS
  Project})}\BibitemShut {NoStop}%
\bibitem [{\citenamefont {Wang}\ \emph {et~al.}(2025)\citenamefont {Wang},
  \citenamefont {Hu}, \citenamefont {He},\ and\ \citenamefont
  {Pang}}]{Wang:2025aov}%
  \BibitemOpen
  \bibfield  {author} {\bibinfo {author} {\bibfnamefont {Xing-Jian}\
  \bibnamefont {Wang}}, \bibinfo {author} {\bibfnamefont {Jing-Fu}\
  \bibnamefont {Hu}}, \bibinfo {author} {\bibfnamefont {Hao-Ning}\ \bibnamefont
  {He}}, \ and\ \bibinfo {author} {\bibfnamefont {Cheng-Qun}\ \bibnamefont
  {Pang}},\ }\bibfield  {title} {\enquote {\bibinfo {title} {{Neutrino and
  Cascade Gamma-Ray Emission from Magnetized Turbulent Coronae in Seyfert
  Galaxies}},}\ }\href@noop {} {\  (\bibinfo {year} {2025})},\ \Eprint
  {http://arxiv.org/abs/2511.00388} {arXiv:2511.00388 [astro-ph.HE]}
  \BibitemShut {NoStop}%
\end{thebibliography}%

\end{document}